\documentclass[12pt]{article}%
\pdfoutput=1
\usepackage[nosort]{cite}
\usepackage{graphicx}
\usepackage{multicol}
\usepackage{amsfonts}
\usepackage{amssymb}
\usepackage{amsmath}
\usepackage{dsfont}
\usepackage{heck}
\usepackage{afterpage}
\usepackage{setspace}
\usepackage{verbatim}
\usepackage{color}
\usepackage{longtable}
\usepackage{float}
\usepackage{subcaption}
\usepackage{epsfig}
\usepackage{epstopdf}
\usepackage{adjustbox}
\newsavebox{\mysavebox}

\usepackage{tikz}
\usepackage[margin=1in]{geometry}
\usepackage{titletoc}%
\usepackage{hyperref}
\hypersetup{colorlinks,%
citecolor=black,%
filecolor=black,%
linkcolor=black,%
urlcolor=black,%
pdftex}
\providecommand{\U}[1]{\protect\rule{.1in}{.1in}}
\usetikzlibrary{decorations.markings}
\newcommand{\C}[1]{\mathbb{C}^{#1}}

\numberwithin{equation}{section}

\def\bR{\mathbb{R}}

\newcommand{\ba}{\begin{eqnarray}}
\newcommand{\ea}{\end{eqnarray}}

\newcommand{\cE}{\mathcal{E}}

\DeclareMathOperator{\SU}{\mathit{SU}}
\DeclareMathOperator{\SO}{\mathit{SO}}

\newcommand{\rep}[1]{\mathbf{#1}}
\newcommand{\Tr}{\, {\rm Tr}}

\newcommand{\be}{\begin{equation}}
\newcommand{\ee}{\end{equation}}

\def\rep#1{{{\boldsymbol{#1}}}}

\tikzstyle{startstop} = [rectangle, rounded corners, minimum width=3cm, minimum height=1cm,text centered, draw=black, fill=blue!10]
\tikzstyle{startstop} = [rectangle, rounded corners, minimum width=3cm, minimum height=1cm,text centered, draw=black, fill=blue!10]

\tikzstyle{io} = [trapezium, trapezium left angle=70, trapezium right angle=110, minimum width=3cm, minimum height=1cm, text centered, draw=black, fill=blue!30]

\tikzstyle{process} = [rectangle, minimum width=3cm, minimum height=1cm, text centered, draw=black, fill=orange!30]
\tikzstyle{decision} = [diamond, minimum width=3cm, minimum height=1cm, text centered, draw=black, fill=green!30]

\tikzstyle{arrow} = [thick,->,>=stealth]

\begin{document}

\date{February 2016}

\title{UV Completions for Non-Critical Strings}

\institution{UNC}{\centerline{${}^{1}$Department of Physics, University of North Carolina, Chapel Hill, NC 27599, USA}}

\institution{COLUMBIA}{\centerline{${}^{2}$Department of Physics, Columbia University, New York, NY 10027, USA}}

\institution{CUNY}{\centerline{${}^{3}$CUNY Graduate Center, Initiative for the Theoretical Sciences, New York, NY 10016, USA}}

\institution{JMU}{\centerline{${}^{4}$Department of Physics, James Madison University, Harrisonburg, VA 22807, USA}}

\authors{Fabio Apruzzi\worksat{\UNC, \COLUMBIA, \CUNY}\footnote{e-mail: {\tt fabio.apruzzi@unc.edu}},
Falk Hassler\worksat{\UNC, \COLUMBIA, \CUNY}\footnote{e-mail: {\tt fhassler@unc.edu}},\\[4mm]
Jonathan J. Heckman\worksat{\UNC, \COLUMBIA, \CUNY}\footnote{e-mail: {\tt jheckman@email.unc.edu}},
and Ilarion V. Melnikov\worksat{\JMU}\footnote{e-mail: {\tt melnikix@jmu.edu}}}

\abstract{Compactifications of the physical superstring to two dimensions provide a
general template for realizing 2D conformal field theories
coupled to worldsheet gravity, i.e. non-critical string theories. Motivated
by this observation, in this paper we determine the quasi-topological 8D theory which governs
the vacua of 2D $\mathcal{N}=(0,2)$ gauged
linear sigma models (GLSMs) obtained from compactifications of type I\ and heterotic
strings on a Calabi-Yau fourfold. We also determine the quasi-topological 6D theory governing the 2D vacua of
intersecting 7-branes in compactifications of F-theory on an elliptically fibered
Calabi-Yau fivefold, where matter fields and interaction terms localize on
lower-dimensional subspaces, i.e. defect operators. To cancel anomalies / cancel tadpoles, these GLSMs must couple
to additional chiral sectors, which in some cases do not admit a known description in terms of a
UV GLSM. Additionally, we find that constructing an anomaly free spectrum can sometimes break
supersymmetry due to spacetime filling anti-branes. We also study various canonical
examples such as the standard embedding of heterotic strings on a Calabi-Yau fourfold
and F-theoretic ``rigid clusters'' with no local deformation moduli of the elliptic fibration.}

\maketitle

\tableofcontents

\enlargethispage{\baselineskip}

\setcounter{tocdepth}{2}

\newpage

\section{Introduction \label{sec:INTRO}}

One of the celebrated facts of string theory is that it defines a consistent
theory of quantum gravity in ten target spacetime dimensions.
At the perturbative level, this is a direct consequence of the restrictions imposed
by coupling a two-dimensional conformal field theory (CFT)\ to worldsheet
gravity. Dualities support this picture and also broaden it in certain
respects. For example, the long distance behavior of M-theory is formulated
in eleven dimensions, and in F-theory, there is still a ten-dimensional
spacetime but one which can be phrased in terms of an underlying twelve-dimensional geometry.

Of course, there are many two-dimensional CFTs with a conformal anomaly
different from that required for the critical superstring. The condition of
conformal invariance means that coupling to worldsheet gravity leads to
spacetime profiles for some of the target space fields of the theory,
including non-trivial profiles for the dilaton and various fluxes
\cite{Polyakov:1981rd, Polyakov:1981re}.  This is a theory of
non-critical strings.

From the viewpoint of effective field theory, one actually expects that the
long distance physics of one-dimensional extended objects will always be
governed by an effective theory of long strings. That is to say, at energies
far below that set by the tension, we expect to have a consistent description
in terms of a non-critical string theory. The theory of effective long strings
has developed over several years, see for example \cite{Polchinski:1991ax,
HariDass:2007dpl, Dubovsky:2012sh, Hellerman:2014cba} and references therein.

Potential applications of non-critical string theory include the ambitious task of
understanding string theory on time dependent backgrounds. Another aspect of working in a
super-critical string theory is that the exponential degeneracy in the ground state leads to a
large number of Ramond-Ramond fluxes, which in turn makes it possible to easily engineer de Sitter vacua.
For some examples along these lines, see e.g. \cite{Silverstein:2001xn, Maloney:2002rr, Hellerman:2004zm, Hellerman:2006nx}.

But as a low energy effective theory, significant care must be taken in any such
approach because once we exit the regime of a perturbative $\alpha^{\prime}$
expansion, higher order effects in the non-linear sigma model beta functions
can make any reliable target space interpretation difficult to maintain,
except in special solvable cases such as linear dilaton backgrounds and
\textquotedblleft quintessential\textquotedblright\ variants such as those
pursued in e.g. \cite{Hellerman:2006nx}. Almost inevitably, there is an energy
scale on the worldsheet above which large gradients in the target space
obscure any conventional spacetime interpretation.

In this paper we provide a general proposal for how to ensure a UV\ complete
starting point for such 2D\ effective string theories. Moreover, the breakdown at high energies
will be understood as the regime in which the 2D\ effective
theory grows into a higher-dimensional theory of quantum gravity, which is in turn UV
completed by the physical superstring!

\begin{figure}[t!]%
\centering
\includegraphics[scale = 0.5, trim={0 4cm 0 6cm}]{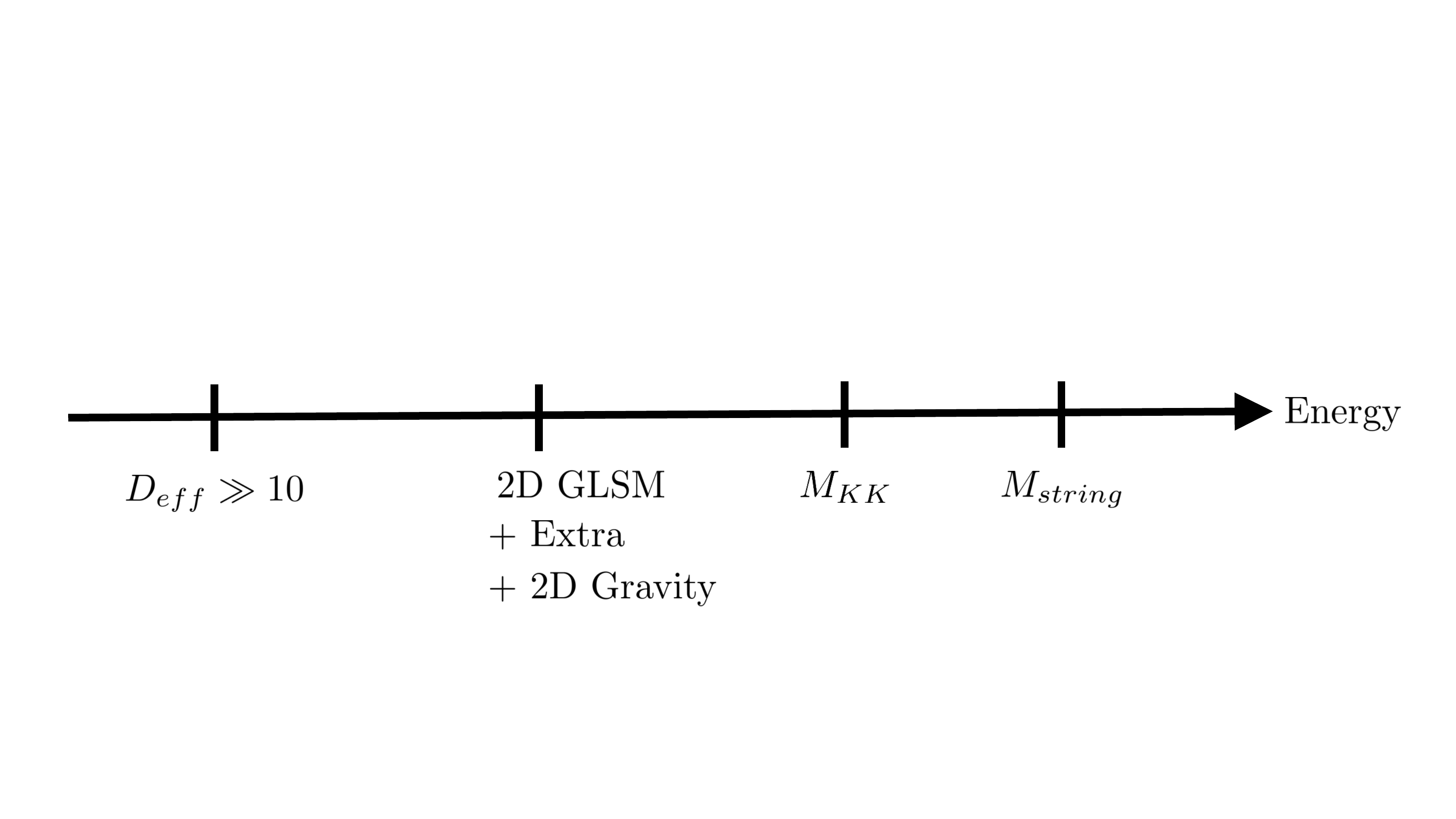}%
\caption{Depiction of energy scales for 2D effective string theories derived from string compactification. In the deep infrared,
we have a 2D conformal fixed point coupled to gravity, leading to an effective string theory. At somewhat higher energy scales, this description passes over to a gauged linear sigma model coupled to extra sectors and gravity, and at even higher energy scales this description
also breaks down and is replaced by a 10D supergravity theory. This is in turn replaced at even higher energies by a corresponding UV completion in string theory.}%
\label{EnergyScales}%
\end{figure}

The basic idea is that we will first begin with a well-known UV\ complete
theory: string theory in ten spacetime dimensions. We shall, however, then
compactify to two dimensions. When decoupled from gravity, this
will provide the basic starting point for a two-dimensional effective quantum field
theory. At low energies, we either enter a gapped phase, or a conformal field theory.
Assuming we flow to a CFT in the IR, coupling to gravity leads to a non-critical
string theory. The important point is that the appearance of a singularity in the
high-momentum behavior of correlators simply tells us that we are exiting the
purely two-dimensional realm, and instead must pass back to the original
worldsheet theory with interpretation in ten spacetime dimensions. See figure
\ref{EnergyScales} for a depiction of the energy scales involved
in the interpretation of our theory. See also reference \cite{Green:1987cw}
for an early discussion of using the low energy limit of string theory to generate
another worldsheet theory.

Aside from these general conceptual motivations to explore the UV\ consistency
of non-critical superstrings, there are additional reasons to be interested in
compactifications of string theory to two dimensions. First of all, in limits
where gravity is decoupled -- as can happen in numerous F-theory
constructions-- we can expect to arrive at a large class of novel
two-dimensional quantum field theories. It is widely expected that general
$(0,2)$ models will be of relevance in a worldsheet formulation of heterotic
flux vacua, though the explicit construction of such models has proven a
remarkably durable obstacle to this programme. Further, we will encounter
particular classes of gauged linear sigma models involving all of the possible
simple gauge groups, including the entire exceptional series. If nothing else,
this provides a much broader arena for constructing candidate vacua.

Additionally, much as in higher dimensional quantum field theories, it is
natural to expect that the geometry of extra dimensions will provide insight
into the strong coupling dynamics of such systems. This has been explored to
some extent in certain cases such as references \cite{Gadde:2013sca,
Kutasov:2013ffl, Franco:2015tna, Franco:2016nwv}. As a final motivation, there is also an
intriguing connection between the supersymmetric quantum mechanics of M-theory
on a (non-singular) K3-fibered Calabi-Yau fivefold and refined Gromov-Witten
invariants of the base Calabi-Yau threefold \cite{Nekrasov:2014nea} (see also \cite{Iqbal:2007ii}).
Owing to the close connection between M-theory on $X$ and F-theory on $S^{1}\times X$,
the effective two-dimensional theories studied here provide an
even further refinement on these general considerations.
The case of M-theory compactified on a smooth Calabi-Yau fivefold was
studied in great detail in reference \cite{Haupt:2008nu}.

With these motivations in mind, our task in this paper will be to lay the
groundwork for all of these potential applications by setting up the general formalism
of string compactification to two dimensions. In particular, we will focus on
the effects of having a non-trivial gauge theory sector, and possibly
additional extra sectors as well. Indeed, to the best of our knowledge, most
of the early literature on string compactification to two dimensions has
focussed on the comparatively simpler class of manifolds with no singularities
and only abelian gauge symmetry. For a guide to this earlier work, see e.g.
\cite{Dasgupta:1996yh, Gukov:1999ya, Gates:2000fj, Font:2004et}.

Compared with these cases, here we expect to have a rich set of quantum field theories with $\mathcal{N} = (0,2)$ supersymmetry
coupled to a 2D $\mathcal{N} = (0,2)$ supergravity theory. Though sharing some similarities
with the structure of the heterotic $\mathcal{N} = 2$ string (see e.g. \cite{Ooguri:1991ie}), there are a few important
differences. For example, generically higher derivative corrections will eliminate any gauged $U(1)$ R-symmetry once we couple to gravity. Additionally, some of the tight constraints typically found in the case of $\mathcal{N} = 2$ strings will be significantly weakened since we shall
only demand that our worldsheet theory make sense as a low energy effective theory.

Now, since part of our aim is to maintain an explicit UV completion of any proposed
2D non-critical string theory, we first treat in detail the cases of
perturbative string theories with a non-abelian gauge theory sector and at
least $(0,2)$ supersymmetry in two dimensions.
This includes compactification on a Calabi-Yau fourfold of the type
I\ superstring, and the heterotic $Spin(32)/%
\mathbb{Z}
_{2}$ and $E_{8}\times E_{8}$ string theories.

For all of the perturbatively realized theories in which we compactify, we
inherit a 2D gauge theory from the dynamics of a spacetime filling 9-brane.
Much as in the case of compactifications in higher dimensions, the low energy
dynamics of this 9-brane is governed by a supersymmetric Yang-Mills theory
wrapped over a Calabi-Yau space. As such, supersymmetric vacua are described
by solutions to an appropriate Hermitian Yang-Mills equation. We determine the
explicit zero mode content for a general supersymmetric background and also
determine the leading order interaction terms for this theory.

Once we proceed to the broader class of non-perturbatively realized vacua, it will prove convenient to
immediately pass to the F-theory formulation of 2D theories where we
compactify on an elliptically fibered Calabi-Yau fivefold. An important aspect
of the latter class of models is that there is typically a limit available
where we decompactify the base of the elliptic model, but some of the
7-branes still wrap compact divisors. This allows us to
decouple our $(0,2)$ quantum field theory sectors from gravity, providing a systematic way to
build up the data of the conformal field theory defined by the intersecting 7-branes of the
compactification.

In our F-theory models there is some geometric localization of the corresponding
zero modes -- They can either descend from bulk modes of a 7-brane or be
localized at the intersection of pairwise intersections of 7-branes.
Additionally, we find that there are
interaction terms localized on subspaces. These can localize on a
K\"ahler threefold, a K\"ahler surface, a Riemann surface and a point. The last
case is somewhat special to two-dimensional theories and comes about from the
intersection of four 7-branes in the compactification. It defines a
quartic interaction term in the two-dimensional effective theory.

In both the heterotic and F-theory constructions, the higher-dimensional theory admits an action
which is supersymmetric on-shell, that is to say, we must impose the equations of motion for the supersymmetry
algebra to fully close. Another aim of our work will be to develop a manifestly off-shell formulation for these theories when treated
as a 2D theory with off-shell $(0,2)$ supersymmetry. In this 2D theory, we explicitly retain all of the Kaluza-Klein modes. This has been successfully carried out for four-dimensional supersymmetric theories, as in reference for 10D Super Yang-Mills theory \cite{Marcus:1983wb} (see also \cite{ArkaniHamed:2001tb}) and in reference
\cite{Beasley:2008dc} for intersecting 7-branes, but as far as we are aware has not been attempted for 2D theories. We find that in the case of the 9-brane action, the 10D Majorana-Weyl spinor constraint can sometimes obstruct the construction of such an off-shell formalism in two dimensions, but that assuming the presence of an additional $\mathbb{Z}_2$ symmetry of the geometry, there is indeed an off-shell formalism for the 9-brane. For intersecting 7-branes in a local F-theory construction, this symmetry is automatically present, and allows us to always construct an off-shell action. An additional benefit of this method of constructing the higher-dimensional theory is that we can then easily read off the zero mode content and leading order interaction terms of the resulting effective theory in two dimensions.

\begin{figure}[t!]%
\centering
\includegraphics[scale = 0.5, trim={0 5cm 0 5cm}]{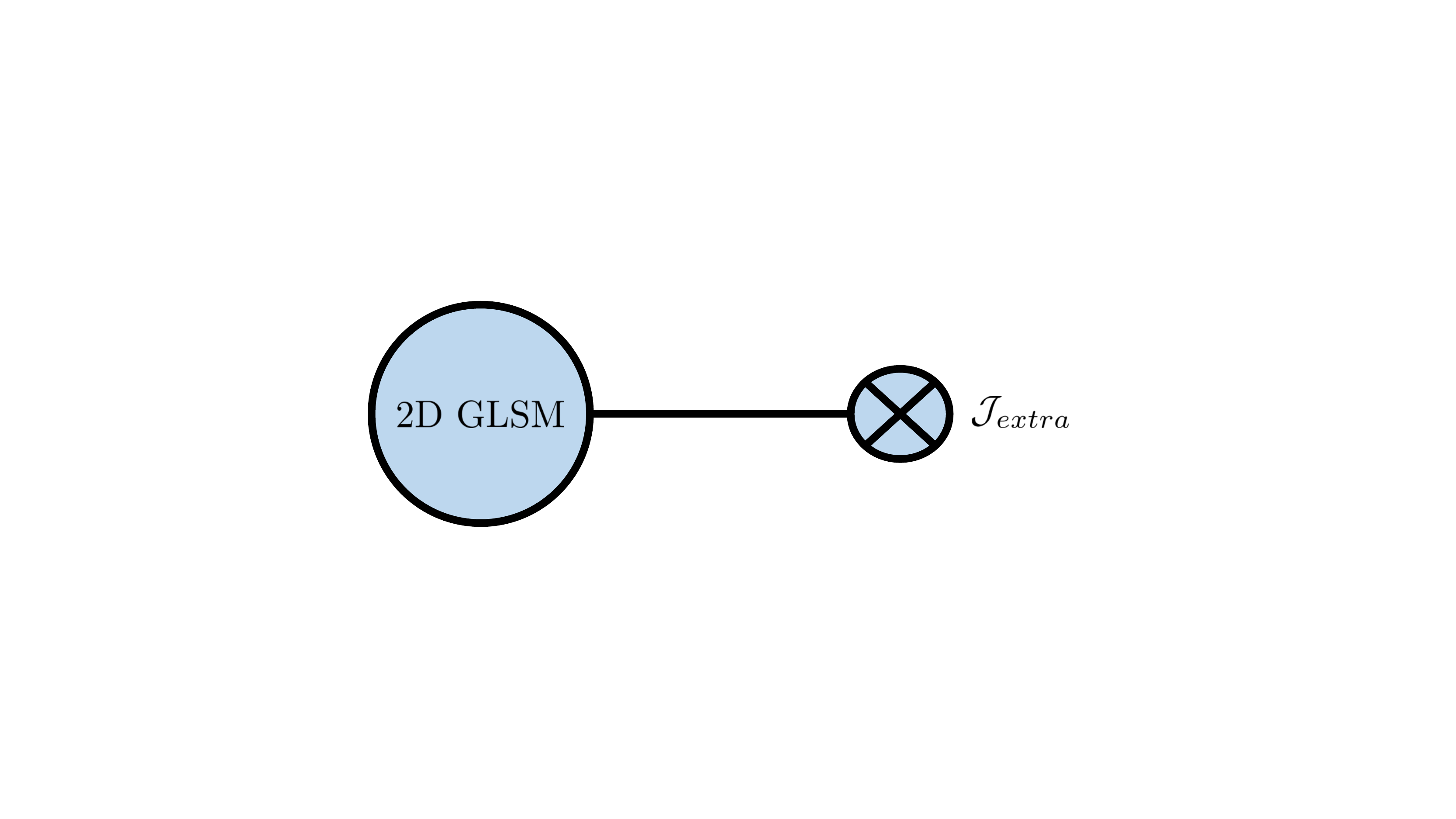}%
\caption{Depiction of the non-gravitational sector of the 2D model obtained from string compactification. Generically, this consists of a
2D gauged linear sigma model (GLSM) coupled to additional extra sectors. These extra sectors can sometimes be strongly coupled conformal field theories in their own right, leading to a rich class of novel 2D theories.}%
\label{QFTsector}%
\end{figure}

In addition to the \textquotedblleft GLSM\ sector\textquotedblright%
\ there are generically other chiral degrees of freedom in the
2D effective theory (see figure \ref{QFTsector} for a depiction).
The necessity of these sectors can be argued for in a few
different ways. First of all, we will see from a detailed calculation of the
zero modes inherited from just the GLSM sector that the spectrum
is typically anomalous. This is a sharp
indication that additional modes must be present to define a consistent gauge
theory. One way that this shows up in a compactification is through the
two-dimensional Green-Schwarz mechanism, i.e. we have a two-dimensional
two-form potential which transforms non-trivially under gauge transformations.
The presence of such a two-form potential also means there is a tadpole in the
effective theory, and this in turn means additional spacetime filling branes
must be included to cancel this tadpole.  By construction, the light degrees of freedom on these
branes have gauge and gravitational anomalies that are just right to cancel
the anomalies from the GLSM\ sector. In most cases, this extra sector is strongly coupled and does not
admit a simple characterization as a GLSM. For example, in a typical
perturbative heterotic string compactification, we will need to introduce some
number of $N$ additional spacetime filling fundamental strings. The limit
where all of these strings are coincident leads us to an additional
sector which is expected to be well-described by the $N$-fold symmetric
orbifold of the usual first quantized heterotic string worldsheet.

For F-theory compactifications, we find that the analogue of these extra
sectors for the perturbative heterotic string are described by spacetime
filling D3-branes wrapped on cycles normal to the directions of the
7-branes. When the point of intersection of the D3-brane and the
7-brane carries an exceptional gauge symmetry, we again generically get a
strongly coupled extra sector. In F-theory, we can also consider D3-branes wrapping two-cycles which
are also common to the 7-branes. The avatar of these contributions
in the type I and heterotic constructions are five-branes wrapped
over four-cycles.

The rest of this paper is organized as follows. In section \ref{sec:EFT} we
give a broader overview of why we expect compactifications of string theory to
two dimensions to give us non-critical string theories. In section
\ref{sec:PERT} we consider the special case of compactifications of
perturbative superstring theories, starting with the case of the type I and
heterotic string theories. This includes a general set of rules for extracting
the zero mode content in the presence of a non-trivial supersymmetric vector
bundle. Next, in section \ref{sec:NPERT}\ we turn to the case of F-theory
compactifications and intersecting 7-branes. Motivated by the successful
analyses of higher-dimensional cases, we shall primarily focus on the local
picture of intersecting 7-branes. In both the perturbative and F-theory
constructions, we will generically encounter gauge theoretic anomalies,
indicating that there are additional degrees of freedom in our models. In
section \ref{sec:TAD} we give a general discussion of tadpole cancellation,
and the prediction that there should be extra sectors coupled to our GLSM. We
follow this in section \ref{sec:EXTRA} with a preliminary analysis of the
dynamics of these extra sectors. After giving these general results, in
section \ref{sec:EXAMPLES} we turn to some explicit examples illustrating the
overall thread of our analysis. For the 9-brane theories we focus on variants of the
\textquotedblleft standard embedding\textquotedblright%
\ constructions. We find that for the $Spin(32) / \mathbb{Z}_2$ heterotic theory,
the addition of spacetime filling fundamental strings (needed for anomaly cancellation) is supersymmetric,
while for the $E_8 \times E_8$ theory, these spacetime filling strings break supersymmetry.
For the F-theory models, we focus on some examples of
\textquotedblleft rigid GLSMs\textquotedblright\ which are the two-dimensional
analogue of non-Higgsable clusters encountered in higher-dimensional F-theory
vacua. Section \ref{sec:CONC}\ contains our
conclusions. We defer a number of technical elements, such as the explicit
construction of the off-shell 2D effective action for the 9-brane and
intersecting 7-brane theories, to a set of Appendices.

\textbf{Note Added:} As we were preparing this work for publication,
reference \cite{Schafer-Nameki:2016cfr} appeared which has some overlap
with the discussion presented here on F-theory compactified on an elliptically fibered Calabi-Yau fivefold.
In some places the holomorphy conventions for the resulting 2D effective theory for intersecting
7-branes are somewhat different. Nevertheless, to the extent we have been able to
compare our results with those found in \cite{Schafer-Nameki:2016cfr}, the broad
conclusions appear to be compatible.

\section{Effective Strings from String Compactification \label{sec:EFT}}

Consider a compactification of a perturbative string theory to $\bR^{1,d-1}$ with $d>2$ spacetime dimensions.  The low energy physics is described by an effective theory of $d$-dimensional (super)gravity coupled to some general quantum field theory.  The effective action, and the vacuum of the effective theory depend on the vacuum expectation values of the moduli fields that encode the choice of background geometry and fluxes.  The presence of the moduli is quite useful to the technically minded string theorist:  it allows for controlled approximations (e.g. string perturbation theory, or a large volume expansion, or both), and the moduli dependence of various physical quantities can shed light on various strong coupling limits.  The resolution of the conifold singularity in type II compactifications to four dimensions with eight supercharges is a beautiful example of the latter.

In the case of $d=2$, the situation is quite different.  We may  obtain the dimensionally reduced action as before, but we must be careful in the interpretation of this action because we no longer have the freedom to work in a fixed vacuum specified by the expectation values of the moduli:  indeed, the ground state wave-function will now be obtained by integrating the non-linear sigma model fields over the full moduli space.  Since the moduli space is typically non-compact and singular, this is a challenging enterprise!  It may be that there are $d=2$ scalar potential terms that fix some (or maybe all) of the moduli and thereby alleviate this particular problem, but based on experience in $d=4$ we might guess that proving this is the case in any particular compactification will not be simple.  Alternatively, we can take a suitable decoupling limit (in essence a decompactification limit) that will allow us to fix moduli to particular values and focus on the gauge theory sector;  F-theory is particularly well-suited for such an approach.

The gauge theory sector will have similar features.  For instance, if there is a Higgs branch in the theory, then we cannot choose a vacuum with some fixed perturbative gauge group; we must integrate over the Higgs branch.  Fortunately, there we are on more familiar ground.  If we focus on the gauge sector, we can describe it as a gauged linear sigma model (GLSM), and in many cases (though by no means all!) we can argue that the resulting path integral leads to a sensible unitary QFT with a normalizable ground state and low energy behavior described by some compact and unitary CFT.  In many GLSMs there is a parameter regime where we can approximate the unitary CFT by a non-linear sigma model over some smooth compact manifold.

Before we can ascertain the low energy dynamics of the GLSM sector, we should be careful to check that our $d=2$ gauge theory is anomaly-free.  This may strike the reader as a pedantic sort of concern:  of course it will be anomaly-free if we made a sensible compactification in the first place.  Here again low-dimensional compactifications provide some (well-known) surprises.  The basic origin of these is the ten-dimensional Green-Schwarz coupling  $\int B_2 \wedge X_8$.   It is indeed the case, that ``if we made a sensible compactification,'' then any anomaly in the GLSM will be cancelled by the dimensional reduction of the GS term, e.g. $\int_{\bR^{1,1}} B_2 \int_{M} X_8$, where $M$ is our compactification manifold.   However, precisely when $\int_{M} X_8 \neq 0$, or equivalently, there is a gauge anomaly in the GLSM sector, the reduced term yields a tadpole for $B_2$.  We must cancel this tadpole by introducing appropriate $\bR^{1,1}$-filling strings or branes, which will then in turn carry extra chiral degrees of freedom, so that the combined anomaly of the GLSM and this ``extra'' sector will vanish.\footnote{Tadpole cancellation might come at the price of breaking spacetime supersymmetry.  We will see below that this is not an idle concern; for instance such breaking takes place for the standard embedding compactification of the $E_8\times E_8$ heterotic string on an irreducible Calabi-Yau fourfold.}

Let us now explain the sense in which we expect our two-dimensional effective
theory to give an effective string theory. First of all, provided we have
engineered a stable string compactification, we have the important feature
that at low energies, the gauge theory sector will flow in the deep infrared
to either a gapped phase or a conformal fixed point. Since we typically can
eliminate the former possibility, we are in some sense guaranteed that the
vast majority of our models flow to some sort of unitary CFT in the infrared.
In the most \textquotedblleft trivial case\textquotedblright\ this will be
some collection of free fields, but even this leads to a rather non-trivial
theory when coupled to worldsheet gravity:\ string theory in flat space! With
this in mind, we generically expect to have a rather rich physical theory on a
target space. In this discussion, it is also important to account for the
fermionic degrees of freedom. Much as in the case of the physical superstring,
these lead to a degeneracy in the ground state which has the spacetime
interpretation of $p$-form potentials in the target space.

It is fairly clear that the resulting theory will be a supercritical string, simply because
the central charge of the complete matter CFT will typically be very large.  Thus, the Weyl mode $\varphi$
of the two-dimensional metric $g = e^{2\varphi} \widehat{g}$, will be a dynamic field and will have the ``wrong sign'' kinetic term, signaling a target-space of signature $(1,D_{\text{eff}}-1)$.  Moreover, there will be a non-trivial dilaton profile, giving an effective string coupling constant $g_{\text{eff}} \sim e^{-\varphi}$, so that we must worry about the strong-coupling dynamics for $\varphi \to -\infty$.  Fortunately, we have an interpretation for this limit:  this is precisely the UV regime for our two-dimensional theory that is acting as the worldsheet for the effective string, and we know that the proper interpretation for that theory is in terms of the physics of our UV complete higher-dimensional theory of gravity!

As we have already mentioned in the Introduction, one of the primary
motivations for studying compactifications of the physical superstring to two
dimensions is that we then get a non-critical string theory propagating on a
more general class of target spaces.   To concretely describe the theory,
we need to specify two ingredients:  the complete matter theory, including
any ``extra'' sectors from space-filling strings or branes, as well as the $(0,2)$
supergravity theory.

The supergravity that perhaps springs most clearly to mind is the $(2,2)$ supergravity
used to construct $\mathcal{N}= 2$ critical string theories
(see e.g. \cite{Ramond:1976qq, Ooguri:1991fp,Ooguri:1991ie}).
However, that cannot be the case in the situation at hand, simply because the
latter involves gauging R-symmetries of the matter theory, and our matter theory
has no R-symmetries to gauge!  The resolution was already discussed in the context
of IIA and IIB compactifications on 4-folds in~\cite{Gates:2000fj}:  there are dilaton
supergravities with $(2,2)$ and $(0,4)$ supersymmetry in two dimensions that do not
involve gauging R-symmetries, and the $(0,2)$ truncation of the latter will be the
appropriate supergravity for our compactifications.  While we will not pursue the
details of this construction here, there is one important aspect of the story:  the
ghost measure for this supergravity has central charges $c_L = -26$ and $c_R = -26+23$.  The factors of $-26$ are the
familiar $bc$ ghosts of diffeomorphisms, while $22=2\times 11$ is the contribution from two
right-moving $\beta\gamma$ ghosts of superdiffeomorphisms, but this is supplemented by a contribution of $+1$ to $c_R$ from a right-moving Weyl fermion that also descends from the gravitino.  Finally, the ten-dimensional dilatino contributes another right-moving Weyl fermion.  All in all, the contribution
to the gravitational anomaly from the dilaton supergravity sector is
\begin{equation}
\Delta (c_L-c_R) = -24~.
\end{equation}

Just as our theory is free of gauge anomalies, it will also be free of the gravitational anomaly: the sum of contributions to $c_L-c_R$ from the matter, ``extra,'' and supergravity sectors will cancel.  Once we have such a generally covariant theory, we can confidently fix to superconformal gauge, where we will obtain a superconformal theory with the Weyl mode, the dilaton, and the dilatinos combining in a $(0,2)$ super-Liouville sector.  This will be the conformal field theory that will act as the worldsheet theory for our effective super-critical string.

We expect that when a target space interpretation will be available for this theory,
it will have non-trivial time dependence (because of the time-like Weyl mode
and $g_{\text{eff}} \sim e^{-\varphi}$) and may well be equipped with gauge
degrees of freedom corresponding to a current algebra in the CFT.

Another interesting feature of our effective string theory is that as we
proceed up in energy scale on the worldsheet, i.e. as we get closer to the
string tension scale, the effective dimension, which will be related to the central charge, will at first appear to grow
before the whole formalism breaks down and we instead replace the effective
string theory by another effective theory:\ that of a compactification of the
physical superstring on a ten-dimensional spacetime. At even higher energy scales,
this in turn must be replaced by the worldsheet description of the model (in the case of the
perturbative type I and heterotic models).

This is certainly far from a complete solution to the dynamics of these effective strings.  To name just one of the issues, our understanding of the starting point theories is certainly not complete in any sense.  However, we believe it is valuable to recognize that many super-critical string theories can be completed in this way at least in principle.  It will be very interesting to see which aspects of strong coupling we can understand based on what we do know about M/F/string theory.  In that sense, we can expect that examples with supersymmetry should help the analysis.  In this work, we take a very minimalist point of view of preserving the smallest amount of two-dimensional supersymmetry where holomorphy can play a powerful role. With that, we turn to a discussion of $(0,2)$ worldsheet supersymmetry, focusing on the gauge sector.

\subsection{Elements of $\mathcal{N} = (0,2)$ Theories}

Let us briefly summarize some of the elements of
$\mathcal{N} = (0,2)$ supersymmetric quantum field theories in two dimensions which we will
be using throughout this work. In Appendix \ref{app:GLSM} we also present the gauged
linear sigma model (GLSM) for theories with non-abelian gauge groups. Perhaps
surprisingly, we have been unable to locate a convenient reference for this
seemingly basic result.

Our aim in this section will be to set our conventions, and in
particular, to emphasize the holomorphic structure we expect to be present in
any candidate effective action. First of all, in building a $(0,2)$ GLSM, we
have modes in a vector multiplet, Fermi multiplet, and chiral
multiplet.\footnote{The real \textquotedblleft vector
multiplet\textquotedblright\ is quite similar to that found in $\mathcal{N} = 1$, $d=4$ superspace.
As there, the field strength lives in a derived fermionic superfield with lowest component a gaugino.}

For the chiral multiplets and Fermi multiplets, we use conventions similar to
those in \cite{Witten:1993yc}:%
\begin{align}
\text{CS}  &  \text{: }\Phi=\phi+\sqrt{2}\theta^{+}\psi_{+}-i\theta
^{+}\overline{\theta}^{+}\partial_{+}\phi\\
\text{F}  &  \text{: }\Lambda=\lambda_{-}-\sqrt{2}\theta^{+}\mathcal{G} -i\theta
^{+}\overline{\theta}^{+}\partial_{+}\lambda_{-}-\sqrt{2}\overline{\theta}^{+}E,
\end{align}
where $E(\Phi)$ is a holomorphic function of the CS\ multiplets, and $\mathcal{G}$ is an auxiliary field.
Following standard terminology, we refer to the $\lambda_-$ as left-movers and $\psi_+$ as right-movers.\footnote{Another delightful confusion involves the spin of these fields:  $\lambda_-$ has spin $+1/2$ and $\psi_+$ has spin $-1/2$.  These recondite issues have to do with ancient preferences and holomorphy conventions.}

An F-term will be represented in terms of a Grassmann integral over half the
superspace, i.e. $\theta^{+}$, and a D-term is given by integrating over both
$\theta^{+}$ and $\overline{\theta}^{+}$. Minimal kinetic terms for the chiral
multiplets and Fermi multiplets are given by the D-terms:
\begin{align}
\text{CS}  &  \text{: } - \frac{i}{2} \int d^2 \theta \text{ } \overline{\Phi} \partial_{-} \Phi\\
\text{F}  &  \text{: } - \frac{1}{2}\int d^2 \theta \text{ } \overline{\Lambda} \Lambda.
\end{align}
In what follows, we shall often have occasion to work with fields
transforming in non-trivial representations and bundles. Then, we shall introduce a
canonical pairing $( \cdot , \cdot)$ to capture this more general possibility.
When we couple to gauge fields, the derivative $\partial_{-}$ is promoted
to a gauge supercovariant derivative $\nabla_{-}$.

An important point to emphasize is that as far as we are aware, there is no simple way to impose a reality condition such as $\Lambda^{\dag} = \Lambda$ on Fermi multiplets and retain a non-trivial kinetic term. We shall instead later show how to obtain a variant of this constraint in some special cases.

One of the items we will be most interested in is the structure of possible F-terms.
These arise from the $E^k(\Phi)$ terms just mentioned, as well as the interaction:%
\begin{equation}
L_{F}=\int d\theta^{+} \, W,
\end{equation}
where we have introduced a quantity $W$ which we shall refer to as the
\textquotedblleft superpotential\textquotedblright:%
\begin{equation}
W = \frac{1}{\sqrt{2}} \Lambda^{k}J_{k}(\Phi),
\end{equation}
where $J_{k}(\Phi)$ is a holomorphic function of the chiral multiplets.

By applying the supercovariant derivative $\overline{\mathcal{D}}_{+}$ (see
Appendix \ref{app:GLSM} for details) we also obtain a necessary condition for off-shell
supersymmetry of our action:%
\begin{equation}
\overline{\mathcal{D}}_{+}W=\underset{k}{\sum}E^{k}(\Phi)J_{k}(\Phi)=0. \label{offshell}%
\end{equation}
This condition needs to be
satisfied for any choice of field configuration and therefore  leads to non-trivial
constraints on the structure of any coupling constants in the theory, i.e.
background parameters.

By expanding out in terms of component fields (and including the kinetic terms
for the various fields), we find that the F-term couplings $J_k$ and $E^k$ lead to
terms in the scalar potential of schematic form $\sum_{k}\left[ |E^k |^2 + |J_k|^2\right]$.
Thus, the F-term conditions for a $d=2$ supersymmetric vacuum are:%
\begin{equation}
J_{k}(\Phi)=0 \,\,\,\text{and}\,\,\,E^k(\Phi)  = 0
\end{equation}
for all $k$. There will also be D-term potential terms from the gauge interactions.

There are a few additional comments we can now make with regards to the absence of a Fermi multiplet with a Majorana-Weyl spinor. The essential difficulty we inevitably face is that there is a clear clash between the requirements of holomorphy, and those of unitarity (i.e. an appropriate reality condition). However, let us suppose that we have a collection of Majorana-Weyl spinors, and with them, a corresponding $\mathbb{Z}_2$ symmetry of the theory. With this in mind, we can at first double the number of degrees of freedom for our fermionic sector to a set of Fermi multiplets $\Lambda^{(even)}$ and $\Lambda^{(odd)}$. We do \textit{not}, however, double any of the other degrees of freedom, and simply decompose for example the term $J(\Phi)$ into an even and odd piece. In this enlarged theory, we can now introduce a formal superpotential term which enforces the holomorphic structure of the theory, and which we refer to as $W_{top}$:
\begin{equation}
W_{top} = \Omega^{(odd)} (\Lambda^{(even)} J^{(odd)} + \Lambda^{(odd)} J^{(even)}),
\end{equation}
and where we take the $E$-fields for both sets of $\Lambda$'s to be trivial. Observe that the F-term equations of motion
are now enforced by independently varying the two $\Lambda$'s. We can be more economical, however, and simply work with the single Fermi multiplet
$\Lambda^{(even)}$, but with a modified $E$-field. The choice of $E$-field is set by the condition that we reproduce the correct holomorphic structure of the vacuum, so we set:
\begin{equation}
E^{(even)} = \frac{1}{\Omega^{(odd)}}\frac{\partial W_{top}}{\partial \Lambda^{(odd)}} = J^{(even)}.
\end{equation}
Then, we are free to eliminate $\Lambda^{(odd)}$ altogether and just use the physical F-term:
\begin{equation}
W = \frac{1}{\sqrt{2}}\Omega^{(odd)} \Lambda^{(even)} J^{(odd)}.
\end{equation}
So in this sense, we still get to set $J^{(even)} = 0$, and this is protected by holomorphy.

For our present purposes, we will be interested in higher-dimensional brane
systems which we wish to represent in terms of an off shell  two-dimensional
effective field theory. In other words, we will try to retain the full
Kaluza-Klein tower of higher dimensional modes, but assembled according to
corresponding $(0,2)$ supermultiplets. We will also demand that
higher-dimensional gauge symmetries are manifest in our formulation. Such
an off-shell formulation will allow us to succinctly state which of our interaction terms are expected to be
protected by supersymmetry (i.e. are holomorphic F-terms) and which are
expected to receive quantum corrections (i.e. the non-holomorphic D-terms).

Indeed, one of the important features of this formulation is that it provides
us with a way to characterize the quasi-topological (in the sense
that it depends on complex structure moduli)\ theory associated with the
internal brane dynamics. From this perspective, the supercharges of the 2D
$(0,2)$ theory can also be interpreted as the BRST\ charges of the topological
theory:%
\begin{equation}
Q_{2D} = Q_{\text{BRST}}.
\end{equation}
The condition that we have an off-shell supersymmetric action then corresponds
to the condition that we have indeed performed the twist correctly. Moreover,
the physical states of the theory, i.e. those in the BRST cohomology simply
label possible ground states of the 2D\ effective theory.

\section{GLSMs from Perturbative String Vacua \label{sec:PERT}}

Motivated by the possibility of constructing UV\ complete non-critical
strings, we now turn to a particularly tractable class of examples obtained
from compactifications to two dimensions of perturbative string theories.
Since we are interested in theories which also admit a gauge theory sector, we
shall primarily focus on compactifications of the type I, and heterotic
superstrings. An important feature of these models is the presence of a
spacetime filling 9-brane with respective gauge group $Spin(32)/\mathbb{Z}_2$,
and $E_{8}\times E_{8}$, so we can expect that upon compactification
this gauge theory sector will give rise to a large class of $(0,2)$
GLSMs.

In this section we will  focus on the 9-brane sector by itself. In later sections we turn to the
effects internal fluxes have on the presence of tadpoles and extra sectors. To this end, we first recall
that in flat space, this theory has gauge group $G$ and $\mathcal{N} = 1$ supersymmetry with a single
Majorana-Weyl spinor transforming in the $\mathbf{16}$ of $Spin(1,9)$.
The action in flat space has two leading order terms:%
\begin{equation}
L_{10D}= \frac{1}{4 g_{YM}^{2}}\int d^{10}x \left( \text{ Tr}F^{IJ}F_{IJ}%
+ 2 i \overline{\chi}\Gamma^{I}D_{I}\chi \right),
\end{equation}
where $D_I$ is the covariant derivative,
$F_{IJ}=\left[  D_{I},D_{J}\right] $ is the non-abelian field strength
for our 10D Yang-Mills theory, and the $\chi$ are the 10D Majorana-Weyl gauginos which
transform in the adjoint representation of $G$. As written, the theory is of
course non-renormalizable, and we should view this as just the
leading order contribution to the full theory (with scattering amplitudes
controlled by the string worldsheet anyway). There is also the
gravitational sector of the theory, which includes the metric, the
Neveu-Schwarz two form potential, and the heterotic dilaton (which controls the
gauge coupling of the Yang-Mills sector).

Suppose now that we compactify this 10D gauge theory to two dimensions.
The simplest way to retain $\mathcal{N} = (0,2)$ supersymmetry
is to compactify on an irreducible Calabi-Yau fourfold $M$ equipped with a principal $G$-bundle $P$.\footnote{By ``irreducible'' we mean that
the smooth compact manifold has a K\"ahler metric with holonomy exactly $\SU(4)$.}
Indeed, doing this enables us to retain a covariantly
constant spinor so that we maintain low energy $(0,2)$ supersymmetry in the
two uncompactified directions. Since we have a manifold of $SU(4)$ holonomy,
the fundamental representation of $SO(8)$ must decompose to the
$\mathbf{4} \oplus \overline{\mathbf{4}}$. This in turn forces the following decomposition of
eight-dimensional representations for $SO(8)$:%
\begin{align}
SO(8)  &  \supset SU(4)\times U(1)\\
\mathbf{8}^{s}  &  \rightarrow\mathbf{1}_{+2}\oplus\mathbf{1}_{-2}%
\oplus\mathbf{6}_{0}\\
\mathbf{8}^{c}  &  \rightarrow\mathbf{4}_{-1}\oplus\overline{\mathbf{4}}%
_{+1}\\
\mathbf{8}^{v}  &  \rightarrow\mathbf{4}_{+1}\oplus\overline{\mathbf{4}}_{-1}.
\end{align}

We now turn to the decomposition of the
supercharges, as well as the mode content of the 10D Super Yang-Mills theory.
First of all, both the 10D gauginos and the supersymmetry parameters transform in the
$\mathbf{16}$ of $SO(1,9)$. Additionally, we have the gauge field which
transforms in the $\mathbf{10}$ of $SO(9,1)$. We begin with the decomposition
expected from compactification on a general eight-manifold:\footnote{Using the triality automorphism, we can
shift the role of the $\mathbf{16}$ and $\mathbf{16'}$. We choose the present chirality convention to conform with
our conventions for $\mathcal{N} = (0,2)$ supersymmetry in two dimensions.}
\begin{align}
SO(9,1)  &  \supset SO(1,1)\times SO(8)\\
\mathbf{16}  &  \rightarrow\mathbf{8}_{-}^{s}\oplus\mathbf{8}_{+}^{c}\\
\mathbf{16'}  &  \rightarrow\mathbf{8}_{+}^{s}\oplus\mathbf{8}_{-}^{c}\\
\mathbf{10}  &  \rightarrow\mathbf{1}_{++}\oplus\mathbf{1}_{--}\oplus
\mathbf{8}_{0}^{v},
\end{align}
where we use the subscripts $+$ and $-$ to indicate a right-moving or
left-moving chiral spinor of $SO(1,1)$, and we double this to indicate the 2D
vector field. Decomposing further into irreducible representations of
$SU(4)\times U(1)$, we have:%
\begin{align}
SO(1,9)  &  \supset SO(1,1)\times SU(4)\times U(1)\\
\mathbf{16}  &  \rightarrow\mathbf{1}_{-,+2}\oplus\mathbf{1}_{-,-2}%
\oplus\mathbf{6}_{-,0}\oplus\mathbf{4}_{+,-1}\oplus\overline{\mathbf{4}%
}_{+,+1}\\
\mathbf{16'}  &  \rightarrow\mathbf{1}_{+,+2}\oplus\mathbf{1}_{+,-2}%
\oplus\mathbf{6}_{+,0}\oplus\mathbf{4}_{-,-1}\oplus\overline{\mathbf{4}%
}_{-,+1}\\
\mathbf{10}  &  \rightarrow\mathbf{1}_{++,0}\oplus\mathbf{1}_{--,0}%
\oplus\mathbf{4}_{0,+1}\oplus\overline{\mathbf{4}}_{0,-1},
\end{align}
so we indeed recognize that descending from the $\mathbf{16}$ there are two singlets under $SU(4)$ which specify
the $\mathcal{N} = (0,2)$ supercharges of our system.

The decomposition we have given for the supercharges also holds for the
10D\ gaugino. Doing so, we see that the 2D\ gauginos descend from the
$\mathbf{1}_{-,+2}\oplus\mathbf{1}_{-,-2}$ as left-movers with $(0,2)$
superpartners $\mathbf{1}_{++,0}\oplus\mathbf{1}_{--,0}$. Additionally, we see
that there are right-movers transforming in the $\mathbf{4}_{+,-1}%
\oplus\overline{\mathbf{4}}_{+,+1}$ with $(0,2)$ superpartners $\mathbf{4}%
_{0,+1}\oplus\overline{\mathbf{4}}_{0,-1}$. A curious feature of working in
two dimensions is that we also recognize left-moving fermions in the
$\mathbf{6}_{-,0}$~, which have no bosonic partners.

An important subtlety with 10D Super Yang-Mills theory is that the $\mathbf{16}$ is actually a
Majorana-Weyl spinor. This issue is reflected in the fact that the $\mathbf{6}_{-,0}$ is actually a real representation
of $SU(4)$. Indeed, counting up the fermionic degrees of freedom, we therefore expect the $\mathbf{6}_{-,0}$ to descend
to a Majorana fermion in two dimensions. This will have important consequences when turn to the construction of
supermultiplets and interaction terms.

In the heterotic models, we must impose a workaround to get everything fully off-shell. One way to do this which is suggested by the
related F-theory models is to assume the presence of a geometric $\mathbb{Z}_{2}$ symmetry for our Calabi-Yau and gauge bundles. Doing so automatically leads to a split of the form content into an equal number of even and odd modes. Turning to the decomposition of the $\mathbf{6}$, we can then take just the even modes, and use these to assemble a Fermi multiplet. When we turn to the construction of the effective action, we will revisit this point in great detail.

In order to respect the structure dictated by the higher-dimensional geometry, we shall find it convenient to view our multiplets in terms of differential $(p,q)$ forms, that is, forms with $p$ holomorphic indices and $q$ anti-holomorphic indices.
More formally, we view them as elements of $\Omega^{p,q}(\text{ad}P )$, that is, as $(p,q)$ forms on $M$ valued in the adjoint bundle associated to $P$. Returning to our decomposition of representations of $Spin(1,9)$ to $SO(1,1) \times SU(4)$, we can now see how to assemble the various modes into superfields which transform as differential forms on the internal space. First of all, we can see that
there is the 2D\ non-abelian vector multiplet. We also introduce a collection of chiral
multiplets valued in $\Omega^{0,1}(\text{ad}P)$ :%
\begin{equation}
\mathbb{D}_{(0,1)}= \overline{\partial}_{A} + \sqrt{2}\theta^{+}%
\psi_{(0,1)}+\ldots.
\end{equation}
where we have used the shorthand $\overline{\partial}_{A}=\overline{\partial
}+A$. The top component is the $(0,1)$ component of the gauge covariant
derivative for the corresponding vector bundle. There is a related chiral
multiplet valued in $\Omega^{0,2}(\text{ad}P)$  that we can construct
from $\mathbb{D}_{(0,1)}$ corresponding to the overall $(0,2)$ field strength:
\begin{equation}
\mathbb{F}_{(0,2)} = F_{(0,2)}+\sqrt{2}\theta^{+}\overline{\partial}_{A}%
\psi_{(0,1)}+\ldots .
\end{equation}

Additionally, we see that there is a Fermi multiplet which transforms as
a $(0,2)$ differential form on the Calabi-Yau fourfold which we refer to as
$\Lambda_{(0,2)}$. Here, we face an issue which leads to some tension in maintaining a purely off-shell formalism for the theory.
The point is that really, we must get out six real rather than six complex degrees of freedom to maintain the 10D Majorana-Weyl spinor condition.
This in turn shows up in our 2D effective theory as the statement
that we expect the Fermi multiplet to contain a Weyl rather than Majorana-Weyl spinor.
Nevertheless, there is a simple (seemingly somewhat ad hoc) workaround for this issue
which is actually automatically implemented in F-theory constructions.

Along these lines, suppose that our geometry also admits a discrete $\mathbb{Z}_2$ symmetry under which the holomorphic four-form transforms as $\Omega \rightarrow - \Omega$ and such that there are an equal number of $\mathbb{Z}_2$ even and odd $(0,2)$ differential forms.\footnote{
To give an explicit example where we expect to have such a $\mathbb{Z}_2$ symmetry, consider the special case of an elliptically fibered Calabi-Yau fourfold $M \rightarrow X$ with $X$ the base. This has a Weierstrass model $y^2 = x^3 + fx + g$, where $f$ and $g$ are sections of $\mathcal{O}_{X}(-4 K_X)$ and $\mathcal{O}_{X}(-6 K_X)$, with $K_X$ the canonical bundle of $X$. We observe that the holomorphic four-form of $M$ can be written as $\Omega_{M} = \frac{dx} {y} \wedge \Omega_{X}$ with $\Omega_{X}$ the (meromorphic) three-form of the base. Now, the defining equation of $M$ enjoys the $\mathbb{Z}_2$ symmetry $y \rightarrow -y$ under which $\Omega_{M} \rightarrow - \Omega_{M}$.}
Then, for example, we can introduce a further splitting as:
\begin{equation}
\mathbb{F}_{(0,2)} \rightarrow \mathbb{F}^{(even)}_{(0,2)} + \mathbb{F}^{(odd)}_{(0,2)},
\end{equation}
in the obvious notation. By a similar token, we can then introduce a $\mathbb{Z}_2$ even Fermi multiplet which transforms as a $(0,2)$ differential form with expansion in components given by:
\begin{equation}
\Lambda^{(even)}_{(0,2)} = \lambda^{(even)}_{-,(0,2)}-\sqrt{2}\theta^{+} \mathcal{G}^{(even)}_{(0,2)}-i\theta
^{+}\overline{\theta}^{+}\partial_{+}\lambda^{(even)}_{-,(0,2)}-\sqrt{2}\overline{\theta}^{+} \mathbb{F}^{(even)}_{(0,2)}.
\end{equation}
Observe that here, we have also specialized the form of the contribution $E$ which is a function of chiral superfields to be that of the even field strength. Indeed, as we will shortly see, to maintain a canonical notion of holomorphy for our 10D action, it will be necessary to shuffle some of the holomorphic data into the $E$-field of the Fermi multiplet, and some into the F-term interactions.
To avoid overloading the notation, we shall sometimes suppress the superscript of even and odd, leaving it implicit.

Let us make an additional comment about the situation where we do not have such a $\mathbb{Z}_2$ symmetry. In such situations, the resulting
effective field theory will still retain $(0,2)$ supersymmetry, but we do not expect a manifestly off-shell formalism in terms of
weakly coupled Fermi multiplets. We leave it to future work to develop an off-shell formalism for this case as well.

Let us now turn to the structure of our 10D gauge theory. At the level of the
F-terms, we expect the superpotential to be invariant under complexified
gauge transformations, i.e. we introduce chiral
multiplets $g=\exp C$ in the complexification of the adjoint representation so that the overall effect
of a gauge transformation is:%
\begin{equation}
\mathbb{D}_{(0,1)}\longmapsto e^{-C}\mathbb{D}_{(0,1)}e^{+C}\text{ \ \ and
\ \ }\Lambda_{(0,2)}\longmapsto e^{-C}\Lambda_{(0,2)}e^{+C}.
\end{equation}
Supersymmetric vacua are parameterized by the F-terms modulo complexified gauge transformations, or equivalently,
by imposing F- and D-terms modulo unitary gauge transformations. In the latter case, the bosonic component of $C$ is
taken to be pure imaginary.

In Appendix \ref{app:HET} we present a complete construction of the 2D off-shell
effective action such that its supersymmetric vacua reproduce the equations of
motion of the 10D\ Super Yang-Mills theory. One term which is not immediately apparent in this
approach is a non-local Wess-Zumino term involving the vector multiplet. It is required in order for our superspace formulation to remain
gauge invariant in arbitrary gauge (i.e. not just Wess-Zumino gauge).
As we shall present all results in Wess-Zumino gauge, we shall
omit this term. For further details on this point, as well as
further discussion of the formulation of 10D Super Yang-Mills theory in 4D $\mathcal{N} = 1$ superspace, see
reference \cite{Marcus:1983wb} (see also \cite{ArkaniHamed:2001tb}). Similar issues also occur for the superspace formulation of intersecting 7-branes.

Modulo these caveats, the superspace formulation provides a quite elegant way to formulate the off-shell content of 10D Super Yang-Mills theory on a
Calabi-Yau fourfold. We begin with the on shell equations of motion, which we obtain by setting the supersymmetric
variation of the 10D gauginos to zero:
\begin{equation}
{\Gamma^{IJ} F_{IJ}}=0.
\end{equation}
Focusing on just the internal degrees of freedom, this becomes:
\begin{equation}
\omega \wedge \omega \wedge \omega \wedge F_{(1,1)}=0\text{ \ \ and \ \ }F_{(0,2)}%
=F_{(2,0)}=0\text{,}%
\end{equation}
where we have introduced the K\"ahler form $\omega$ for the Calabi-Yau fourfold $M$, and
decomposed the $2$-form field-strength according to type.   These are of course
the Hermitian Yang-Mills equations.  When the principal bundle $P$ is associated to
some complex vector bundle $\mathcal{V}$, then the second condition is the statement that
$V$ is a holomorphic vector bundle.  The DUY theorem~\cite{Uhlenbeck:1986ym}
then implies that the first condition is satisfied if and only if $\mathcal{V}$ is stable with
respect to $\omega$.  We show in Appendix \ref{app:HET} that the first constraint arises from a D-term of
the 2D theory, while the second constraint is a holomorphic F-term constraint.
Indeed, while we expect the stability conditions for vector bundles to receive
various quantum corrections as we pass to small volume, the purely holomorphic terms are
protected by $(0,2)$ supersymmetry. This fact is neatly
summarized by the corresponding F-term interaction:%
\begin{equation}
W_{M}= - \frac{1}{\sqrt{2}}\frac{1}{g^{2}_{YM}}\underset{M}{\int}\Omega\wedge\text{Tr}(\Lambda^{(even)}
_{(0,2)}\wedge\mathbb{F}^{(odd)}_{(0,2)}), \label{WBULK}%
\end{equation}
where $\Omega$ is the holomorphic four-form of the Calabi-Yau fourfold. The
F-term equations of motion (obtained by varying with respect to $\Lambda$)
then give the condition $F^{(odd)}_{(0,2)}=0$, while the condition $F^{(even)}_{(0,2)} = 0$ comes about
from the condition that the $E$-field of the Fermi multiplet vanishes.

We can also write the D-terms for our system. In this case, we must exercise some caution since we can expect terms
not protected by holomorphy to receive large quantum corrections. However, at least at large volume we can deduce the form of these
interaction terms. Summarizing the contributions from Appendix \ref{app:HET}, we have:
\begin{align}
  S_\mathrm{tot} & = S_{D} + S_{F} \\
  S_{D} & =  - \frac{1}{g^{2}_{YM}} \int d^2 y d^{2} \theta \int_M \, \Tr \left( \frac{1}{8} * \overline{\Upsilon} \wedge \Upsilon
   - \frac{i}{2} * \overline{\mathbb{D}_{(0,1)}} \wedge [\nabla_- , \mathbb{D}_{(0,1)}] - \frac{1}{2} * \overline{\Lambda_{(0,2)}^{(even)}} \wedge \Lambda_{(0,2)}^{(even)} \right) \\
  S_{F} & = - \frac{1}{\sqrt{2}}\frac{1}{g^{2}_{YM}} \int d^2 y d \theta^{+} \int_M \, \Tr \left( \Omega\wedge\Lambda^{(even)}_{(0,2)}\wedge\mathbb{F}^{(odd)}_{(0,2)} \right) + h.c.
\end{align}

Both the F- and D-term constraints directly follow from the bosonic potential obtained from integrating out
all auxiliary fields. This is given by:
\begin{equation}
U_{\mathrm{Bosonic}} = \frac{1}{4 g^{2}_{YM}} \int_{M} \left(\vert \vert F_{(1,1)} \vert \vert^{2} + \vert \vert F_{(0,2)} \vert \vert^{2} \right),
\end{equation}
where the norm on the differential forms includes a Hodge star and complex conjugation operation. So for a supersymmetric vacuum where $U_{\mathrm{Bosonic}} = 0$, we need both $F_{(0,2)}$ and $\omega \wedge \omega \wedge \omega \wedge{F_{1,1}}$ to vanish.

Here, we have clearly made use of the fact that we have a $\mathbb{Z}_2$ symmetry (obtained by tuning moduli of the fourfold) which allows us to split up the mode content and retain a Fermi multiplet. If we suspend the conditions of unitarity, we do not need this additional constraint, and we can also write the conditions arising from the holomorphic interactions again in terms of a single overall superpotential, now involving a $(0,2)$ differential form:
\begin{equation}\label{WTOP}
W_{top} = - \underset{CY_{4}}{\int}\Omega\wedge\text{Tr}(\Lambda
_{(0,2)}\wedge\mathbb{F}_{(0,2)}),
\end{equation}
and where we set the $E$-field of Fermi multiplet $(0,2)$ differential form to zero. Again, this generates the holomorphic equation $F_{(0,2)} = 0$. The price one pays for doing this, however, is that the resulting theory should really be treated as a topological one, since the unitarity condition imposed by the 10D Majorana-Weyl constraint is now absent.

Lastly, we can ask to what extent we expect our off-shell presentation of the 10D theory to really remain decoupled from the
gravitational degrees of freedom of the system. Indeed, we
observe that the off-shell variation of the
superpotential term gives us:%
\begin{equation}
\overline{\mathcal{D}}_{+}W_{M}= - \frac{1}{\sqrt{2}}\frac{1}{g^{2}_{YM}} \underset{M}{\int}\Omega\wedge
\text{Tr}(\mathbb{F}^{(even)}_{(0,2)}\wedge \mathbb{F}^{(odd)}_{(0,2)}),
\end{equation}
which does not vanish off-shell, a priori. Indeed, the condition $F_{(0,2)}=0$ is an
on-shell constraint. Even so, the structure of this term is topological
(essentially a holomorphic analogue of an instanton density) of the type
introduced by Donaldson and Thomas in reference \cite{DTgeom}.

What this term tells us is that there must be geometric moduli
coupled to our theory. From the standpoint of string compactification, it is clear that this
must be so, because there will necessarily be moduli fields associated with
the complex structure and K\"{a}hler deformations of the geometry. Indeed,
from that perspective our present split into \textquotedblleft gauge theory
+\ everything else\textquotedblright\ is somewhat artificial in a
two-dimensional model. Including these contributions from the moduli, we
indeed expect our on-shell action to vanish. If we view the complex structure
moduli as parameters of our gauge theory, then the condition $\overline{\mathcal{D}}%
_{+}W_{\text{gauge}}=0$ means that we must tune these parameters to be
vanishing for any choice of vector bundle (really an on-shell condition). The
perhaps surprising point is that even for intersecting 7-branes in
F-theory, a similar phenomenon will be encountered so there is no complete
decoupling limit:\ Some remnant of the geometric moduli must always be included.

Now, in practice, we of course would like to restrict our attention to the
gauge theory sector. To do so, it is convenient to introduce a Fermi multiplet
which functions as a Lagrange multiplier. Along these lines, we introduce a
Fermi multiplet $\Pi$ a $(4,0)$ form on the Calabi-Yau fourfold such that the
complex structure moduli appears via:%
\begin{equation}
\overline{\mathcal{D}}_{+}\Pi = \Omega.
\end{equation}
The superpotential is then of the form:%
\begin{equation}
W=W_{M}+W_{\text{bkgnd}}= - \frac{1}{\sqrt{2}}\frac{1}{g^{2}_{YM}} \underset{M}{\int}\Omega
\wedge\text{Tr}(\Lambda^{(even)}_{(0,2)}\wedge\mathbb{F}^{(odd)}_{(0,2)}) + \frac{1}{\sqrt{2}}\frac{1}{g^{2}_{YM}} \underset{M}{\int}\Pi \wedge\text{Tr}(\mathbb{F}^{(even)}_{(0,2)}\wedge\mathbb{F}^{(odd)}_{(0,2)}).
\end{equation}
and then $\overline{\mathcal{D}}_{+}W=0$ off-shell (by construction).

\subsection{Zero Mode Spectrum}

Suppose then, that we have succeeded in constructing a stable holomorphic
vector bundle on our Calabi-Yau fourfold. We would now like to determine the
corresponding zero mode spectrum for our system. We begin by assuming that we have a solution to
the Hermitian Yang-Mills equations, and with it, a corresponding vector bundle
with structure group $K$ with commutant $H$ inside of the parent gauge group
$G$. Starting from the principal $G$ bundle $adP$, we decompose according to
representations of $H$ and bundles of $K$:%
\begin{equation}
adP\rightarrow\underset{i}{\oplus}(\tau_{i},\mathcal{E}_{i}),
\end{equation}
i.e. for each representation $\tau_{i}$ of $H$, there is a corresponding
bundle $\mathcal{E}_{i}$.

Expanding around this background, we now see that for a stable vector bundle,
we get precisely one vector multiplet and gaugino --as expected-- in the
adjoint representation of $H$. Additionally, we have the fluctuations of the
supermultiplets $\mathbb{D}_{(0,1)}$ and $\Lambda_{(0,2)}$. Consider first
the zero mode fluctuations for $\mathbb{D}_{(0,1)}$. These are
counted by appropriate bundle valued cohomology groups, and they
naturally pair up between the representation $\tau_{i}$ and $\tau_{i}^{\ast}$
\begin{align}
\delta\mathbb{D}_{(0,1)}^{(\tau_{i})}  &  \in H_{\overline{\partial}}%
^{1}(CY_{4},\mathcal{E}_{i})\\
\delta\mathbb{D}_{(0,1)}^{(\tau_{i}^{\ast})}  &  \in H_{\overline{\partial}%
}^{1}(CY_{4},\mathcal{E}_{i}^{\vee})\simeq H_{\overline{\partial}}^{3}%
(CY_{4},\mathcal{E}_{i})^{\ast},
\end{align}
in the obvious notation.

Let us now turn to the zero modes for the Fermi multiplets. As we have already mentioned, for the
bulk 10D theory, the tension between the 10D Majorana-Weyl constraint and holomorphy in the 2D $\mathcal{N} = (0,2)$
theory means that to get a truly off-shell action, we need to assume an accidental $\mathbb{Z}_2$ symmetry, and work in terms of $\Lambda_{(0,2)}^{(even)}$. However, once we pass to the zero mode content, there is always a $\mathbb{Z}_2$ symmetry present given by passing from the representation $\tau$ to its dual $\tau^{\ast}$. With this in mind, we can either view the Fermi multiplets as transforming in the representation $\tau$ or $\tau^{\ast}$. For ease of presentation, we shall only write the modes in representation $\tau$:
\begin{equation}
\delta\Lambda_{(0,2)}^{(\tau_{i})}  \in H_{\overline{\partial}}%
^{2}(CY_{4},\mathcal{E}_{i}).
\end{equation}
Note that by Serre duality, we also have $H_{\overline{\partial}}^{2}(CY_{4},\mathcal{E}_{i}) = H_{\overline{\partial}}^{2}(CY_{4},\mathcal{E}^{\vee}_{i})^{\ast}$, so we could have alternatively
counted the Fermi multiplets in terms of the representation $\tau_{i}^{\ast}$. This prescription is sufficient provided that $\tau$ and $\tau^{\ast}$ are distinct representations. However, when they are not, we also see that the bundle $\mathcal{E}$ is self-dual. Under the conjugation map: $\mathcal{E} \rightarrow \mathcal{E}^{\vee}$ defined by the canonical pairing on these representations, we can therefore restrict to the $\mathbb{Z}_2$ even sector of this map. We shall present an explicit example of this type when we count the Fermi multiplets associated with vector bundle moduli, and also when we consider the standard embedding for the $E_8 \times E_8$ heterotic string.

We can also assemble the count of zero modes into an overall holomorphic Euler
characteristic:%
\begin{align}
\chi\left( CY_4 , \mathcal{E}_{i}\right)   &  =h^{0}(\mathcal{E}_{i})-h^{1}%
(\mathcal{E}_{i})+h^{2}(\mathcal{E}_{i})-h^{3}(\mathcal{E}_{i})+h^{4}%
(\mathcal{E}_{i})\\
&  =-h^{1}(\mathcal{E}_{i})+h^{2}(\mathcal{E}_{i})-h^{3}(\mathcal{E}_{i}),
\end{align}
where in the second line we used the fact that for a stable vector
bundle, $h^{0}(\mathcal{E}_{i})=h^{4}(\mathcal{E}_{i})=0$. So, more
explicitly, we have:%
\begin{align}
\chi\left( CY_4 , \mathcal{E}_{i}\right)   &  =-\#(\delta\mathbb{D}_{(0,1)}%
^{(\tau_{i})})+\#(\delta\Lambda_{(0,2)}^{(\tau_{i})})-\#(\delta\mathbb{D}%
_{(0,1)}^{(\tau_{i}^{\ast})})\\
\chi\left( CY_4 ,  \mathcal{E}_{i}^{\vee}\right)   &  =-\#(\delta\mathbb{D}%
_{(0,1)}^{(\tau_{i}^{\ast})})+\#(\delta\Lambda_{(0,2)}^{(\tau_{i})})-\#(\delta\mathbb{D}_{(0,1)}^{(\tau_{i})}).
\end{align}
A helpful method for calculating such holomorphic Euler characteristics is in
terms of the Hirzebruch-Riemann-Roch index formula:%
\begin{equation}
\label{eq:HRR}
\chi\left(  CY_{4},\mathcal{E}_{i}\right)  =\underset{CY_{4}}{\int}%
\text{ch}(\mathcal{E}_{i})\text{Td}(CY_{4}),
\end{equation}
where ch$(\mathcal{E}_{i})$ is the Chern character class for $\mathcal{E}_{i}$
and Td$(CY_{4})$ is the Todd class of the tangent bundle for the Calabi-Yau fourfold.
See Appendix \ref{app:CHERN} for further details.

\subsection{Gauge Anomalies}

Having determined the zero mode content which descends from our 2D effective
field theory, it is natural to ask whether our 2D GLSM is free of
anomalies. Actually, we expect that in general the gauge theory will be
anomalous. The reason is that in the effective action there is a term of the
form $B_{2}\wedge X_{8}(F,R)$ where $X_{8}(F,R)$ depends on the characteristic
classes of the gauge bundle and tangent bundle.  Since $B_2$ transforms
under gauge transformations, and there is no a priori reason for $\int_{M} X_8$ to vanish,
we see that we should expect the GLSM sector to have an equal and opposite anomaly.
 We will revisit these terms in section \ref{sec:TAD}, where we will discuss tadpole
 cancellation.  For now we will derive the overall
contribution to the anomalies from the GLSM\ sector. In particular, our plan
in this subsection will be to repackage these contributions in terms of
topological quantities.

To this end, we study the contribution to the gauge
anomalies from matter fields transforming in a representation $\tau$ of the
unbroken gauge group $G$, as well as their \textquotedblleft
partners\textquotedblright\ transforming in the dual representation
$\tau^{\ast}$. The zero mode content is then controlled by the Dolbeault
cohomology for some holomorphic vector bundle $\mathcal{E}$ and its dual $\mathcal{E}%
^{\vee}$. Overall, we have the contribution to the gauge anomaly from
modes in the representations $\tau$ and $\tau^{\ast}$:
\begin{equation}
I_{\text{gauge}}(\tau~\text{and}~\tau^\ast) =-\text{Ind}(\tau
)\times\chi(CY_{4},\mathcal{E}).
\end{equation}
where here, Ind($\tau$) refers to the index of a representation, which in the conventions of the present paper are
such that the fundamental representation of $SU(N)$ has index one.

This counts anomaly contributions from Weyl fermions in the corresponding representations.
By including both $\tau$ and $\tau^\ast$ from the decomposition we ensure that our counting
correctly takes into account the Weyl repackaging of the fermions corresponding to self-dual
bundles.   Summing over all representations, we therefore obtain a manifestly topological formula for the
gauge theory anomaly.

%

A priori, there is no reason for these contributions to vanish, and we will
see that in a string compactification, this anomaly can either be viewed as
being cancelled by a non-local Green-Schwarz term or by the contributions
from an extra sector.

\subsection{Gravitational Anomalies}

Having discussed the gauge anomalies generated by the GLSM\ sector, we now
turn to the gravitational anomalies. First of all, there will be a
contribution from the GLSM\ sector of our theory. Its form is roughly similar
to that already discussed for the gauge anomalies, so we simply summarize with
the relevant formula. With our normalization, where a Weyl fermion contributes $+1$ to
the central charge, this is given by:
\begin{equation}
I_{\text{gravity}}(\tau~\text{and}~\tau^{\ast})= \frac{\text{dim}%
(\tau)}{12}\times\chi(CY_{4},\mathcal{E}),
\end{equation}
with notation as in the previous subsection.

Consider next the \textquotedblleft gauge singlet
sectors.\textquotedblright\ The gauge singlets of the model consists of
possible moduli fields coming from integrating a two-form potential over an
internal two-cycle (counted by $h^{1,1}$), as well as from the complex
structure moduli (counted by $h^{3,1}$) and vector bundle moduli (counted by
$h^{1}($End$\mathcal{V})$). By $(0,2)$ supersymmetry, all of these
contributions assemble into chiral multiplets with corresponding right-moving
fermionic superpartners. Additionally, we can expect there to be Fermi
multiplets which must be accounted for as well. To count these contributions,
we now observe that if we interpret gravity as a gauging of translations, then
we can count the various superpartners of the gravitino which assemble into
Fermi multiplets in a way quite similar to the method used in the context of
the Yang-Mills sector. Doing so, we see that instead of modes descending from
bundle valued $(0,1)$ gauge fields, we instead get fermions descending
from bundle valued $(0,2)$-forms. This is all to the good, because it
means the net contribution will again assemble into a set of topological
indices. Summarizing, we get the full gauge neutral contribution to the
anomaly as:%
\begin{align}
I_{\text{gravity}}(\text{neutral})  &  = - \frac{1}{12}\left(  h^{1,3}-h^{1,2}+h^{1,1}\right)  - \frac{1}{12}\left(  h^{1}(\text{End}%
\mathcal{V})-\frac{1}{2}h^{2}(\text{End}\mathcal{V})\right) \\
&  =\frac{1}{12}\left(\chi_{1}+\frac{1}{2}\chi\left(  \text{End}\mathcal{V}\right) \right)  .
\end{align}
Again, the factor of $1/2$ is taking into account the repackaging of the fermions
into Weyl representations.

\subsection{Zero Mode Interactions}

Now that we have arrived at a general formula for the zero modes in the
presence of our background vector bundle, it is natural to ask what are the
resulting interactions. These can be canonically split according to
interactions which descend from 10D\ Super Yang-Mills theory, stringy corrections, and those which
arise from non-perturbative instanton effects coming from 1-branes and
5-branes wrapping two-cycles and respectively six-cycles of the internal
geometry. At least at large volume and weak string coupling, these interaction terms are expected to be
a subleading effect, but they can be important in the IR phase of the 2D theory.
So, the best we can hope for is to deduce possible interaction terms
compatible with the symmetries of our effective field theory in two dimensions.

Along these lines, we shall focus on the dominant contributions coming from
expanding the F-term $W_{M}$ of equation (\ref{WBULK}) around our
fixed background. At this point we encounter an important subtlety in listing the F-term interactions. The key issue
is that again, we need to make sure our Fermi multiplets valued as $(0,2)$ differential forms are counted correctly, that is, we can just write the Fermi multiplets as transforming in a representation $\tau$, but not in the dual representation $\tau^{\ast}$. Rather, we absorb these would-be interaction terms into the $E$-field for $\Lambda^{(\tau)}_{(0,2)}$. The procedure for deducing these interaction terms is actually quite conveniently summarized by first writing down the fluctuations around the topological F-term $W_{top}$ of equation (\ref{WTOP}).
Doing so, we clearly get cubic F-term interactions of the form:
\begin{equation}
W_{\text{top,cubic}}=\underset{CY_{4}}{\int}\Omega\wedge(f_{\alpha\beta\gamma
}\text{ }\delta\Lambda_{(0,2)}^{(\alpha)}\wedge\delta\mathbb{D}_{(0,1)}%
^{(\beta)}\wedge\delta\mathbb{D}_{(0,1)}^{(\gamma)})
+ \underset{CY_{4}}{\int}\Omega\wedge(f_{\alpha^{\ast}\beta^{\ast}\gamma^{\ast}
}\text{ }\delta\Lambda_{(0,2)}^{(\alpha^{\ast})}\wedge\delta\mathbb{D}_{(0,1)}%
^{(\beta^{\ast})}\wedge\delta\mathbb{D}_{(0,1)}^{(\gamma^{\ast})}),
\end{equation}
where we have integrated the zero mode profiles over the internal Calabi-Yau
fourfold directions. Here, $\alpha$, $\beta$ and $\gamma$ are appropriate
representations of the unbroken gauge group $H$, and $f_{\alpha\beta\gamma}$
is a Clebsch-Gordan coefficient for the decomposition descending from the
adjoint of $G$. So for the physical theory, we instead just write the contribution from the representation $\alpha$, and not its dual,
but where we have to adjust the value of the $E$-field for $\delta\Lambda_{(0,2)}^{(\alpha)}$
as per our discussion in section \ref{sec:EFT}.

Now, in a higher-dimensional setting, we would stop at this cubic interaction
term since higher order interactions define irrelevant interaction terms
suppressed by the cutoff. However, in a general 2D model, such power counting
arguments do not apply since formally speaking, a free scalar has scaling
dimension zero. From this perspective, we must expect that integrating out
Kaluza-Klein modes of the higher-dimensional system will lead to additional
correction terms. Let us illustrate this point by focussing on quartic
interactions. Using the propagator $1/(\Omega\wedge\overline{\partial}%
_{A}^{\prime})$, i.e. where we omit the zero modes from the inverse, we see
that an exchange diagram involving Kaluza-Klein excitations generates the
interaction term:%
\begin{equation}
W_{top,\text{quartic}}=\underset{CY_{4}}{\int}h_{\alpha\beta\gamma\delta}\text{
}(\Omega\wedge\delta\Lambda_{(0,2)}^{(\alpha)}\wedge\delta\mathbb{D}%
_{(0,1)}^{(\beta)})\wedge\frac{1}{\Omega\wedge\overline{\partial}_{A}^{\prime
}}\wedge(\Omega\wedge\delta\mathbb{D}_{(0,1)}^{(\gamma)}\wedge\delta
\mathbb{D}_{(0,1)}^{(\delta)}),
\end{equation}
i.e. we have a Massey product for the resulting cohomology groups. We can
continue iterating this process to include ever higher order Massey products.
Indeed, we can continue on to higher order intersection pairings by again
contracting one of the Kaluza-Klein modes from $\mathbb{D}_{(0,1)}$ with one
of the fluctuations from $\Lambda_{(0,2)}$. Doing so, we get the more general
class of interaction terms of the schematic form:%
\begin{equation}
W_{top,(n)}=\underset{CY_{4}}{\int}(\Omega\delta\Lambda\delta\mathbb{D}%
)\wedge\left(  \frac{\Omega\delta\mathbb{D}\delta\mathbb{D}}{\overline
{\partial}_{A}^{\prime}}\right)  ^{n-1}\wedge(\Omega\delta\mathbb{D}%
\delta\mathbb{D}).
\end{equation}
As before, we can read off the physical F-term by starting with the topological F-term, differentiating with respect to modes
in the dual representation, and using that to set the value for the $E$-field for the physical Fermi multiplets.

Based on this, one might naturally ask whether there is any suppression
mechanism at all for these higher order interaction terms. Indeed, there is:
It is factors of $g_{\text{string}}$ for the string theory. Each successive
power of $\Omega\delta\mathbb{D}\delta\mathbb{D}/\overline{\partial}%
_{A}^{\prime}$ comes from exchange of a massive gauge boson of the 10D theory, and since
each such propagator comes with an additional factor of $g_{\text{string}}$, we consequently find
an additional power of $g_{\text{string}}$ so that
$W_{(n)}\sim(g_{\text{string}})^{n-3}$. When we pass to the F-theory
realization of these interactions, we will effectively resum these
contributions, resulting in a leading order quartic coupling.

\section{GLSMs from Intersecting 7-Branes \label{sec:NPERT}}

One of the interesting features of the perturbative string vacua encountered
earlier is that the dynamics of the GLSM are inevitably tied up with those of
the gravitational sector of the 2D model. Additionally, we saw one awkward feature of the 10D Majorana-Weyl condition and
the constraints it imposes on assembling the mode content into Fermi multiplets.
With lower-dimensional branes we expect that most of these issues can be bypassed. Our plan in this setion will be
to construct a 2D GLSM describing intersecting 7-branes coming from F-theory compactified on a Calabi-Yau fivefold.

Recall that to reach a two-dimensional supersymmetric Minkowski vacuum, we
consider F-theory compactified on a Calabi-Yau fivefold $Y\rightarrow B$ with
base $B$ a K\"ahler fourfold. The geometry is described in Minimal Weierstrass
form by the equation:%
\begin{equation}
y^{2}=x^{3}+fx+g
\end{equation}
where $f$ and $g$ are respectively sections of $\mathcal{O}_{B}(-4K_{B})$ and
$\mathcal{O}_{B}(-6K_{B})$. There are 7-branes localized along
components of the discriminant locus $\Delta=0$ where:%
\begin{equation}
\Delta=4f^{3}+27g^{2},
\end{equation}
that is, these are 7-branes wrapped over K\"ahler threefolds. Additionally,
there can be intersections between these 7-branes along complex surfaces.
At such intersections, we expect additional localized matter, as well as
interaction terms which couple the localized matter to the bulk modes. Triple
intersections of 7-branes occur along Riemann surfaces, i.e. complex curves.
Along these triple intersections, it is natural to expect additional
interaction terms to be localized. Four 7-branes can also form a quartic
intersection at points of the geometry, leading to additional interaction
terms between our matter fields.

In addition to these geometric intersections, there can also be various gauge
field fluxes switched on along the worldvolume of the branes, which sometimes appear in combination with
\textquotedblleft T-brane vacua\textquotedblright\ controlled by non-abelian
intersections of 7-branes \cite{Donagi:2003hh, Cecotti:2010bp} (see also
\cite{Donagi:2011jy, Anderson:2013rka, Heckman:2010qv, Collinucci:2014qfa,
Collinucci:2014taa}).

So, compared with the case of the heterotic models just studied, there are
necessarily a few additional geometric ingredients to specify, such as where
various fields and interaction terms localize. An important benefit of this
local approach, however, is that it is far more straightforward to then
evaluate possible wave function overlaps, i.e. to explicitly evaluate possible
interaction terms in the model.

With this in mind, our plan in this section will be to determine the low
energy effective action for intersecting 7-branes wrapped on K\"{a}hler
threefolds. A decoupling limit is available when the K\"ahler threefold $X$
is Fano, i.e. $-K_X > 0$ and the normal bundle has negative first Chern class.
We organize our analysis according to the corresponding
codimension, proceeding first with the bulk theory, and then proceed to
effects localized on lower-dimensional subspaces.

\subsection{Partial Twist on a K\"{a}hler Threefold}

Since we are interested in models which preserve $\mathcal{N} = (0,2)$ supersymmetry, our
first task is to understand the partial twist necessary for our bulk
7-brane wrapped on $X$ to preserve supersymmetry in the uncompactified
directions. In some sense, we have already accomplished this task via our
study of 9-branes wrapped on a Calabi-Yau fourfold. We shall therefore
pursue two routes to determine the twist. First, we explain how the heterotic
results already obtained dictate the structure of the twist and bulk interaction terms. Second, we
perform an \textquotedblleft intrinsic\textquotedblright\ computation which
makes no reference to a possible heterotic dual. Our procedure will be similar
to that used for $\mathcal{N} = 4$ Super Yang-Mills theory on a K\"ahler surface \cite{Vafa:1994tf},
and for 7-branes wrapped on a K\"ahler surface \cite{Beasley:2008dc} (see also \cite{Bershadsky:1997zs, Donagi:2008ca}).
For some discussion of 6D topological gauge theory on a Calabi-Yau threefold see reference see reference \cite{Iqbal:2003ds}.

Our primary goal is to make sure that all of the modes and interaction terms of the
eight-dimensional Yang-Mills theory can be organized according to two-dimensional $(0,2)$
supersymmetry. As in the case of the 9-brane on a Calabi-Yau fourfold, there is one non-local Wess-Zumino type
term which must be included to really maintain supersymmetric gauge invariance. This term is obtained by
reduction of the term given in reference \cite{Marcus:1983wb} (see also \cite{ArkaniHamed:2001tb}). We note, however, that in Wess-Zumino gauge (i.e. the gauge used
throughout this paper) this term vanishes.

Having dispensed with this caveat, let us recall that in flat space, eight-dimensional
super Yang-Mills theory with gauge group $G$ consists of an eight-dimensional
vector boson, a complex scalar, and fermions that transform in the $\mathbf{8}^{s}_{+1}\oplus\mathbf{8}^{c}_{-1}$ of $SO(1,7)\times U(1)_{R}$ where
$U(1)_{R}$ is the symmetric group of rotations transverse to the location of
the 7-brane. All of these fields transform in the adjoint representation of $G$, and under the $U(1)_{R}$ the complex scalar has charge $+2$.

Let us first use our results from the heterotic analysis to derive the bulk mode
content and the structure of the bulk interactions. The key point is that although our
K\"{a}hler threefold $X$ may embed in a base $B$ which is not Calabi-Yau, the
partial twist operates by essentially altering the spin content of the various
fields so that they are effectively living in the local space $\mathcal{O}%
_{X}(K_{X})\rightarrow X$ which is Calabi-Yau. With this in mind, suppose that
we specialize our discussion of the 9-brane action to this particular
Calabi-Yau fourfold. Reduction of the bulk heterotic modes, and contracting
with the holomorphic four-form and the metric in the directions normal to $X$,
we see that our bulk 9-brane modes $\Lambda_{(0,2)}$ and $\mathbb{D}%
_{(0,1)}$ now decompose as:%
\begin{align}
\text{9-brane}  &  \rightarrow\text{7-brane}\\
\Lambda_{(0,2)}  &  \rightarrow\Lambda_{(0,2)}\oplus\Lambda_{(3,1)}\\
\mathbb{D}_{(0,1)}  &  \rightarrow\mathbb{D}_{(0,1)}\oplus\Phi_{(3,0)}.
\end{align}
The 10D Majorana-Weyl constraint reduces to the constraint that $\Lambda_{(0,2)}$ and $\Lambda_{(3,1)}$
are not actually independent degrees of freedom, but instead, are describing a single Fermi multiplet's worth of degrees of freedom.
Observe that here, we naturally have achieved the desired $\mathbb{Z}_2$ symmetry used in the heterotic model to keep the bulk action
off shell. Here, this is reflected in the fact that the local holomorphic four-form
on the total space $\mathcal{O}(K_X) \rightarrow M \rightarrow X$ is:
\begin{equation}
\Omega_M = dz \wedge \Omega_{X}
\end{equation}
and the $\mathbb{Z}_2$ symmetry acts as $z \rightarrow - z$.

By a similar token, we can read off the physical F-terms for the 7-brane theory using the bulk topological F-term used for the heterotic
theory, in which we absorb the interaction terms involving $\Lambda_{(3,1)}$ into the $E$-field for $\Lambda_{(0,2)}$.
The bulk topological interaction terms for the heterotic model now descend to:%
\begin{equation}
W_{top,X}= - \underset{X}{\int}\text{Tr}(\Lambda_{(0,2)}\wedge\mathbb{D}_{(0,1)}%
\Phi_{(3,0)}) - \underset{X}{\int}\text{Tr}(\Lambda_{(3,1)}\wedge\mathbb{F}%
_{(0,2)}),
\end{equation}
so in the absence of any other interaction terms, we get the bulk F-term
equations of motion by varying with respect to the Fermi multiplets:%
\begin{equation}
F_{(0,2)}=F_{(2,0)}=0\text{ \ \ and \ \ }\overline{\partial}_{A}\phi=0\text{.}%
\end{equation}
When we want to work with a manifestly off-shell formalism which also preserves unitarity, we instead just have $\Lambda_{(0,2)}$ with
$E$-field set by:
\begin{equation}
E_{(0,2)} = \frac{\partial W_{top,X}}{\partial \Lambda_{(3,1)}} = \mathbb{F}_{(0,2)},
\end{equation}
and the associated physical F-term is just:
\begin{equation}
W_{X}= - \frac{1}{\sqrt{2}} \underset{X}{\int}\text{Tr}(\Lambda_{(0,2)}\wedge\mathbb{D}_{(0,1)}\Phi_{(3,0)}).
\end{equation}
This also follows directly from the bosonic potential for the bulk modes,
from whence we get the BPS equations $F_{(0,2)} = F_{(2,0)} = \overline{\partial}_{A} \phi = 0$.

Let us now explain how to derive this same set of modes and interactions directly using
the partial topological twist intrinsic to the 7-brane theory itself.
Recall that we are interested in a 7-brane wrapping a K\"{a}hler threefold
$X$. The structure group of the tangent bundle is $U(3)$, so we need to
further decompose our representations according to the subgroup $SO(1,1)\times
U(3)\times U(1)_{R}$. Our task is to pick a homomorphism $U(1)_{R}\rightarrow
U(3)$ such that the resulting spin content of the model organizes into
manifest $(0,2)$ supermultiplets.

With this discussion in mind, let us now turn to the explicit partial twist
for the 7-brane theory wrapped on a K\"ahler threefold. We begin with the
decomposition of $Spin(1,7)\times U(1)_{R}$ to $Spin(1,1)\times Spin(6)\times
U(1)_{R}$:%
\begin{align}
Spin(1,7)\times U(1)_{R}  &  \supset Spin(1,1)\times Spin(6)\times U(1)_{R}\\
\mathbf{8}_{+1}^{s}  &  \rightarrow\mathbf{4}_{+,+1}\oplus\overline
{\mathbf{4}}_{-,+1}\\
\mathbf{8}_{-1}^{c}  &  \rightarrow\overline{\mathbf{4}}_{+,-1}\oplus
\mathbf{4}_{-,-1}\\
\mathbf{8}_{0}^{v}  &  \rightarrow\mathbf{1}_{--,0}\oplus\mathbf{1}%
_{++,0}\oplus\mathbf{6}_{0,0}\\
\mathbf{1}_{+2}  &  \rightarrow\mathbf{1}_{0,+2}.
\end{align}
Decomposing further according to the subgroup $SU(3)_{X}\times U(1)_{X}/%
\mathbb{Z}
_{3}\subset Spin(6)$, we have:%
\begin{align}
Spin(1,7)\times U(1)_{R}  &  \supset Spin(1,1)\times SU(3)\times
U(1)_{X}\times U(1)_{R}\\
\mathbf{8}_{+1}^{s}  &  \rightarrow\mathbf{1}_{+,+\frac{3}{2},+1}%
\oplus\mathbf{3}_{+,-\frac{1}{2},+1}\oplus\mathbf{1}_{-,-\frac{3}{2},+1}%
\oplus\overline{\mathbf{3}}_{-,+\frac{1}{2},+1}\\
\mathbf{8}_{-1}^{c}  &  \rightarrow\mathbf{1}_{+,-\frac{3}{2},-1}%
\oplus\overline{\mathbf{3}}_{+,+\frac{1}{2},-1}\oplus\mathbf{1}_{-,+\frac
{3}{2},-1}\oplus\mathbf{3}_{-,-\frac{1}{2},-1}\\
\mathbf{8}_{0}^{v}  &  \rightarrow\mathbf{1}_{--,0,0}\oplus\mathbf{1}%
_{++,0,0}\oplus\mathbf{3}_{0,+1,0}\oplus\overline{\mathbf{3}}_{0,-1,0}\\
\mathbf{1}_{+2}  &  \rightarrow\mathbf{1}_{0,0,+2}\\
\mathbf{1}_{-2}  &  \rightarrow\mathbf{1}_{0,0,-2}.%
\end{align}
Our goal in specifying a twist is that the resulting $U(1)$ charge for a
spinor will then be a scalar on the K\"ahler threefold. The twist is given by
the generator:%
\begin{equation}
J_{\text{top}}=J_{X}+\frac{3}{2}J_{R}.
\end{equation}
With respect to this choice, the charge assignments for the various modes are:%
\begin{align}
Spin(1,7)\times U(1)_{R}  &  \supset Spin(1,1)\times U(3)_{X}\times
U(1)_{\text{top}}\\
\mathbf{8}_{+1}^{s}  &  \rightarrow\mathbf{1}_{+,+3}\oplus\mathbf{3}%
_{+,+1}\oplus\mathbf{1}_{-,0}\oplus\overline{\mathbf{3}}_{-,+2}\\
\mathbf{8}_{-1}^{c}  &  \rightarrow\mathbf{1}_{+,-3}\oplus\overline
{\mathbf{3}}_{+,-1}\oplus\mathbf{1}_{-,0}\oplus\mathbf{3}_{-,-2}\\
\mathbf{8}_{0}^{v}  &  \rightarrow\mathbf{1}_{--,0}\oplus\mathbf{1}%
_{++,0}\oplus\mathbf{3}_{0,+1}\oplus\overline{\mathbf{3}}_{0,-1}\\
\mathbf{1}_{+2}  &  \rightarrow\mathbf{1}_{0,+3}\\
\mathbf{1}_{+2}  &  \rightarrow\mathbf{1}_{0,-3}.%
\end{align}
At this point, we can begin to assemble our modes into appropriate vector,
chiral and Fermi multiplets. Along these lines, we observe that the fermions
of the Fermi multiplets have opposite chirality to those of the chiral
multiplets. Additionally, for the fermions of the chiral multiplets, we should
expect that both the representation under $U(3)_{X}\times U(1)_{\text{top}}$
and its dual both show up in the multiplet. Taking all of this into account,
we therefore obtain:%
\begin{align}
\text{V \ \ }V_{(0,0)}  &  :\mathbf{1}_{--,0}\oplus\mathbf{1}_{++,0}%
\oplus\mathbf{1}_{-,0}\oplus\mathbf{1}_{-,0}\\
\text{CS \ \ }\Phi_{(3,0)}  &  :\mathbf{1}_{+,+3}\oplus\mathbf{1}_{0,+3}\\
\overline{\text{CS}} \text{ \ \ } \overline{\Phi}_{(0,3)}  &  :\mathbf{1}_{+,-3}\oplus\mathbf{1}_{0,-3}\\
\text{CS \ \ }\mathbb{D}_{(0,1)}  &  :\overline{\mathbf{3}}_{+,-1}\oplus\overline{\mathbf{3}}_{0,-1}\\
\overline{\text{CS}} \text{ \ \ } \overline{\mathbb{D}}_{(1,0)}  &  :\overline{\mathbf{3}}_{+,-1}\oplus\overline{\mathbf{3}}_{0,-1}\\
\text{F \ \ } \Lambda_{(0,2)}  &  :\mathbf{3}_{-,-2}\\
\text{F \ \ } \Lambda_{(3,1)}  &  :\overline{\mathbf{3}}_{-,+2}.
\end{align}
Note that in the above, we have split up the contribution into the CS multiplets and their complex conjugates. Additionally,
the $(3,1)$ differential form is not an independent degree of freedom
separate from the $(0,2)$ differential form. The reason is that as we have already remarked, the remnant of the
10D Majorana-Weyl constraint in the 7-brane theory
means these are not really independent degrees of freedom.
Nevertheless, for the purposes of writing out possible F-term
interactions, it is helpful to keep it in mind, in particular
for determining the correct value of the $E$-field for $\Lambda_{(0,2)}$.

Summarizing, we have the supermultiplets transforming as differential forms of
the internal space. Including the possibility of a non-trivial principal $G$
bundle, these modes are sections of the following bundles:
\begin{align}
\Phi_{(3,0)}  &  \in\mathcal{O}_{X}(K_{X})\otimes adP\\
\mathbb{D}_{(0,1)}  &  \in\Omega_{X}^{(0,1)}\otimes adP\\
\Lambda_{(0,2)}  &  \in\Omega_{X}^{(0,2)}\otimes adP\\
\Lambda_{(3,1)}  &  \in\Omega_{X}^{(0,1)}(K_{X})\otimes adP.
\end{align}

Consider next the bulk equations of motion. As explained in detail in Appendix
\ref{app:FTH}, the bulk BPS\ equations of motion for the internal fields are:\footnote{Readers familiar
with the similar equations of motion for 7-branes on a K\"ahler surface found in reference \cite{Beasley:2008dc} will note
the absence of a factor of $1/2$ in our commutator for $[\phi , \overline{\phi}]$. This pre-factor can be altered by an overall rescaling of the metric in the directions normal to the 7-brane. This is due to our conventions for normalization of all fields which follows the ones commonly used in $\mathcal{N} = (0,2)$ supersymmetric models.}
\begin{align}
\text{D-terms}  &  \text{: } \omega \wedge \omega \wedge F_{(1,1)}+ \left[
\phi,\overline{\phi}\right]  =0\\
\text{F-terms}  &  \text{: }F_{(0,2)}=F_{(2,0)}=\overline{\partial}_{A}\phi=0.
\end{align}
In Appendix \ref{app:FTH} we give the full off-shell 2D equation which reproduces these
equations of motion. In particular, the F-term equations of motion directly
follow from the bulk superpotential:%
\begin{equation}
W_{top,X}= - \underset{X}{\int}\text{Tr}(\Lambda_{(0,2)}\wedge\mathbb{D}_{(0,1)}%
\Phi_{(3,0)}) - \underset{X}{\int}\text{Tr}(\Lambda_{(3,1)}\wedge\mathbb{F}%
_{(0,2)}).
\end{equation}
Observe that this is also compatible with the supersymmetric structure of the
F-terms obtained on the heterotic side. Indeed, by an appropriate reduction of
10D\ Super Yang-Mills on $\mathcal{O}_{X}(K_{X})\rightarrow X$, we realize
precisely this structure. Again, to reach the physical superpotential, we instead have a non-trivial value for the $E$-field
in $\Lambda_{(0,2)}$ given by $E_{(0,2)} = \mathbb{F}_{(0,2)}$ and simply have the superpotential:
\begin{equation}
W_{X}= - \frac{1}{\sqrt{2}} \underset{X}{\int}\text{Tr}(\Lambda_{(0,2)}\wedge\mathbb{D}_{(0,1)}%
\Phi_{(3,0)}).
\end{equation}

\subsubsection{Bulk Zero Modes}

Much as in the case of compactifications of the heterotic string on a
Calabi-Yau threefold, we can consider the zero modes associated with a vacuum
solution to the F- and D-terms described above. For simplicity, we shall
assume that the Higgs field is switched off. Then, we simply need to solve the
Hermitian Yang-Mills equations on a K\"ahler threefold $X$. Assuming we have
done so, we can consider a decomposition of the structure group for $adP$ as
$G\supset H\times K$ where we assume the gauge field fluxes define a vector
bundle with structure group $K$, with commutant $H$. Decomposing the adjoint
representation into irreducible representations of $H\times K$, we then have:%
\begin{equation}
adP\rightarrow\underset{i}{\oplus}(\tau_{i},\mathcal{E}_{i}).
\end{equation}
Hence, for a zero mode fluctuation in a representation $\tau_{i}$, the total
number are counted as:%
\begin{align}
\delta\Phi_{(3,0)}^{(\tau_{i})} &  \in H_{\overline{\partial}}^{0}%
(X,K_{X}\otimes\mathcal{E}_{i})\\
\delta\mathbb{D}_{(0,1)}^{(\tau_{i})} &  \in H_{\overline{\partial}}%
^{1}(X,\mathcal{E}_{i})\\
\delta\Lambda_{(0,2)}^{(\tau_{i})} &  \in H_{\overline{\partial}}%
^{2}(X,\mathcal{E}_{i}).
\end{align}
Additionally, the matter fields in the dual representation are:%
\begin{align}
\delta\Phi_{(3,0)}^{(\tau_{i}^{\ast})} &  \in H_{\overline{\partial}}%
^{0}(X,K_{X}\otimes\mathcal{E}_{i}^{\vee})\\
\delta\mathbb{D}_{(0,1)}^{(\tau_{i}^{\ast})} &  \in H_{\overline{\partial}%
}^{1}(X,\mathcal{E}_{i}^{\vee})\\
\delta\Lambda_{(0,2)}^{(\tau_{i}^{\ast})} &  \in H_{\overline{\partial}}%
^{2}(X,\mathcal{E}_{i}^{\vee}).
\end{align}
Observe that in the above, we have not included the contribution from the differential form $(3,1)$,
since the physical degrees of freedom are already fully accounted for by the $(0,2)$ differential form.

We can assemble these zero mode counts into a pair of indices, i.e.
holomorphic Euler characteristics for the bundle $\mathcal{E}_{i}$ and its
dual:%
\begin{align}
\chi(\mathcal{E}_{i}) &  =-\#(\delta\mathbb{D}_{(0,1)}^{(\tau_{i})}%
)+\#(\delta\Lambda_{(0,2)}^{(\tau_{i})})-\#(\delta\Phi_{(3,0)}^{(\tau
_{i}^{\ast})})\\
\chi(\mathcal{E}_{i}^{\vee}) &  =-\#(\delta\mathbb{D}_{(0,1)}^{(\tau_{i}%
^{\ast})})+\#(\delta\Lambda_{(0,2)}^{(\tau_{i}^{\ast})})-\#(\delta\Phi
_{(3,0)}^{(\tau_{i})}),
\end{align}
where in the above, we have used the fact that for a stable vector bundle
$h^{0}(\mathcal{E}_{i})=0$. This can in turn be written in terms of
characteristic classes defined on $X$ using the Hirzebruch-Riemann-Roch index formula.
In the above, we implicitly assumed that $\mathcal{E} \neq \mathcal{O}_{X}$. In the special case
where we have the trivial bundle, we have $h^{0}(\mathcal{E} = 1$, which
counts the contributions from the gauginos.

\subsection{Matter Localized on a Surface}

Much as in the case of higher-dimensional F-theory vacua, there can be various
lower dimensional subspaces where the elliptic fibration becomes more
singular. Our plan in this section will be to deduce the matter content and
interactions which localize along a collision of two 7-branes, each
localized on a K\"ahler threefold, respectively $X_{1}$ and $X_{2}$.

First of all, we see that such an intersection takes place along a K\"ahler
surface, i.e. a complex dimension two subspace:%
\begin{equation}
S=X_{1}\cap X_{2}\text{.}%
\end{equation}
We would like to understand what sort of matter fields localize at this
intersection. The structure is exactly the same as in six dimensional vacua,
as well as four-dimensional vacua. The modes will be charged
under non-trivial representations of the bulk gauge groups $G_{1}$ and $G_{2}%
$, and so we can view these modes as generalized bifundamentals. Additionally,
we can determine the geometric content of these localized modes, i.e. what
sort of differential forms we expect to have localized on $S$.

To accomplish this, we shall follow a procedure similar to the one spelled out
in \cite{Beasley:2008dc}. We can model the
intersection in terms of a parent gauge group $G_{\text{parent}}\supset
G_{1}\times G_{2}$. By activating a background  for the adjoint valued
$(3,0)$ form $\phi$, we get modes which are naturally trapped along a
lower-dimensional subspace. Expanding around this background, we also see that
there will be interactions between the bulk modes and the modes trapped at
the intersection. Essentially, we just take $W_{X}$ from the bulk and view two of
the three fields in the interaction terms as localized fluctuations. When we
do so, however, we need to ensure that our mode content and interactions
respect all symmetries of the system.

To actually deduce the form content for the modes, we now observe that in flat
space, these modes need to fill out a four-dimensional $\mathcal{N}=2$
hypermultiplet \cite{Beasley:2008dc}. This can in turn be organized as two
four-dimensional $\mathcal{N}=1$ chiral multiplets, so upon decomposing to
$(0,2)$ multiplets, we learn that we should expect two chiral multiplets
$Q\oplus Q^{c}$, and two Fermi multiplets $\Psi\oplus\Psi^{c}$, where the
superscript \textquotedblleft$c$\textquotedblright\ serves to remind us that
these modes transform in the conjugate (i.e. dual) representation of the gauge
group $G_{1}\times G_{2}$. The matter fields will transform as differential forms valued in
the bundles $\mathcal{R}_1 \times \mathcal{R}_2$ for
$Q$ and $\Psi$, and in the dual bundle for $Q^{c}$ and $\Psi^{c}$.
An important point is that when we package the Fermi multiplets into differential forms, we must ensure
that just as in the context of the heterotic models, that we properly count the total number of dynamical
degrees of freedom. This is the remnant of the 10D Majorana-Weyl constraint,
but now for localized modes.\footnote{We thank T. Weigand
for alerting us to a previous misstatement on the mode counting.}

What sort of differential forms should we expect our localized modes to be?
The answer comes by tracking down the effects of a vev for the scalars in the
$Q\oplus Q^{c}$. When we do so, we trigger a modification in the
BPS\ equations of motion for the bulk $(3,0)$ form:%
\begin{equation}
\overline{\partial}_{A}\phi=\delta_{S}\wedge \langle \langle Q^{c},Q \rangle \rangle_{ad P}, \label{poleequation}%
\end{equation}
where we have introduced an outer product $\langle \langle \cdot , \cdot \rangle \rangle_{ad P}$ with values in
$K_{S}\otimes adP$, and $\delta_{S}$ is a $(1,1)$-form delta function
distribution with support along our surface $S$. There is a related source term
equation of motion for the bulk gauge fields:
\begin{equation}
\omega \wedge \omega \wedge F_{(1,1)} + [ \phi , \overline{\phi} ] = \omega \wedge \omega \wedge \delta_S \left( \mu(\overline{Q} , Q) - \mu(Q^c , \overline{Q^c}) \right)
\end{equation}
with another outer product (i.e. moment map) $\mu( \cdot , \cdot )$ specified by a choice of unitary structure on the bundle
$K_S^{1/2} \otimes \mathcal{R}_1 \otimes \mathcal{R}_2$.

By inspection, then, we see that $Q\oplus Q^{c}$ transform as sections of bundles:%
\begin{equation}
Q\in K_{S}^{1/2}\otimes\mathcal{R}_{1}\otimes\mathcal{R}_{2}\text{ \ \ and
\ \ }Q^{c}\in K_{S}^{1/2}\otimes\mathcal{R}_{1}^{\vee}\otimes\mathcal{R}%
_{2}^{\vee},
\end{equation}
where in the above, we have introduce a choice of square-root for the
canonical bundle on $S$. Strictly speaking, all we really need is a $spin_{C}$
structure on $S$, which can be twisted by an overall line bundle contribution
descending from the gauge bundles $\mathcal{R}_{1}$ and $\mathcal{R}_{2}$.
Indeed, sometimes such contributions are inevitable due to the presence of the
Minasian-Moore-Freed-Witten anomaly \cite{Minasian:1997mm, Freed:1999vc}.
Returning to equation (\ref{poleequation}), we see that this equation of
motion comes about provided we couple the pullback of the bulk mode
$\Lambda_{(0,2)}$ to the $Q$'s:%
\begin{equation}
W_{S}\supset \frac{1}{\sqrt{2}} \underset{S}{\int} \langle Q^{c},\Lambda_{(0,2)}Q \rangle.
\end{equation}
where $\langle \cdot , \cdot \rangle$ is a canonical pairing between
$K_{S}^{1/2} \otimes \mathcal{R}^{\vee}_1 \otimes \mathcal{R}^{\vee}_2$ and $K_{S}^{1/2} \otimes \mathcal{R}_1 \otimes \mathcal{R}_2$.

Consider next the Fermi multiplets which also localize on $S$. Just as for the bulk modes encountered
previously, in this case we expect there to be a reduction in the dynamical degrees of freedom from the 10D Majorana-Weyl constraint.
Indeed, in flat space we expect to have a 4D hypermultiplet's worth of degrees of freedom present. So, it will again
be necessary to introduce the device $W_{top}$ in our discussion of the mode content as well as the interaction terms. To deduce the
F-term interactions for these modes, we observe that a necessary equation of
motion is:%
\begin{equation}
\overline{\partial}_{A}Q=0\text{ \ \ and \ \ }\overline{\partial}_{A}Q^{c} = 0,
\end{equation}
i.e. that the bulk gauge field from $X$ can couple to these modes at all. For
this to be so, we must have F-term couplings of the form:
\begin{equation}
\frac{1}{\sqrt{2}} \underset{S}{\int}\left\langle  \Psi^{c},\left(  \overline{\partial}+\mathbb{A}%
_{1}+\mathbb{A}_{2}\right)  Q\right \rangle  +\left \langle  Q^{c},\left(  \overline
{\partial}+\mathbb{A}_{1}+\mathbb{A}_{2}\right)  \Psi\right \rangle  ,
\end{equation}
where $\mathbb{A}_{i}$ corresponds to the pullback of the chiral multiplet which
transforms as a $(0,1)$ gauge field on each bulk 7-brane.

This in turn fixes the form content of the modes. We must have:
\begin{equation}
\Psi\in\Omega_{S}^{(0,1)}(K_{S}^{1/2}\otimes\mathcal{R}_{1}\otimes
\mathcal{R}_{2})\text{ \ \ and \ \ }\Psi^{c}\in\Omega_{S}^{(0,1)}(K_{S}%
^{1/2}\otimes\mathcal{R}_{1}^{\vee}\otimes\mathcal{R}_{2}^{\vee}).
\end{equation}
In this case, we also see that there is only one physically independent Fermi multiplet.
In what follows, we take it to be $\Psi$ rather than $\Psi^c$.

To summarize then, along each intersection, we have localized matter fields,
and these fields transform in the following representations:%
\begin{align}
Q  &  \in K_{S}^{1/2}\otimes\mathcal{R}_{1}\otimes\mathcal{R}_{2}\\
Q^{c}  &  \in K_{S}^{1/2}\otimes\mathcal{R}_{1}^{\vee}\otimes\mathcal{R}%
_{2}^{\vee}\\
\Psi &  \in\Omega_{S}^{(0,1)}(K_{S}^{1/2}\otimes\mathcal{R}_{1}\otimes
\mathcal{R}_{2})\\
\Psi^{c}  &  \in\Omega_{S}^{(0,1)}(K_{S}^{1/2}\otimes\mathcal{R}_{1}^{\vee
}\otimes\mathcal{R}_{2}^{\vee}).
\end{align}
We also have interaction terms between one bulk mode (i.e. its pullback onto
the surface $S$) and two matter fields:%
\begin{equation}
W_{top,S}= \underset{S}{\int} \langle Q^{c},\Lambda_{(0,2)}Q \rangle + \left \langle  \Psi^{c},\left(
\overline{\partial}+\mathbb{A}_{1}+\mathbb{A}_{2}\right)  Q\right \rangle  +  \left \langle
Q^{c},\left(  \overline{\partial}+\mathbb{A}_{1}+\mathbb{A}_{2}\right) \Psi \right \rangle  .
\end{equation}

It is also of interest to work out the resulting contribution to the bosonic potential. This leads to modified kinetic terms for the internal degrees of freedom, as reflected in the bulk and surface localized contributions to the action. The energy density $\mathcal{U}$
localizes in the internal directions as:
\begin{align}
\mathcal{U}_{\text{Bulk+Surface}} &  = \left\Vert F_{(0,2)}\right\Vert_{X}
^{2}+\left\Vert \overline{\partial}_{A}\phi-\delta_{S}\left\langle
\left\langle Q^{c},Q\right\rangle \right\rangle _{adP}\right\Vert_{X}^{2}\\
&  +\left\Vert \omega \wedge \omega \wedge F_{(1,1)}+ [\phi,\overline{\phi}]- \omega \wedge
\omega \wedge\delta_{S}\left(  \mu(\overline{Q},Q) - \mu(Q^{c},\overline{Q^{c}})\right)  \right\Vert_{X}^{2}\\
&  +\left\Vert \overline{\partial}_{A_{1}+A_{2}}Q\right\Vert _{S}%
^{2}+\left\Vert \overline{\partial}_{A_{1}+A_{2}}Q^{c}\right\Vert _{S}^{2}%
\end{align}
let us also note that there are additional corrections to this structure once we include interactions localized along Riemann surfaces and points.

\subsubsection{Localized Zero Modes}

We can also use the analysis presented above to determine the zero mode
content of our localized zero modes. These are counted by the following
cohomology groups:%
\begin{align}
\delta Q  &  \in H_{\overline{\partial}}^{0}(K_{S}^{1/2}\otimes\mathcal{R}%
_{1}\otimes\mathcal{R}_{2})\\
\delta Q^{c}  &  \in H_{\overline{\partial}}^{0}(K_{S}^{1/2}\otimes
\mathcal{R}_{1}^{\vee}\otimes\mathcal{R}_{2}^{\vee})\\
\delta\Psi &  \in H_{\overline{\partial}}^{1}(K_{S}^{1/2}\otimes
\mathcal{R}_{1}\otimes\mathcal{R}_{2})\\
\delta\Psi^{c}  &  \in H_{\overline{\partial}}^{1}(K_{S}^{1/2}\otimes
\mathcal{R}_{1}^{\vee}\otimes\mathcal{R}_{2}^{\vee}).
\end{align}
Here, we note that much as in our heterotic models, the modes $\Psi$ and $\Psi^c$ are not
independent degrees of freedom.

We can in turn introduce an index formula which counts appropriate
combinations of these zero modes:%
\begin{equation}
\chi(S, K_{S}^{1/2}\otimes\mathcal{R}_{1}\otimes\mathcal{R}_{2})    =\#(\delta
Q)-\#(\delta\Psi)+\#(\delta Q^{c}).
\end{equation}
Of course by Serre duality we could equivalently count the zero modes using the
dual representation.

\subsection{Interactions Localized on a Curve and a Point}

Geometrically, we can also see that three components of the discriminant locus
can intersect along a Riemann surface. Although this is a
\textquotedblleft non-generic\textquotedblright\ intersection inside of
a\ K\"ahler threefold, it is rather natural in the context of an F-theory
compactification because we reach such configurations by Higgsing a parent 7-brane gauge theory. That is to say,
we can locally expand $\phi_{(3,0)}$ around a non-zero value, and the breaking patterns will include a cubic interaction
between localized fluctuations trapped on pairwise intersections. This follows the same analysis presented for
example in references \cite{Beasley:2008dc, Beasley:2008kw, Cecotti:2009zf, Cecotti:2010bp}.
Assuming, therefore, that we have three K\"ahler surfaces $S_{1}$, $S_{2}$ and $S_{3}$ inside of $X$, we
would like to determine what sorts of couplings will be present between three
such fields at the common locus of intersection, which we denote by $\Sigma$.

Owing to the structure of $\mathcal{N} = (0,2)$ F-term interactions, we must couple a Fermi
multiplet with some number of chiral multiplets. To set conventions, we
suppose that we are given three K\"ahler surfaces with fields in the following
bundle assignments on corresponding surfaces $S_{i}$:%
\begin{align}
\Psi_{1}  &  \in\Omega_{S_{1}}^{(0,1)}(K_{1}^{1/2}\otimes\mathcal{V}_{1})\\
Q_{2}  &  \in K_{2}^{1/2}\otimes\mathcal{V}_{2}\\
Q_{3}  &  \in K_{3}^{1/2}\otimes\mathcal{V}_{3}.
\end{align}
Since we are assuming we have a gauge invariant interaction anyway, we can
also assume that tensor product of the restrictions of the $\mathcal{V}_{i}$
are trivial:
\begin{equation}
\mathcal{V}_{1}|_{\Sigma}\otimes\mathcal{V}_{2}|_{\Sigma}\otimes
\mathcal{V}_{3}|_{\Sigma}=\mathcal{O}_{\Sigma}.
\end{equation}
This leaves us with the task of studying the bundle:%
\begin{equation}
\mathcal{B}=\Omega_{S_{1}}^{(0,1)}(K_{1}^{1/2})|_{\Sigma}\otimes K_{2}%
^{1/2}|_{\Sigma}\otimes K_{3}^{1/2}|_{\Sigma}.
\end{equation}
Now, by the adjunction formula, we can write:%
\begin{equation}
K_{1}^{1/2}|_{\Sigma}\otimes K_{2}^{1/2}|_{\Sigma}\otimes K_{3}^{1/2}%
|_{\Sigma}=K_{\Sigma}^{3/2}\otimes\left(  \mathcal{N}_{1}\otimes
\mathcal{N}_{2}\otimes\mathcal{N}_{3}\right)  ^{-1/2},
\end{equation}
where the $\mathcal{N}_{i}$ denotes the normal bundle for $\Sigma$ in the
surface $S_{i}$. On the other hand, the very fact that we have a triple
intersection of K\"ahler surfaces inside our threefold in the first place means
that $\mathcal{N}_{1}\otimes\mathcal{N}_{2}\otimes\mathcal{N}_{3}\simeq
K_{\Sigma}$. So, we therefore learn that:%
\begin{equation}
\mathcal{B}=\Omega_{\Sigma}^{(0,1)}(K_{\Sigma}),
\end{equation}
i.e. the triple intersection defines a $(1,1)$ volume form which can be
integrated over the Riemann surface.

As a brief aside, we note that such interaction terms should be expected:   if
we specialize to the case of $X=T^{2}\times S$, we have the dimensional
reduction of a 4D $\mathcal{N}=1$ theory on a $T^{2}$, and it is well known
that cubic Yukawa interactions localize at points of such constructions
\cite{Beasley:2008dc}.

Consider next the possibility of intersections localized at a point of the
K\"ahler threefold $X$. In the Calabi-Yau fivefold geometry, this originates
from a quartic intersection of components of the discriminant locus. The
novelty with the present situation is that because of the Higgsing patterns
available in an intersecting 7-brane configuration, we can expect four
K\"{a}hler surfaces to intersect at a point.\footnote{Strictly speaking, this
will actually involve a branched cover of the original threefold $X$, and it
is the different sheets of the cover which are forming the quartic
intersection.\ This is also true for cubic intersections. This point has been
explained in detail, for example, in references \cite{Cecotti:2010bp, Hayashi:2009ge}.}
Indeed, as we already mentioned in the context of heterotic constructions,
such interactions are expected to be present upon integrating some
Kaluza-Klein modes. The novelty in F-theory is that these interactions appear
to be geometrically localized at a point. For this reason, we can write the
general form of such interactions, assuming of course that a gauge invariant
interaction term is possible at all:%
\begin{equation}
W_{top,p}= h_{\alpha\beta\gamma\delta}\text{ }\left(  \delta\Psi^{(\alpha)}\delta
Q^{(\beta)}\delta Q^{(\gamma)}\delta Q^{(\delta)}\right)  |_{p}.
\end{equation}

\subsection{Summary of Interaction Terms}

Compared with the relatively concise form of the interaction terms presented
for the heterotic models, for the F-theory models we see that there are
various matter fields and interaction terms localized along subspaces of the
K\"ahler threefold. We now collect the relevant F-term interactions in one
place. The full $W_{top}$ is given by:%
\begin{equation}
W_{top}=W_{top,X}+\underset{S}{\sum}W_{top,S}+\underset{\Sigma}{\sum
}W_{top,\Sigma}+\underset{p}{\sum}W_{top,p}%
\end{equation}
where:%
\begin{align}
W_{top,X}  &  = - \underset{X}{\int}\text{Tr}(\Lambda_{(0,2)}\wedge\mathbb{D}%
_{(0,1)}\Phi_{(3,0)}) - \underset{X}{\int}\text{Tr}(\Lambda_{(3,1)}%
\wedge\mathbb{F}_{(0,2)})\\
W_{top,S}  &  = \underset{S}{\int} \langle Q^{c},\Lambda_{(0,2)}Q \rangle +\left\langle  \Psi^{c},\left(
\overline{\partial}+\mathbb{A}_{1}+\mathbb{A}_{2}\right)  Q\right \rangle  +\left \langle
Q^{c},\left(  \overline{\partial}+\mathbb{A}_{1}+\mathbb{A}_{2}\right)
\Psi\right \rangle \\
W_{top,\Sigma}  &  = \underset{\Sigma}{\int}f_{\alpha\beta\gamma}\text{ }\delta
\Psi^{(\alpha)}\delta Q^{(\beta)}\delta Q^{(\gamma)}\\
W_{top,p}  &  = h_{\alpha\beta\gamma\delta}\text{ }\left(  \delta\Psi^{(\alpha
)}\delta Q^{(\beta)}\delta Q^{(\gamma)}\delta Q^{(\delta)}\right)  |_{p}.
\end{align}
Where in the above, we have omitted the implicit construction of $W$ and the $E$-fields which follows
from $W_{top}$.

Including the D-term interactions, we can assemble the full action:
\begin{align}
  S_\mathrm{total} & = S_{D} + S_{F} \\
  S_{D} & =  \int d^2 y d^{2} \theta \int_{X} \; \biggl( \frac{1}{8} \left( \overline{\Upsilon} , \Upsilon\right) - \frac{1}{2} \left(\overline{\Lambda_{(0,2)}}, \Lambda_{(0,2)}\right) \\
        & -\frac{i}{2} \left( \,\overline{\Phi_{(3,0)}}\,,\, [  \nabla_-, \Phi_{(3,0)}]\,\right)  - \frac{i}{2}\left(\overline{\mathbb D_{(0,1)}} , [\nabla_-,\mathbb{D}_{(0,1)}]\right)   \\
&+ \delta_S \wedge \left( - \frac{i}{2} \left( \overline{Q} , \nabla_-Q \right) - \frac{i}{2}  \left(\overline{Q^c} , \nabla_- Q^c \right)
- \frac{1}{2} \left(  \overline{\Psi}, \Psi \right) \right) \biggr)\\
  S_{F} & = \int d^{2} y d^{2} \theta^{+} W + h.c.
\end{align}
where the gauge coupling of the 2D GLSM is set by:
\begin{equation}
\frac{1}{e^2} = \mathrm{Vol}(X),
\end{equation}
in 10D Planck units.

Finally, much as in the case of the heterotic models encountered previously,
we observe that the condition for off-shell supersymmetry will be violated,
i.e. $\overline{\mathcal{D}}_{+}W \neq0$, even
though on-shell we have satisfied all supersymmetric equations of motion. Just
as in the heterotic context, the condition here is the same: We must couple
our model to the geometric moduli of the system so that this off-shell
condition is retained. From the perspective of our local gauge theory
construction, one way to ensure this is to introduce an appropriate Fermi
multiplet Lagrange multiplier. Concretely, these can be extracted by
following through the dimensional reduction of the 9-brane action with
superpotential term $(\Omega-\Pi)\wedge$Tr$(\Lambda_{(0,2)}\wedge
\mathbb{F}_{(0,2)})$ and tracking the descent of $\Pi$ into the
intersecting 7-brane action.

\subsection{Anomalies}

In our discussion above, we have focussed on elements which can be calculated
in various local patches of an F-theory model.\ It is also of interest to
study the question of whether our resulting spectrum of states is indeed
anomaly free. To address this question, we strictly speaking need a more
global picture on the contribution to both gauge anomalies and gravitational
anomalies. For some global F-theory vacua, this can be addressed using the
spectral cover construction (see e.g. \cite{Hayashi:2009ge, Donagi:2009ra,
Marsano:2009ym}), though in some cases even this tool is unavailable (i.e. if
all interaction terms do not descend from the unfolding of a single globally
defined $E_{8}$ singularity). For this reason, in this section we shall focus
on the contribution to anomalies from the local model (see also \cite{Schafer-Nameki:2016cfr}).

With this in mind, let us calculate the contribution to the gauge anomalies
due to the bulk zero modes and the zero modes localized on a K\"{a}hler
surface. Adopting similar notation to that used in our analysis of 9-brane
actions, we assume we have a zero mode transforming in a representation $\tau$
and which transforms as a section of the bundle $\mathcal{E}$. From the bulk
zero modes, we get the contribution:
\begin{equation}
I_{X}(\tau \,\,\mathrm{and}\,\, \tau^{\ast}) = -\text{Ind}(\tau)\times(\chi(X,\mathcal{E}%
)+\chi(X,\mathcal{E}^{\vee}))~.
\end{equation}
Observe that in contrast to the 9-brane theory studied in the previous section,
the counting of bulk Fermi multiplets is slightly different.
For matter fields localized on a K\"ahler surface transforming in a non-trivial
representation $r$ and as a section of the bundle $\mathcal{R}$, we also
find a contribution to the gauge anomaly, now given by:%
\begin{equation}
I_{S}(r \,\,\mathrm{and}\,\, r^{\ast}) = -\text{Ind}(r)\times \chi(S,K_{S}^{1/2}\otimes\mathcal{R})~.
\end{equation}

For the gravitational anomalies, we must include numerous fields in the
reduction, which will in turn require us to globally correlate the
contributions from various fluxes. We therefore defer a full treatment
of such cases to particular examples, and also refer the interested reader to
reference \cite{Schafer-Nameki:2016cfr}.

\section{Anomalies and Tadpoles \label{sec:TAD}}

In the previous sections we focussed on the GLSM sector generated by either a
9-brane or a configuration of intersecting 7-branes. One of the
interesting features of working in two dimensions is that we have seen that a
priori, there is no reason for the GLSM we have so constructed to be anomaly
free. Indeed, when we turn to explicit examples, we will typically find that
in isolation, the GLSM suffers from an anomaly.

From the perspective of a two-dimensional effective field theorist, there are
two quite related ways one might attempt to \textquotedblleft
repair\textquotedblright\ such an anomalous gauge theory. One way is to
simply introduce additional degrees of freedom. By 't Hooft anomaly matching, these contributions
can in turn be captured by simply adding a non-local two-form potential which
transforms under a gauge transformation with parameter $\varepsilon$ as:%
\begin{equation}
\delta_{\varepsilon}B\sim\text{Tr}(\varepsilon\cdot dA).
\end{equation}
Indeed, this is simply the dimensional reduction of the famous Green-Schwarz
mechanism to two dimensions.

Of course, these two ways of cancelling anomalies are actually quite closely
related. For example, in the context of the perturbative type I and heterotic
theories, we have a coupling in the ten-dimensional action of the form:%
\begin{equation}
S_{\text{Green-Schwarz}}\propto\int B_{2}\wedge X_{8}(F,R),
\end{equation}
where $X_{8}(F,R)$ depends on the 9-brane gauge field strengths as well as
the background curvatures of the model. When there is a non-zero background
value for $X_{8}$, we can integrate it over our eight-manifold on which we
have compactified. Doing so, we generate a term given by integrating $B_{2}$
over our 2D spacetime.

Now, as has been appreciated in other contexts for some time (see e.g.
\cite{ Dasgupta:1996yh, Vafa:1995fj, Sethi:1996es}), this in turn generates a
tadpole for the two-form potential which must be cancelled by introducing
additional branes which couple to this potential. For the type I\ theory,
these are spacetime filling D1-branes, and for the perturbative heterotic
theories these are fundamental strings. These brane theories each enjoy an
effective flavor symmetry from the ambient 9-brane, and as such, we expect
them to contribute matter fields to the GLSM\ sector. More precisely, we
expect there to be additional 2D currents which contribute to the gauge
theory.

Turning next to F-theory, we can also see that we should in general expect
there to be a tadpole which will now be cancelled by D3-branes wrapped on
two-cycles. An interesting feature of these models is that we can have a D3-brane wrapping
a two-cycle which either intersects a 7-brane at a point, or we can have D3-branes
wrapped over a two-cycle which is also common to the 7-brane.
In the former case, we get the F-theory analogue of the spacetime
filling 1-branes seen in the type I\ and heterotic models. In the latter
case, we get the F-theory analogue of five-branes of these models. One can of
course incorporate such ingredients also in our theories based on 9-branes.

Our plan in this section will therefore be to give a general
discussion of the contribution from tadpoles in the perturbative string models
encountered previously. We then consider the analogous contribution in
F-theory models.

\subsection{Perturbative Vacua}

In this subsection we consider anomaly cancellation and induced tadpoles for
perturbative vacua with 9-branes, i.e. we assume we have compactified the
perturbative type I, or heterotic string. Since we shall assume a perturbative
vacuum, we exclude the presence of five-branes. As we explain, these can be
incorporated in a straightforward manner.

To start, we recall that the choice of gauge group implies that the anomaly
polynomial for $d=10$ heterotic supergravity factorizes as:%
\begin{equation}
I_{12}=Y_{4}X_{8}%
\end{equation}
where (see e.g. \cite{Green:1987mn, Polchinski:1998rr}):
\begin{align}
Y_{4}  &  =\operatorname{tr}_{d,(d)}R^{2}-\frac{1}{30}\operatorname{Tr}F^{2}\\
X_{8}  &  =\operatorname{tr}R^{4}+\frac{1}{4}\left(  \operatorname{tr}%
R^{2}\right)  ^{2}-\frac{1}{30}\operatorname{Tr}F^{2}\operatorname{tr}%
R^{2}+\frac{1}{3}\operatorname{Tr}F^{4}-\left(  \frac{1}{30}\operatorname{Tr}%
F^{2}\right)  ^{2}~.
\end{align}
The first trace, $\operatorname{tr}_{d,(d)}%
$, is in the fundamental representation of $\SO(d)$.\footnote{For us
$d=10$ is the starting point, but we will also be interested in $d=8$ when we
compactify.} $\operatorname{Tr}$ is defined as follows: for a simple Lie
algebra it is the trace in the adjoint representation normalized so that the
longest root has length squared $2$; for any semi-simple Lie algebra like
$\mathfrak{e}_{8}\oplus\mathfrak{e}_{8}$ it is given by a sum of the traces in
the simple pieces. When it is not likely to cause confusion, we will drop the
qualifications on the traces.

In a compactification to two dimensions, the anomaly polynomial is given by
taking $I_{12}$ and integrating over an eight-manifold. That is, we wind up
with a formal four-form (as appropriate for anomalies in two dimensions). In
particular, we expect the structure of the anomaly polynomial to be controlled
by topological data of the internal manifold.\ Along these lines, for a real vector
bundle $E$, introduce the Pontryagin classes:%
\begin{equation}
\operatorname{tr}F^{2}=-2(2\pi)^{2}p_{1}(E)~,\qquad\operatorname{tr}F^{4}%
=2(2\pi)^{4}(p_{1}^{2}(E)-2p_{2}(E)).
\end{equation}
We also introduce related Pontryagin classes for the tangent bundle, which we
write as $p_{i}$. Then, the resulting form of $X_{8}$ for the perturbative
theories with gauge group $Spin(32)/%
\mathbb{Z}
_{2}$ and $E_{8}\times E_{8}$ are:%
\begin{align}
\frac{1}{(2\pi)^{4}}X_{8}^{SO(32)}  &  =3p_{1}^{2}-4p_{2}-4p_{1}\times
p_{1}(E)+16p_{1}(E)^{2}-32p_{2}(E)\\
\frac{1}{(2\pi)^{4}}X_{8}^{E_{8}\times E_{8}}  &  =3p_{1}^{2}-4p_{2}-4\left(
p_{1}(E_{1})+p_{1}(E_{2})\right)  p_{1}+8\left(  p_{1}(E_{1})^{2}+p_{1}%
(E_{2})^{2}-p_{1}(E_{1})p_{1}(E_{2})\right)  ~.
\end{align}
These expressions can be simplified by using the solution to the Bianchi identity,
which requires (without 5-branes)
\begin{align}
p_1(M_8) = p_1(E)~
\end{align}
in SO(32) theories and
\begin{align}
p_1(M_8) = p_1(E_1) + p_1(E_2)
\end{align}
in the $E_8\times E_8$ theory.  Using these simplifications and specializing further to the case of holomorphic vector bundles where $E\otimes \C ~=~ \cE\oplus\overline{\cE}$, so that 
\begin{equation}
p_{i}\left(  E\right)  =(-1)^{i}c_{2i}(\mathcal{E}),
\end{equation}
we write the eight-forms as:%
\begin{align}
\frac{1}{(2\pi)^{4}}X_{8}^{SO(32)}  &  =8\left(  -\chi(M_{8})+3c_{2}%
(M_{8})^{2}-8c_{4}(\mathcal{E})\right) \\
\frac{1}{(2\pi)^{4}}X_{8}^{E_{8}\times E_{8}}  &  =8\left(  -\chi
(M_{8})+3c_{2}(M_{8})^{2}-12c_{2}(\mathcal{E}_{1})c_{2}(\mathcal{E}%
_{2})\right)
\end{align}
where in the above, $\chi(M_{8})=c_{4}(M_{8})$ is the Euler class on a
complex manifold. For additional details on Chern class manipulations
see Appendix \ref{app:CHERN}.

If $M_8$ is an irreducible CY, then we can obtain the integrated versions of
these classes:
\begin{align}
\label{eq:X8M8}
\frac{1}{192(2\pi)^{4}} \int_{M_8} X_{8}^{SO(32)}  &= 60 -\frac{1}{3} \int_{M_8} c_4(\cE)~, \nonumber\\
\frac{1}{192(2\pi)^{4}} \int_{M_8} X_{8}^{E_{8}\times E_{8}}  & =  60 -\frac{1}{2} \int_{M_8} c_2(\cE_1) c_2(\cE_2)~.
\end{align}
Thus, for generic  stable vector bundles on our eight manifold $X_{8}$
will integrate to a non-zero number. So, a two-form potential term is
inevitable, and its participation in the Green-Schwarz mechanism is also
required. This also means there is a tadpole which must be cancelled by some
number of spacetime filling 1-branes. In a perturbative vacuum, we determine
the total number of such branes by integrating $X_{8}$ over our eight
manifold. In Appendix \ref{app:NORM} we determine the precise normalization
factor in the effective action, finding:%
\begin{equation}
N_{\text{1-branes}}=-\frac{1}{192(2\pi)^{4}}\int_{M_{8}}X_{8}~.
\label{Nonebranes}%
\end{equation}
An important feature of this constraint is that in a supersymmetric vacuum,
$N_{\text{1-branes}}\geq0$. So in other words, we get a non-trivial
 restriction on the topology of the compactification manifold and bundle. This
is very much as in higher-dimensional models, except that here it occurs in
very standard constructions (like the standard embedding).

We might also wonder not only about signs but also integrality of $N$.  As we will
see below, $c_4(\cE)$ will be divisible by six.\footnote{This is familiar in the case
of the tangent bundle from~\cite{Sethi:1996es}.}  Thus, in the $SO(32)$ case there
is no issue with integrality.  On the other hand, in the case of the $E_8\times E_8$ string
 it is unclear to us whether $c_2(\cE_1)c_2(\cE_2)$ is necessarily an even class.

\subsection{Non-Perturbative Vacua}

Let us now turn to a similar analysis for non-perturbatively realized vacua.
One mild way to extend the above results is to consider non-perturbative vacua
in which for the type I$\ $and heterotic models, $Y_{4}$ is not
cohomologically trivial. In these cases, we also have spacetime five-branes
wrapped over four-cycles, and our models are best viewed as some limit of
heterotic M-theory. Since we are then inevitably dealing with a
non-perturbatively realized vacuum, there seems little point in not simply
passing directly to the F-theory realizations of this and related models.

Along these lines, we can consider the issue of anomaly cancellation for these
models which is now accomplished through the presence of spacetime filling
D3-branes. It is straightforward to determine the homology class wrapped by
the D3-branes. We simply consider F-theory on the background $S^{1}\times
CY_{5}$, and pass to the dual M-theory model on a Calabi-Yau fivefold. There,
the D3-branes are instead represented by spacetime filling M2-branes. So, we
can simply tally up the total homology class wrapped by these M2-branes. This
follows from the terms $C_{3}\wedge G_{4}\wedge G_{4}$ and $C_{3}\wedge
X_{8}(R)$. The end result is that the two-cycle wrapped by the D3-branes is:%
\begin{equation}
\lbrack\Sigma_{D3}]=\frac{1}{2} \left(\frac{G_{4}}{2 \pi}\wedge \frac{G_{4}}{2 \pi} \right) - \frac{1}{48}\left(
p_{2}(CY_5)-\frac{1}{4}p_{1}(CY_5)^2 \right)  .
\end{equation}
In general, we see that there can be D3-branes which wrap two-cycles also
wrapped by 7-branes, and we can also have D3-branes which only intersect
at a point. This gives rise to different types of extra sectors.

\section{Extra Sectors \label{sec:EXTRA}}

So far, our discussion has focussed on the physics associated with
higher-dimensional 9-branes (and for F-theory, 7-branes). We have also
seen that an inevitable feature of these models is the appearance of a tadpole
for the two-form potential which is necessarily cancelled by the presence of
additional spacetime filling branes. By inspection, these branes must couple
to the relevant two-form potential, and as such, we can expect an additional
``extra sector'' in addition to the 2D\ GLSM\ sector realized by the
higher-dimensional branes.

In this section we switch perspective, and focus on the physics of the
extra sector, treating the higher-dimensional brane as a flavor symmetry for
this sector. First, we
consider the special case of extra sectors in compactifications of type I string theory. Here, the
extra sectors are realized by probe D1-branes which fill the 2D spacetime and sit at a point of
the Calabi-Yau fourfold. We then turn to generalizations of these extra sectors
for both perturbative heterotic vacua and F-theory vacua.  With these examples in mind,
we then make some general remarks about a curious tension between cancelling
anomalies and preserving supersymmetry.

\subsection{Perturbative Type I\ Models}

Consider first the case of compactifications of perturbative type I\ strings
on a Calabi-Yau fourfold. In this case, there are no spacetime filling
five-branes, but the tadpole for the RR two-form indicates that there are
$N$ spacetime filling D1-branes. The spectrum of this theory has
been studied for example in \cite{Polchinski:1995df}, and in T-dual form has
also been considered in detail in \cite{Bachas:1997kn}.

Let us first recall the worldvolume theory for $N$ D1-branes in flat space in
type IIB\ string theory. First, we observe that the worldvolume theory has
$\mathcal{N}=(8,8)$ worldvolume supersymmetry. Recall that in type IIB\ string
theory, the bosonic mode content for $N$ D1-branes in flat space consists of a
$U(N)$ gauge theory, with eight real scalars $X^{I}$ in the adjoint of $U(N)$
transforming in the $\mathbf{8}^{v}$ representation of $SO(8)$. We also have
sixteen Majorana-Weyl fermions $\Psi^{A}\oplus\widetilde{\Psi}^{A^{\prime}}$ transforming in the $\mathbf{8}^{s}%
\oplus \mathbf{8}^{c}$ and the adjoint of $U(N)$.

Consider next the worldvolume theory of the D1-brane in the type I\ theory.
Owing to the orientifold projection, the worldvolume theory now has $(0,8)$
worldvolume supersymmetry. In addition to the $1-1$ strings, we also have $9-1$
strings stretched from the spacetime filling 9-branes with gauge group
$Spin(32)/%
\mathbb{Z}
_{2}$ to the stack of D1-branes. Additionally, the mode content of the
D1-brane theory will be different due to the presence of the orientifold
projection. The $1-1$ strings for the gauge fields will now
organize according to an $O(N)$ gauge theory.\footnote{The fact that the gauge
group is $O(N)$ rather than $SO(N)$ is due to the presence of an overall
global $\mathbb{Z}_{2}$ Wilson line which can be activated in the type
I\ theory. Indeed, this $\mathbb{Z}_{2}$ discrete gauge symmetry implements the type
I\ analogue of the GSO\ projection for the heterotic fundamental string \cite{Polchinski:1995df}.}
Moreover, the $X^{I}$ and $\Psi^{A}$ transform in $\text{Sym}^2 \mathbf{N}$, and the $\widetilde{\Psi}^{A^{\prime}}$
transform in the $\wedge^2\rep{N} = \text{adjoint}$ representation and are the $(0,8)$ gauginos. Finally, we also
have the $9-1$ strings $\gamma$. These are left-moving fermions which transform in the bifundamental representation $(\mathbf{F},\mathbf{N})$, where here we have indicated the \textquotedblleft
flavor\textquotedblright\ 9-brane as an $\SO(F)$ gauge group with fundamental
representation of dimension $F$.

To proceed further, it will be helpful to organize the various multiplets
according to a holomorphy convention compatible with $\mathcal{N} = (0,2)$ supersymmetry.
Along these lines, we decompose the fields according to the subalgebra $\mathfrak{su}%
(4)\subset\mathfrak{so}(8)$. By the same logic applied to the bulk field
theory, we can trace through the effects of the twisting operation on these
representations. Doing so, we find that the fields of our extra sector now
combine---as expected---into various $(0,2)$ supermultiplets.
The real scalars $X^{I}$ and fermions $\Psi^{A}$ form a
chiral multiplet $\mathbb{X}_{(0,1)}$ which transforms in the $\mathbf{4}$ of
$SU(4)\subset SO(8)$ and the two-index symmetric representation of $O(N)$. We
also have a Fermi multiplet $\Lambda^{(even)}_{(0,2)}$ which transforms in the $\mathbf{6}$ of
$SU(4)$, and the adjoint representation of $O(N)$. Again, here the 10D Majorana-Weyl constraint
effectively halves the degrees of freedom which would have been present for a $(0,2)$ differential
form. The remaining light $1-1$ strings are the gauge fields
and gauginos valued in the adjoint representation of $O(N)$.

We also have the $9-1$ strings which transform in the bifundamental representation $(\mathbf{F},\mathbf{N})$ of $\SO(F)\times O(N)$.  At first sight, it appears difficult to write an off-shell (0,2) action for these Majorana-Weyl fermions.  However, the key is that these are left-moving degrees of freedom without any potential terms.  Thus, for all intents and purposes we can treat them as a left-moving current algebra gauged by the $\SO(F)\times O(N)$ gauge fields.  Treated in this form, we can write the requisite supersymmetric couplings.
A WZW presentation of this structure was explored in~\cite{Distler:2007av}.
In what follows, we will not delve into such an off-shell presentation.  Instead, we will just
discuss the free fermion presentation of this current algebra, so that in this sector our supersymmetry
will only close on-shell.

Let us now turn to interaction terms between the various modes of our extra
sector. We primarily focus on the F-terms, as they are protected by
supersymmetry. To begin, consider the interactions just involving the $1-1$
strings. The bulk interaction terms presented earlier allow us to write a
corresponding F-term. For ease of exposition, we present this using the topological version of the
superpotential, and use the prescription outlined in section \ref{sec:EFT} to read off the physical superpotential:
\begin{equation} \label{Wone}
W^{1-1}_{(top)}= - \Omega \wedge\text{tr}\left(  \Lambda
_{(0,2)}\wedge\mathbb{X}_{(0,1)}\wedge\mathbb{X}_{(0,1)}\right)  ,
\end{equation}
in the obvious notation (in particular the $\tr$ is in the fundamental representation of $O(N)$).  Since the adjoint
representation is just the two-index anti-symmetric representation, this is gauge invariant.

Consider next the interactions which involve the $9-1$ string $\gamma$. This is a Majorana-Weyl spinor
transforming in the bifundamental representation $(\mathbf{F} , \mathbf{N})$ of $SO(F) \times O(N)$. We can clearly construct a
bilinear $\gamma^{A} \gamma^{B}$ with $A$ and $B$ indices in the fundamental of $SO(F)$, and with $O(N)$ indices contracted.  These currents can then be easily coupled to an $SO(F)$ gauge field while preserving on-shell (0,2) supersymmetry.
In the compactified theory, the vector multiplet arises from the pullback of
the 9-brane gauge field to the two-dimensional worldvolume of the D1-branes.

What sorts of interactions can the $\gamma$ have with the remaining D1 degrees of freedom?
Since  the $\gamma$ are left-moving fermions, as are the $\Lambda_{(0,2)}$, it is easy to see that there are no direct Lorentz-invariant and gauge-invariant terms that couple the $\gamma$ to $\Lambda_{(0,2)}$ or $\mathbb{X}_{(0,1)}$ at the two-derivative level.

An important feature is that even after compactifying the 9-brane theory,
we should still expect $F=32$ in compactifications of the type I\ theory. The
reason is that the D1-brane is a pointlike object in the compactified space
and as such, does not experience the effects of the flux in the same way that
bulk 9-brane modes do. More concretely, we see that the couplings to the bulk 9-brane
modes do not induce a mass term for the $9-1$ strings.
The result of compactification is therefore to gauge a subgroup $H\subset
Spin(F)$. The residual flavor symmetry is then given by the commutant. More
precisely, we realize a coset space $Spin(F)/H$.

\subsubsection{Anomalies}

Having discussed some aspects of the zero mode spectrum as well as the
interaction terms, let us now turn to anomalies associated with this extra
sector.  To make the computation, we tabulate the fermions of the D1-brane sector.
All of these are Majorana-Weyl and fall into the following representations
\begin{center}
\begin{tabular}{l l l }
fermion	& chirality 		& rep. of $\SO(8)\times\SO(F)\times O(N)$ \\
$\Psi^A$	& right-moving	& $ (\mathbf{8^s}, \mathbf{1}, \text{Sym}^2 \mathbf{N})$ \\
$\widetilde{\Psi}^{A'}$ & left-moving	& $ (\mathbf{8^c}, \mathbf{1}, \wedge^2\mathbf{N})$ \\
$\gamma$ & left-moving	& $ (\mathbf{1}, \mathbf{F}, \mathbf{N} ) $
\end{tabular}
\end{center}
Recall that for $O(N)$ $\text{ind}(\mathbf{N}) = 2$, $\text{ind}(\wedge^2\mathbf{N}) = 2 (N-2)$, and
$\text{ind}(\text{Sym}^2\mathbf{N}) = 2(N+2)$.  With that, we evaluate the anomalies.

First, we have the $O(N)$ anomaly.  Including an overall factor of $1/2$ for the Majorana-Weyl spinors, we obtain
\begin{align}
I_{D1}= \frac{1}{2} \left[-8 \times 2 (N+2) + 8 \times 2 (N-2) + 2 F\right] = (F-32)~.
\end{align}
Thus, as we expect, the only sensible choice is $F=32$.  Fortunately, that is precisely the
choice we need for our application to type I theories.

Now we reconsider the anomaly of our full theory, including the D1-brane sector.  Let us
denote the ``old'' D9-brane anomaly that we found from the analysis above by $I_{\text{old}}$
The ``new'' anomaly, which includes the contribution from the 1--9 strings in the bifundamental $(\mathbf{32}, \mathbf{N})$, is then given by
\begin{align}
I_{\text{new}} = I_{\text{old}} - \frac{1}{2} \times 2 \times N = I_{\text{old}} - N~.
\end{align}
This is a satisfying answer. Tracing through the logic which led us to
consider an extra sector in the first place, the term $B_{2}\wedge
X_{8}$ leads to a tadpole which can be cancelled by introducing spacetime
filling D1-branes. From a gauge theory perspective, we can alternatively
cancel the anomaly by introducing ``by hand'' an extra set of weakly coupled
states, namely those of the D1-brane.

We also compute the gravitational anomaly on the D1-brane world-volume:
\begin{align}
\frac{c_L-c_R}{12} = \frac{1}{12}\times \frac{1}{2} \left[ -8 \frac{N(N+1)}{2} + 8 \frac{N(N-1)}{2} + 32 N\right] = N~.
\end{align}
This is just right to cancel the ``old'' gravitational anomaly.

\subsection{Perturbative Heterotic Extra Sectors}

Consider next the extra sectors associated with perturbative heterotic
compactifications. Just as in the case of the type I theory, the presence of the term
$B_{2}\wedge X_{8}$ indicates that there will
generically be spacetime filling fundamental strings in addition to the GLSM
sector generated by the original compactification.

Now, the worldvolume theory of a single fundamental string is extremely
well-known. It consists of a set of left-moving currents which couple to the
pullback of the 9-brane gauge field. Additionally, we have the standard
embedding coordinates for the heterotic string in the Calabi-Yau fourfold.
This can also be given a rather explicit character using GLSM\ techniques.
Additionally, because multiple fundamental heterotic strings do not form a bound
state, we can also determine the net contribution to the conformal anomalies
for multiple coincident heterotic strings:%
\begin{equation}
(c_{L},c_{R})=(24N,12N).
\end{equation}
In spite of this, the explicit microscopic characterization of multiple
heterotic strings is still somewhat subtle. Nevertheless, one can expect that
at least in the large $N$ limit, a holographic dual description may emerge
\cite{Lapan:2007jx}.  At any rate, we at least observe that the gravitational anomaly
matches that found for $N$ D1-branes above.

\subsection{F-theory Extra Sectors}

Finally, we come to the case of F-theory extra sectors. As opposed to the
constructions encountered previously, in F-theory we should not expect a
spacetime filling D1-brane or F1-brane string to play the role of such an
extra sector. One reason for this is that in the corresponding tadpole
cancellation conditions of F-theory, it is really the four-form potential
rather than a two-form potential which plays the key role in any analysis of
anomaly inflow. Additionally, the very notion of spacetime filling 1-branes in F-theory is
rather special and only holds for special configurations of the axio-dilaton.
In general, $SL(2,\mathbb{Z})$ covariance obstructs the presence of such
objects.

Based on this, we must seek the presence of such extra sectors in the form of
D3-branes wrapped on various cycles of an F-theory compactification.\ For
two-cycles which are common to a 7-brane, the analogous contribution in
heterotic and type I is a non-perturbative five-brane. As such, the
two-dimensional theories defined by these theories are expected to be somewhat
subtle. However, there are also two-cycles transverse to the 7-branes.
These are the F-theory analogues of the 1-branes encountered in other
duality frames. Indeed, these D3-branes have eight Neumann-Dirichlet mixed
boundary conditions, and so will contribute a comparable zero mode content to
that of the probe D1-branes encountered in compactifications of the type
I\ string. Additionally, by considering the orientifold limit of an F-theory
compactification, we can see that a D3-brane wrapped in the normal direction
to a 7-brane becomes -- after applying two T-dualities in the two
directions transverse to the 7-brane -- a D1-brane, while the 7-brane
becomes a 9-brane. So, we see that the structure of this
theory is actually quite close to that encountered in the type I construction.

There are also some important differences between these two constructions.
Perhaps the most significant is that in the limit where gravity is decoupled
from the intersecting 7-brane configuration, the gauge theory dynamics of
the D3-brane must also necessarily decouple. This is simply because it is
wrapping a non-compact curve of infinite volume, so it instead behaves as a
corresponding ``flavor sector'' for the 2D GLSM\ defined by the
7-branes. Nevertheless, at the point of intersection between the D3- and
7-branes, there are additional localized currents. These are the
analogue of the $9-1$ strings encountered in the type I\ construction.

Though more challenging to study, we can also consider the effects of moving
the D3-brane to special points of the intersecting 7-branes. For example,
at various points of the internal geometry, the elliptic fibration may become
more singular, i.e. there is symmetry restoration along a subspace. When this
occurs, there is a corresponding change in the D3-brane sector. It would be
interesting to determine further details of these models in future work.

Now, in addition to these extra sectors, we can also in general expect
D3-branes to wrap two-cycles also common to a 7-brane. Even so, they may
still be separated away from the 7-brane, and so in this sense can be
decoupled (the 3-7 strings being massive). When these D3-branes are nearby a
7-brane, the 3-7 strings become light, and we get another source of an
extra sector. In the perturbative vacua studied previously, these are
associated with five-branes wrapped over a four-cycle (as such, they would not
really be perturbative vacua if we included them). In the flat space limit,
these theories are given by a possibly strongly coupled $\mathcal{N}=2$
supersymmetric system in four dimensions. What we are doing is taking this
strongly coupled system and wrapping it over a curve common to the
7-branes and D3-branes. Again, this leads to a rather rich class of extra
sectors which interact with our GLSM\ sector. We defer a more
complete analysis of these models to future work.

\subsection{Anomalies Versus Supersymmetry}\label{sec:ANOMS}

One of the general features of our 2D\ GLSMs is that in general, we do not
expect the gauge theory sector to be anomaly free by itself. Observe, however,
that the gauginos and Fermi multiplets contribute with one sign to the gauge
anomaly, while the chiral multiplets contribute with the opposite sign. This
leads to a general question about whether the zero mode sector can cancel
anomalies supersymmetrically.

First of all, we can see that in perturbatively realized vacua, the
contribution from the extra sector currents contributes to the gauge anomaly
with the same sign as Fermi multiplets. That means that we
can only use this sector to cancel an anomaly provided the 2D\ GLSM\ sector
has a sufficient number of chiral multiplets. Otherwise, we would need to add anti-branes
instead, breaking supersymmetry.

Now, in non-perturbatively realized vacua, we
can in principle get another contribution to the anomaly. In heterotic
M-theory, this would be given by M5-branes wrapped over a four-cycle, and in
F-theory it is given by D3-branes wrapped over a two-cycle common to a 7-brane. In general,
Consider the case of D3-brane modes which are also non-trivially charged under a representation
of the 7-brane gauge group. In general, these degrees of freedom will be part of a strongly coupled
extra sector, but we can nonetheless count them via anomaly matching considerations.

In the special case where the 7-brane gauge group is perturbatively
realized (i.e., it is of $SU$, $SO$ or $Sp$ type) more can be determined.
For example, in flat space, these modes must organize according
to 4D $\mathcal{N}=2$ hypermultiplets. Upon reduction to two dimensions, the mode content
will organize according to $\mathcal{N} = (0,4)$ hypermultiplets and Fermi multiplets
(see e.g. \cite{Tong:2014yna, Gadde:2015tra} for some recent discussions).
So when we wrap on a curve, we can count the net contribution to the
anomaly via the bundle valued cohomology groups:
\begin{equation}
Q \oplus Q^{c \dag}  \in H^{0}(\Sigma, \mathcal{R})\text{, \ \ }\Psi \oplus \Psi^{c \dag} \in
H^{1}(\Sigma, \mathcal{R}).
\end{equation}
for some bundle on the curve $\Sigma$, and where the $Q$'s denote $\mathcal{N} = (0,2)$ chiral multiplets and the $\Psi$'s denote $\mathcal{N} = (0,2)$ Fermi multiplets. Observe that by an appropriate choice of bundle $\mathcal{R}$, we can get more chiral
multiplets than Fermi multiplets. So in principle, such non-perturbative sectors can also participate in anomaly cancellation.

\section{Examples \label{sec:EXAMPLES}}

In the previous sections we introduced a general formalism for extracting
two-dimensional $\mathcal{N} = (0,2)$ quantum field theories from a string compactification.
In particular, we expect that in most cases, these theories will flow to a
fixed point (though it may be one in which all fields are free). Our aim in
this section will be to give a few examples illustrating these general ideas.
We of course expect there to be a non-trivial target space interpretation of
the resulting theories since in many cases we will reach a super-critical
string theory with a large target space dimension.

With this aim in mind, we first begin with examples of $(2,2)$ supersymmetry,
and explain how starting from such a locus we can reach a special class of
$(0,2)$ models. This is a common strategy in the $(0,2)$ literature. Next, we
turn to examples constructed from compactifications of perturbative strings on
a Calabi-Yau fourfold. We focus on the case of the \textquotedblleft standard
embedding\textquotedblright\ i.e. where we embed the spin connection of the
Calabi-Yau fourfold in the gauge group of the ten-dimensional Yang Mills
theory. In particular, we give a global count of the number of degrees of
freedom and also verify that all gauge and gravitational anomalies have indeed
cancelled. Quite strikingly, we find that for the $E_8 \times E_8$ heterotic theory,
anomaly cancellation with a rank four gauge bundle always leads to supersymmetry breaking.

After this, we turn to some examples from F-theory. Using methods from
the spectral cover construction of vector bundles, we can of course produce
very similar structures to that already seen on the heterotic side (see e.g.
\cite{Schafer-Nameki:2016cfr} for some examples). We shall, however, aim to
focus on some cases which are more \textquotedblleft unique\textquotedblright%
\ to F-theory in the sense that the results are more transparent in that
duality frame. To this end, we consider the 2D analogue of \textquotedblleft
non-Higgsable clusters\textquotedblright\ encountered in previous work in six
\cite{Morrison:2012np} and four \cite{Morrison:2014lca} dimensions. In two
dimensions, such structures are better viewed as \textquotedblleft
rigid\textquotedblright\ clusters since the notion of Higgsing a symmetry in
two dimensions is more subtle. We mainly focus on examples, deferring a full
classification to future work.

\subsection{$\mathcal{N} = (2,2)$ Models}

To give some examples, we begin with two-dimensional models with $(2,2)$
supersymmetry. A straightforward way to engineer such examples is to start
with a four-dimensional $\mathcal{N}=1$ supersymmetric field theory.
Compactifying on a further $T^{2}$ then leads to $(2,2)$ supersymmetry. The
structure of interactions is then inherited from four dimensions. However, the
IR\ dynamics can be somewhat different as there are now non-trivial solitonic
excitations which can wrap along the cycles of the $T^{2}$.

From the perspective of string compactification, we get such examples by
specializing to the case of $T^{2}\times CY_{3}$ for type I\ and heterotic
models, and to $T^{2}\times CY_{4}$ for F-theory models. In these cases, we
also see that the Fermi multiplets and chiral multiplets combine to give
$(2,2)$ chiral multiplets. Some detailed analyses of this special case has
appeared for example in \cite{Greiner:2015mdm} to which we refer the
interested reader for additional discussion. Amusingly enough, we can take
well-known constructions of the Standard Model of particle physics obtained in
previous work and simply reduce to two dimensions.

One important feature of all these models is that now, the Green-Schwarz
mechanism plays a less prominent role. In field theory terms, this is because
now, all of our left-movers naturally pair up with right-movers owing to
$(2,2)$ supersymmetry.\ Indeed, returning to our actual computation of the
integrated $X_{8}$ for perturbative type I\ and heterotic models, we see that
in the special case of perturbative models on $T^{2}\times CY_{3}$ the
integral of $X_{8}$ always vanishes. For F-theory models, there is a related
constraint, although now, we expect there to still be spacetime filling
D3-branes wrapped over two-cycles which are also common to 7-branes.\ That
is, we expect there to typically be four (and not eight) Neumann-Dirichlet
boundary conditions for open strings stretched between D3-branes and
7-branes in models with $(2,2)$ supersymmetry.

Let us make few additional qualitative remarks. First
of all, we can see that in the F-theory constructions, the cubic Yukawa
couplings localized at points are now localized over the $T^{2}$. This is as
expected from our general considerations, where we saw that the triple
intersection of three K\"{a}hler surfaces in the K\"{a}hler threefold should
lead to such localized interactions. We also see, however, that there are
generically no quartic intersections. If we consider a mild
tilting of the 7-branes (say by activating a flat Wilson line along the
$T^{2}$), we can engineer such structures as well.

As a particularly simple class of models, we can also see how the $(2,2)$
supersymmetric $\mathbb{CP}^{N}$ model arises in these sorts of constructions.
Recall that this is described by a $U(1)$ gauge theory with $N+1$ chiral
multiplets of charge $+1$. Additionally, there is a Fayet-Iliopoulos parameter
which controls the overall size of the manifold.

Now, a curious feature of this model is that from a four-dimensional
perspective, it would appear to define an anomalous gauge theory in
four-dimensions. What is really going on in a string theory construction is
that if we attempt to engineer a $U(1)$ gauge theory with $(N+1)$ chiral
multiplets, there will inevitably be an axionic multiplet which functions as
an additional chiral multiplet of charge $-(N+1)$, and serves as a
\textquotedblleft field dependent FI\ parameter.\textquotedblright

With this in mind, let us now engineer an example of this type. We start in
F-theory compactified to four dimensions with a pairwise intersection between
an $SU(2)$ 7-brane wrapped on a del Pezzo surface $S$, and a non-compact
$I_{1}$ factor of the discriminant which intersects the $SU(2)$ locus along a $\mathbb{P}^1$.
As is well-known from earlier work on 4D compactifications, we can activate a supersymmetric bulk
flux which breaks $SU(2)$ to $U(1)$, and which (for a suitable choice of del Pezzo surface and bulk fluxes)
does not generate any bulk zero modes. Restricting the flux
to a matter curve will then give us our zero modes for the GLSM.

At the intersection curve, we have an
enhancement to an $SU(3)$ locus, so we expect to have a hypermultiplet's worth
of degrees of freedom transforming in the fundamental representation of
$SU(2)$. Now, by a suitable choice of flux through the $SU(2)$ factor, we
break to $U(1)$, and have no zero modes from the bulk. Restricting the flux
onto the curve, and activating the flux from the $I_{1}$ flavor brane, we now
see that the number of charge $+1$ and $-1$ fields is given by a line bundle
cohomology group:%
\begin{align}
\text{charge}+1\text{ chirals}  &  \text{: }H^{0}(\mathbb{P}^{1}%
,\mathcal{O}(+N - 1))\\
\text{charge}-1\text{ chirals}  &  \text{: }H^{0}(\mathbb{P}^{1}%
,\mathcal{O}(-N-1)).
\end{align}
So provided $N > 0$, we just have charge $+1$ chiral multiplets localized.
If this is all the matter fields, we necessarily find that the $U(1)$ gauge theory is anomalous.
This is acceptable in the present context, because we will have a coupling to
the dynamical FI\ parameter anyway. Compactifying to two dimensions, we
therefore obtain our $\mathbb{CP}^{N}$ model. Similar constructions can of course be
performed in heterotic models as well.

Even at the level of zero modes, the effective dimension of the target space
depends on the energy scale at which we analyze the effective string theory.
Observe that at higher energy scales, we cannot treat the FI\ parameter $\xi$ (and its axionic partner)
as fixed. Doing so, we get a new geometric interpretation: a non-compact complexified cone over $\mathbb{CP}^N$.

Starting from this construction, we can also consider activating non-trivial
tiltings / fluxes on the $T^{2}$ factor of the compactification. This
corresponds in the $\mathcal{N} = (2,2)$ model to an operator deformation which moves us to
a more general $\mathcal{N} = (0,2)$ model.

\subsection{Perturbative Models with Rank Four Bundles}

Consider next some examples from perturbative strings on a Calabi-Yau fourfold. Canonical examples of this type are given by the
``standard embedding,'' i.e. where we embed the spin connection in the gauge connection. More generally,
a particularly simple class of solutions are obtained by picking a stable
holomorphic rank $4$ bundle $\cE$ over our manifold $M$.   These of course
include the standard embedding, where we take $\cE = T_M$.  In this section
we will study some aspects of the resulting compactifications
for both the type I\ and heterotic string. As some aspects
of the analysis are different, we split our discussion
up according to whether the 10D gauge group is $Spin(32)/%
\mathbb{Z}
_{2}$ or $E_{8}\times E_{8}$.

However, before we get into that, we tabulate a few simple computations regarding the topology of the bundle.  We restrict attention to $\cE$ with $c_i(\cE) = c_i(T_M)$ for $i=1,2$.   This leads to a significant simplification of various characteristic classes and results
\begin{align}
\chi(\cE)  &= 8 - \frac{c_4(\cE)}{6} ~,&
\chi(\wedge^2\cE) & = 12 + \frac{2c_4(\cE)}{3} ~,&
\chi(\cE\otimes\cE^\vee) & = 512 + \frac{ c_4(M)-4 c_4(\cE)}{3}~.
\end{align}

\subsubsection{$Spin(32)/%
\mathbb{Z}
_{2}$ Models}

To begin, we consider  the type I\ string theory or
heterotic $Spin(32)/%
\mathbb{Z}
_{2}$ string on a Calabi-Yau fourfold equipped with a rank $4$ stable holomorphic bundle $\cE$. In this case, we decompose the adjoint
representation according to the branching rule:%
\begin{align}
SO(32)  &  \supset SO(24)\times SO(8)\\
\text{adj}_{SO(32)}  &  \rightarrow(\text{adj}_{SO(24)},\mathbf{1})\oplus
(\mathbf{1},\text{adj}_{SO(8)})\oplus(\mathbf{24},\mathbf{8}^{v})
\end{align}
for compactification on a generic eight-manifold. Specializing to manifolds
with $SU(4)$ holonomy for the metric, we have:%
\begin{align}
SO(32)  &  \supset SO(24)\times SU(4)\times U(1)\\
\text{adj}_{SO(32)}  &  \rightarrow(\text{adj}_{SO(24)},\mathbf{1})\oplus
(\mathbf{1},\text{adj}_{SU(4)})\\
&  \oplus(\mathbf{1},\mathbf{6}_{+1})\oplus(\mathbf{1},\overline{\mathbf{6}%
}_{-1})\oplus(\mathbf{1},\mathbf{1}_{0})\\
&  \oplus(\mathbf{24},\mathbf{4}_{+1/2})\oplus(\mathbf{24},\overline{\mathbf{4}%
}_{-1/2}).
\end{align}
Associating $\mathbf{4}$ to forms valued in $\cE$ and $\mathbf{6}$ to forms valued in $\wedge^2\cE$, we find the following massless spectrum.  Note that all fermions are counted as Weyl.
\begin{enumerate}
\item Moduli.
\begin{itemize}
\item CS:  $h^1(T)+ h^1(T^\vee) + h^1 (\cE\otimes \cE^\vee)$;
\item Fermi:  $h^2(T) + \frac{1}{2} h^2(\cE\otimes\cE^\vee)$.  This is an integer by the result above.
\item $c_L-c_R = \chi(T) +\frac{1}{2} \left[\chi(\cE\otimes\cE^\vee) -2\right] = 263-\frac{2}{3} c_4(\cE)$.\footnote{The $-2$ factor in the square bracket subtracts off $h^0(\cE\otimes \cE^\vee)=h^4(\cE\otimes\cE^\vee) = 1$.}
\end{itemize}
\item $SO(24)\times U(1)$--charged fields:  $\rep{1}_{+1}$.
\begin{itemize}
\item CS:  $h^1(\wedge^2 \cE) + h^3(\wedge^2 \cE)$;
\item Fermi:  $h^2(\wedge^2 \cE)$.
\item $c_L-c_R = \chi(\wedge^2\cE) = 12+\frac{2}{3} c_4(\cE)$.
\end{itemize}
\item $SO(24)\times U(1)$--charged fields:  $\rep{24}_{+1/2}$.
\begin{itemize}
\item CS:  $h^1(\cE) + h^3(\cE)$;
\item Fermi:  $h^2(\cE)$.
\item $c_L-c_R = \chi(\cE) = 8 - \frac{1}{6}c_4(\cE)$.
\end{itemize}
\item left-moving gauginos:  $c_L-c_R = 277$.
\item $d = 2$ gravity : $c_L-c_R = -24$.  As we discussed above, this is the contribution from the gravitational sector of the bulk theory.
\end{enumerate}
We now use
\begin{equation}
\text{Ind}(\text{adj}_{SO(24)})=44\text{, \ \ Ind}(\mathbf{24})=2\text{,}%
\end{equation}
and find that the $SO(24)\times U(1)$ and gravitational anomalies are all proportional to each other.  Namely,
\begin{align}
I_{\SO(24)} &= -44 -2 \chi(\cE) = -60 +\frac{c_4(\cE)}{3}~, \nonumber\\
I_{U(1)} & = -\chi(\wedge^2\cE) - 24\times \frac{1}{4} \chi(\cE) =  -60 +\frac{c_4(\cE)}{3}~,\nonumber\\
\frac{c_L-c_R}{12} & = 60 -\frac{c_4(\cE)}{3}~.
\end{align}
These vanish if and only if $c_4(\cE) = 180$.

The non-vanishing anomaly indicates that there must be additional
degrees of freedom present in the model. These are readily accounted for in
the type I\ picture by introducing a suitable number of D1-branes. Based on
our local gauge theory analysis, we see that the $9-1$ strings are Fermi
multiplets in the fundamental representation of the gauge group $SO(24)$.
So, given $N$ D1-branes of type I string theory, we expect a contribution to the gauge
theory anomaly:%
\begin{equation}
I_{\text{gauge}}(D1\text{'s})=-\text{Ind(}\mathbf{24}\text{)}\times\frac{1}{2}\times
N_{D1}= - N_{D1}\text{.}%
\end{equation}
So, the net contribution to the gauge anomaly is:%
\begin{equation}
I_{\text{gauge}}(\text{9-brane})+I_{\text{gauge}}(D1\text{'s})=-60+\frac{1}%
{3}\chi(CY_{4})-N_{D1}.
\end{equation}
On the other hand, returning to equation (\ref{Nonebranes}), we have that:
\begin{equation}
N_{D1}=-\frac{1}{192(2\pi)^{4}}\underset{M}{\int}%
X_{8}=-60+\frac{1}{3}c_4(\cE)~.
\end{equation}
So we cancel the anomaly, as expected, and we will preserve supersymmetry if $c_4(\cE) \ge 180$.

Though we have phrased the calculation in terms of type I\ string theory, it
is clear that there is a very similar calculation for the S-dual heterotic
model.\ There, the additional contribution to the gauge anomaly comes about
from $N$ spacetime filling fundamental strings. This is again a chiral theory
and its currents directly couple to the 9-brane.

\subsubsection{$E_{8}\times E_{8}$ Models}

Let us now turn to the related calculation for a rank $4$ bundle compactification of the
$E_{8}\times E_{8}$ heterotic string. Here, we shall encounter an interesting subtlety having to
do with tadpole cancellation: We will find that in these models, cancelling anomalies requires us
to add spacetime filling anti-fundamental strings. That is to say, these models will break
$\mathcal{N} = (0,2)$ supersymmetry. Exploring the target space interpretation of this case would clearly be especially interesting.

Let us begin by analyzing the zero mode content of the theory. In this case,
we embed the structure group $\SU(4)$ of $\cE$
 in one of the $E_{8}$ factors. Since
the other $E_{8}$ is a spectator, we will primarily focus on the
\textquotedblleft visible sector.\textquotedblright Of course, the net anomaly
contribution will depend on matter coming from both sectors, and if it is to be cancelled by
the ``extra sector,'' the anomalies in various symmetries must be proportional, just as we observed above in the $\SO(32)$ example.  The spectator $E_{8}$ is the simplest anomaly to evaluate.  Since $\text{Ind}(\mathbf{248}) = 60$, we simply have
\begin{align}
I_{E_8} = - 60~.
\end{align}

To examine the matter spectrum further consider the branching rules for the decomposition of the adjoint
representation:%
\begin{align}
E_{8}  &  \supset SO(10)\times SU(4)\\
\mathbf{248}  &  \rightarrow(\text{adj}_{SO(10)},\mathbf{1})\oplus
(\mathbf{1,}\text{adj}_{SU(4)})\oplus(\mathbf{16,4})\oplus(\overline
{\mathbf{16}}\mathbf{,}\overline{\mathbf{4}})\oplus(\mathbf{10,6}).
\end{align}
In this case, the relevant degrees of freedom transforming in a representation
of the unbroken $SO(10)$ gauge group are counted by the Hodge numbers of the Calabi-Yau fourfold.\ In
particular, we can count the number of CS\ multiplets and Fermi multiplets in
the various representations. In this case, there are some additional subtleties having to do
with the fact that we have Fermi multiplets which transform in self-dual bundles, i.e. $\mathcal{E} = \mathcal{E}^{\vee}$.
This fact means that there is an involution operation $\mathcal{E} \rightarrow \mathcal{E}^{\vee}$, and so we can
split up the modes according to whether they are even or odd. As in our previous discussions of packaging 2D Majorana-Weyl fermions in terms
of Fermi multiplets, this means to properly count these degrees of freedom, we only retain the even sector. In practice, this means we have to
divide by two in tallying up the contribution to a gauge anomaly. With this caveat dispensed with, we have:
\begin{equation}%
\begin{tabular}
[c]{|l|l|l|l|l|}\hline
Multiplet%
$\backslash$%
Representation & $(\text{adj}_{SO(10)},\mathbf{1})$ & $(\mathbf{16,4})$ &
$(\overline{\mathbf{16}}\mathbf{,}\overline{\mathbf{4}})$ & $(\mathbf{10,6}%
)$\\\hline
Fermi & $0$ & $h^{2}(\mathcal{E})$ & dual count & $h_{(even)}^{2}(\wedge
^{2}\mathcal{E})$\\\hline
Chiral Multiplet & $0$ & $h^{1}(\mathcal{E})$ & $h^{1}(\mathcal{E}^{\vee})$ &
$h^{1}(\wedge^{2}\mathcal{E})$\\\hline
\end{tabular}
\ \ .
\end{equation}
where in the above, the terminology ``dual count''
means the modes in the $(\overline{\mathbf{16}},\overline{\mathbf{4}})$ have already
been accounted for by the modes in the $(\mathbf{16},\mathbf{4})$.

We can also calculate the contribution to the various gauge and gravitational
anomalies. For purposes of exposition, we choose to focus on one particular
case, i.e. that of the non-abelian gauge anomalies for $SO(10)$.\ Summing up
the net contribution from the matter charged in various representations, and
using the formulae:%
\begin{equation}
\text{Ind}(\text{adj}_{SO(10)})=16\text{, \ \ Ind}(\mathbf{10})=2\text{,
\ \ Ind}(\mathbf{16})=4
\end{equation}
we get that the total gauge anomaly is:%
\begin{align}
I_{\SO(10)}(\text{9-brane})  &  =-16-4\chi(\cE)-2\times\frac{1}{2}\chi(\wedge^2\cE) = -60~.
\end{align}
This matches the anomaly in the spectator $E_8$, as it had to do.

So, the Green-Schwarz term will
certainly render the theory anomaly free, and we can introduce spacetime-filling strings to solve the tadpole.
However, there is also a crucial difference from the $\SO(32)$ case.  From the general
formulas for $X_8$ given above in~(\ref{eq:X8M8}), we see that
for any $E_8\times E_8$ compactification that leaves the second $E_8$ factor untouched and
uses a holomorphic bundle with $c_2(\cE_1) = c_2(M)$ and $c_1(\cE_1) = 0$
\begin{align}
N_{\text{strings}} =-\frac{1}{192(2\pi)^{4}} \int_M X_8 = -60~.
\end{align}
Thus, a solution of the tadpole necessarily breaks supersymmetry.

\subsection{F-theory Models}
In this subsection we consider compactifications of F-theory on an elliptically fibered
Calabi-Yau fivefold. One of the advantages of F-theory based models is that in many
cases, there is a limit available in which the effects of gravity can be
decoupled. To realize a local model, we must consider a 7-brane wrapping a
Fano threefold in a local geometry such that the normal bundle has negative
first Chern class.

Before proceeding to specific examples, let us recall our general discussion
given in section \ref{sec:ANOMS}. There, we observed that to have an anomaly free theory
which preserves supersymmetry, we typically need to introduce extra sectors from D3-branes wrapping two-cycles in the
base geometry. For D3-branes wrapped on a curve normal to a 7-brane, we get additional Fermi multiplets, while
for D3-branes wrapped on a curve common to a 7-brane, we get the possibility of additional chiral and Fermi multiplets.

For example, for an isolated 7-brane wrapped on a $\mathbb{P}^{3}$, we cannot activate a
supersymmetric background value for the gauge fields and Higgs field of the
model. So, the only zero mode contribution is from the gauginos of the model.
As this is a negative contribution to the anomaly, we conclude that to cancel anomalies supersymmetrically,
we need to introduce D3-branes wrapping a curve of the $\mathbb{P}^3$.

In some sense, this example is not that representative since in general, a
Fano threefold will have non-trivial solutions to the Hermitian Yang-Mills
equations (and their generalization involving non-trivial profiles for the
Higgs fields). An example of this type is given by $X=\mathbb{P}^{1}%
\times\mathbb{P}^{1}\times\mathbb{P}^{1}$.

More generally, we expect that just as in the analysis of 6D\ and 4D\ vacua ,
the special case of local models with no local complex structure deformations
(see e.g. \cite{Morrison:2012np, Morrison:2014lca}) will provide useful
building blocks for constructing more elaborate F-theory models. These are
often referred to as \textquotedblleft non-Higgsable
clusters\textquotedblright\ in higher dimensions, though in two dimensions we
shall instead use the term \textquotedblleft rigid clusters\textquotedblright%
\ since the notion of a Higgs branch is more subtle in two dimensions.

Our plan in the remainder of this subsection will be to discuss in greater
detail the specific examples of $X=\mathbb{P}^{3}$ and $X=\mathbb{P}%
^{1}\times\mathbb{P}^{1}\times\mathbb{P}^{1}$.

\subsubsection{Local $\mathbb{P}^{3}$ Model}

One way to construct a $\mathbb{P}^{3}$ model is to take a decoupling limit
involving F-theory with base a $\mathbb{P}^{1}$ bundle over $\mathbb{P}%
^{3}$. The explicit characterization is given by a toric construction which is
itself described by a $(2,2)$ GLSM. This has a $U(1)_{1}\times U(1)_{2}$ gauge
theory and fields $u_{i},v_{i}$ of respective charges:%
\begin{equation}%
\begin{tabular}
[c]{|l|l|l|l|l|l|l|}\hline
& $u_{1}$ & $u_{2}$ & $u_{2}$ & $u_{3}$ & $v_{1}$ & $v_{2}$\\\hline
$U(1)_{1}$ & $+1$ & $+1$ & $+1$ & $+1$ & $n$ & $0$\\\hline
$U(1)_{2}$ & $0$ & $0$ & $0$ & $0$ & $+1$ & $+1$\\\hline
\end{tabular}
\ \ \ .
\end{equation}
The moment map constraints are then $D_{1}=D_{2}=0$ modulo $U(1)_{1}\times
U(1)_{2}$ gauge transformations, where:%
\begin{align}
D_{1} &  =\left\vert u_{1}\right\vert ^{2}+\left\vert u_{2}\right\vert
^{2}+\left\vert u_{3}\right\vert ^{2}+\left\vert u_{4}\right\vert
^{2}+n\left\vert v_{1}\right\vert ^{2}-\xi_{1}\\
D_{2} &  =\left\vert v_{1}\right\vert ^{2}+\left\vert v_{2}\right\vert
^{2}-\xi_{2}.
\end{align}
Without loss of generality, we can restrict to the case $n\geq0$.
Geometrically, the local description is given by a geometry of the form
$\mathcal{O}_{\mathbb{P}^{3}}(nH)\rightarrow\mathbb{P}^{3}$, where $H$ is the
hyperplane class divisor of $\mathbb{P}^{3}$. An F-theory model over this base
is given in minimal Weierstrass form as:
\begin{equation}
y^{3}=x^{3}+f(u,v)xz^{4}+g(u,v)z^{6}\ ,
\end{equation}
where  $[z,x,y]$ are homogeneous coordinates of the weighted projective space
$\mathbb{P}_{[3,2,1]}^{2}$, and $f(u,v)$ and $g(u,v)$ respectively sections of
$\mathcal{O}_{B}(-4K_{B})$ and $\mathcal{O}_{B}(-6K_{B})$. From the GLSM
presentation, we also have:
\begin{equation}
-[K_{B}]=2[S]+(4+n)[F]\ ,
\end{equation}
where $[S]$ is the divisor class associated to $v_{2}=0$ and $[F]$ is the
class associated to the divisor $u_{1}=0$. We can now write the general
expression for $f$ and $g$:%
\begin{subequations}
\begin{equation}
f=\sum_{k=-4}^{k=M}v_{1}^{k+4}v_{2}^{4-k}f_{16-kn}(u)\text{ \ \ and
\ \ }g=\sum_{j=-6}^{j=J}v_{1}^{j+6}v_{2}^{6-j}g_{24-jn}(u)\ ,
\end{equation}
where $f_{16-kn}$ and $g_{24-jn}$ are homogeneous polynomials in the $u_{i}$
with degree indicated by their subscript. Moreover, the bounds $M$ and $J$ in
the sums are given by $16-kn\geq0$ and $24-jn\geq0$. This means that
\end{subequations}
\begin{equation}
M=\mathrm{Min}(4,[16/n])\ ,\qquad J=\mathrm{Min}(6,[24/n])\ .
\end{equation}

Let us now discuss some general features of this model and the associated
geometry. There are two generic locations where we
expect 7-branes to localize, i.e. at $v_{1}=0$ and $v_{2}=0$. These are
roughly speaking the remnants of the two stacks of 9-branes present in the
heterotic construction, now realized in terms of the corresponding factors.
From the general structure of perturbative anomaly cancellation, we also see
that there will be D3-branes which will wrap the $\mathbb{P}^{1}$ fiber
direction and sit at points of the $\mathbb{P}^{3}$.\ In general, we also
expect there to be D3-branes wrapped over two-cycles of the $\mathbb{P}^{3}$,
which are in turn counted by the class $H \cdot H$. In the dual heterotic
M-theory description, the D3-branes wrapped over the $\mathbb{P}^{1}$ fiber
translate to spacetime filling M2-branes which also wrap the interval between
the two $E_{8}$ factors. Additionally, we have M5-branes wrapped over the
elliptic fiber and a two-cycle of the $\mathbb{P}^{3}$. The total number of
such M5-branes is the parameter $n$.

Let us now see show how to realize a rigid cluster for appropriate $n$. To this
end, it is enough to study the structure of the minimal
Weierstrass model. For example, we see that the value of $n$ is bounded as:%
\begin{equation}
0 \leq n\leq24.
\end{equation}
The upper bound comes about because we require that the elliptic fiber remain in Kodaira-Tate
form. For the case $n = 24$, we have:
\begin{align}
f  &  =v_{1}^{4}v_{2}^{4}f_{16}(u)+v_{1}^{3}v_{2}^{5}f_{40}(u)+...+v_{2}%
^{8}f_{112}(u)\\
g  &  =v_{1}^{7}v_{2}^{5}g_{0}(u)+v_{1}^{6}v_{2}^{6}g_{24}(u)+v_{1}^{5}%
v_{2}^{7}g_{48}(u)+...+v_{2}^{12}g_{168}(u).
\end{align}
So by inspection, we see two $E_{8}$ factors, one at $v_{2}=0$ which is a
rigid cluster, and another at $v_{1}=0$ which can be maximally unfolded. We
interpret this as the situation in which we activate a generic vector bundle
on the non-rigid $E_{8}$ factor. An interesting feature of this construction
is that the F-theory model provides us with a rather direct way to count the
vector bundle moduli on the heterotic side. Indeed, we can also recognize the
dual geometry wrapped by the heterotic 9-brane. It is given by an elliptically
fibered Calabi-Yau fourfold with $\mathbb{P}^{3}$ base:%
\begin{equation}
y^{2}=x^{3}+f_{16}(u)x+g_{24}(u).
\end{equation}

Proceeding in this fashion down to lower values of $n$, we can also track the
singular fiber for all of the remaining cases. We shall refer to a ``rigid
cluster'' as one for which the local geometry is:
\begin{equation}
\mathcal{O}(-n H) \rightarrow\mathbb{P}^{3},
\end{equation}
with $n > 0$ (i.e. we can decouple gravity) and in which the elliptic
fibration over the threefold is always singular. This occurs in the range:
\begin{equation}
5 \leq n \leq24.
\end{equation}
It is interesting to contrast this with the higher-dimensional case for 6D
vacua studied in reference \cite{Morrison:2012np}. There, the local geometry
is $\mathcal{O}(-n) \rightarrow\mathbb{P}^{1}$ with $3 \leq n \leq12$.

For each value of $n$, we can also deduce the order of vanishing for $f$, $g$
and $\Delta$. Consequently, we can also read off the expected matter structure
for these models. As in higher dimensions, however, the fiber type does not
directly translate to the realized gauge symmetry of the 2D model, because of
possible quotients by outer automorphisms of a larger algebra (i.e.
monodromy).

So far, we have focussed on the geometric data associated with this model. We
can now ask to what extent we expect the resulting 2D theory to preserve
supersymmetry.\ To give a simple example, let us focus on the case of
$\mathcal{O}(-24H)\rightarrow\mathbb{P}^{3}$, in which case there is an
$E_{8}$ 7-brane wrapped over an isolated $\mathbb{P}^{3}$. Now, the key point
for us is that in the BPS\ equations of motion,
\begin{equation}
\omega\wedge\omega\wedge F_{(1,1)}+[\phi,\overline{\phi}]=0\text{,
\ \ }\overline{\partial}_{A}\phi=0\text{, \ \ }F_{(0,2)}=0,
\end{equation}
the positivity of the associated Lichnerowicz operator makes it impossible for
us to find a non-trivial vacuum solution.\ That is to say, the only zero mode
content is an $E_{8}$ vector multiplet. This contribution is negative, and so
we can already anticipate that to cancel anomalies supersymmetrically, we would need
a D3-brane wrapped on a curve in the $\mathbb{P}^3$. Note also that this D3-brane can be interpreted as
being generated by a non-trivial flux of the bulk 7-brane theory. We leave a more complete discussion of such sectors
for future work.

\subsubsection{Local $\mathbb{P}^{1}\times\mathbb{P}^{1}\times\mathbb{P}^{1}$
Model}

We now turn to a class of examples with $X=\mathbb{P}^{1}\times\mathbb{P}%
^{1}\times\mathbb{P}^{1}$. For starters, we setup
some notation. The homology ring for $X$ is generated by the three divisor
classes $[S_{1}]$, $[S_{2}]$, $[S_{3}]$, where we label the three
$\mathbb{P}^{1}$ factors of $X$, and $S_{i}=\mathbb{P}_{(k)}^{1}%
\times\mathbb{P}_{(l)}^{1}$ with $i\neq k$ and $i\neq l$. In what follows we
omit the square brackets from all divisor classes. In this notation, the
canonical class for $X$ is:%
\begin{equation}
K_{X}=-2S_{1}-2S_{2}-2S_{3}.
\end{equation}
The local geometry in the base is then captured by a general normal bundle
which we take to be:%
\begin{equation}
\mathcal{N}=\mathcal{O}(-n_{1}S_{1}-n_{2}S_{2}-n_{3}S_{3}).
\end{equation}
To have a decoupling limit, we require $n_{i}>0$ for all $i$. Since we are
primarily interested in examples, we shall specialize to the case $n_{i}$ all
equal to some $n>0$.

Now, following the related discussion in e.g. \cite{Morrison:2012np}, we can
calculate the order of vanishing on $X$ for $f$ and $g$ of the Weierstrass
model. Along these lines, we assume that they do vanish (i.e. we have a rigid cluster) and
write the canonical class for the non-compact
base $B$ given by the total space $\mathcal{N} \rightarrow X$ as:%
\begin{equation}
-K_{B}=\gamma X+D,
\end{equation}
where $\gamma$ is a positive rational number and $D$ is an effective divisor such that
$X\cdot D\geq0$. Now, by adjunction, we have:%
\begin{equation}
K_{X}=X\cdot X+K_{B}\cdot X,
\end{equation}
so:%
\begin{equation}
-K_{X}=(\gamma-1)X\cdot X+D\cdot X
\end{equation}
or:%
\begin{equation}
2S_{1}+2S_{2}+2S_{3}=n(1-\gamma)\left(  S_{1}+S_{2}+S_{3}\right)  +D\cdot X.
\end{equation}
Since we have assumed $D\cdot X\geq0$, we can solve for $\gamma$ to find:%
\begin{equation}
\gamma=\frac{n-2}{n}.
\end{equation}
As the order of vanishing for $f$, $g$ and $\Delta$ is simply given by the
restriction of $-4K_{B}$, $-6K_{B}$ and $-12K_{B}$, we can now read off the
order of vanishing on $X$ for each of these sections:%
\begin{equation}
\text{ord}_{X}f=\left[  \frac{4(n-2)}{n}\right]  ,\text{ \ \ ord}_{X}g=\left[
\frac{6(n-2)}{n}\right]  \text{, \ \ ord}_{X}\Delta=\left[  \frac{12(n-2)}%
{n}\right]  .
\end{equation}
So in this case, the range of possible values for $n$ is:%
\begin{equation}
0\leq n\leq12,
\end{equation}
and a rigid cluster is obtained for $n>2$ (i.e. a singular elliptic fiber must
occur over $X$).

Specializing now to the case of $n=12$, we have an isolated $E_{8}$ 7-brane.
To get a supersymmetric vacuum, we now need to switch on an internal flux
through $X$. One choice is given by activating a $U(1)$ valued flux in the
Cartan of the $SU(2)$ factor of $\left(  E_{7}\times SU(2)\right)  /%
\mathbb{Z}
_{2}\subset E_{8}$. To get a choice of flux compatible with the BPS\ equations
of motion, we take the line bundle:%
\begin{equation}
\mathcal{L}=\mathcal{O}(kS_{1}-kS_{2}),
\end{equation}
where $k>0$ is taken to be an integer. We can verify that this satisfies the
stability condition $\omega\wedge\omega\wedge F_{(1,1)}=0$ by noting that when
the K\"{a}hler class is aligned as a multiple of $S_{1}+S_{2}+S_{3}$,
$kS_{1}-kS_{2}$ clearly has trivial intersection number with the two-cycle
represented by $\omega\wedge\omega$.

Let us now analyze the zero mode content in the presence of this abelian flux.
To this end, we need to consider the breaking pattern:%
\begin{align}
\mathfrak{e}_{8}  & \supset\mathfrak{e}_{7}\times\mathfrak{su}(2)\\
\mathbf{248}  & \mathbf{\rightarrow}\left(  \mathbf{133},\mathbf{1}\right)
\oplus(\mathbf{1,3)\oplus(56,2).}%
\end{align}
So decomposing further to the $\mathfrak{u}(1)$ factor, we have:%
\begin{align}
\mathfrak{e}_{8}  & \supset\mathfrak{e}_{7}\times\mathfrak{u}(1)\\
\mathbf{248}  & \mathbf{\rightarrow}\mathbf{133}_{0}\oplus\mathbf{1}_{0}%
\oplus\mathbf{1}_{-2}\oplus\mathbf{1}_{+2}\oplus\mathbf{56}_{+1}%
\oplus\mathbf{56}_{-1}\mathbf{.}%
\end{align}
To count the number of chiral multiplets and Fermi multiplets,
we will need to evaluate various line bundle cohomologies.
At this point, it is helpful to recall:%
\begin{equation}
H^{m}(X,\mathcal{O}(q_{1}S_{1}+q_{2}S_{2}+q_{3}S_{3}))=\underset{k_{1}%
+k_{2}+k_{3}=m}{\oplus}H^{k_{1}}(\mathbb{P}_{(1)}^{1},\mathcal{O}%
(q_{1}))\times H^{k_{2}}(\mathbb{P}_{(2)}^{1},\mathcal{O}(q_{2}))\times
H^{k_{3}}(\mathbb{P}_{(3)}^{1},\mathcal{O}(q_{3})).
\end{equation}
Now in our specific case, we always have $q_{3}=0$ and $q_{1}q_{2}<0$. For
example, if $q_{1}>0$ and $q_{2}<0$, we need $m=1$ and $k_{1}=0$, $k_{2}=1$,
and $k_{3}=0$. Now, we next observe that:%
\begin{equation}
H^{1}(X,\mathcal{O}(tS_{1}-tS_{2}))=H^{0}(\mathbb{P}^{1},\mathcal{O}(t))\times
H^{1}(\mathbb{P}^{1},\mathcal{O}(-t)),
\end{equation}
which has dimension $(t+1)(t-1)=(t^{2}-1)$. Returning to the specific mode
content of our model, we note that the fluctuations from the chiral multiplets
come from the Hodge numbers $h^{1}$ and $h^{3}$, while that of the Fermi
multiplets comes from $h^{2}$. So, we never get a contribution to the Fermi
multiplets, and only get a contribution to the chiral multiplets. The specific
number in each representation of $\mathfrak{e}_{7}\times\mathfrak{u}(1)$ is:%
\begin{equation}%
\begin{tabular}
[c]{|l|l|l|l|l|}\hline
Rep: & $\mathbf{1}_{-2}$ & $\mathbf{1}_{+2}$ & $\mathbf{56}_{+1}$ &
$\mathbf{56}_{-1}$\\\hline
\#CS: & $4k^{2}-1$ & $4k^{2}-1$ & $k^{2}-1$ & $k^{2}-1$\\\hline
\end{tabular}
.
\end{equation}
For $k>1$, we indeed have a positive number of chiral multiplets. For example,
the total contribution to the $E_7$ gauge anomaly is:%
\begin{equation}
I_{E_7}=-\text{Ind}(\mathbf{133})+\text{Ind}%
(\mathbf{56})\times\left(  2k^{2}-2\right)  =24k^{2}-60,
\end{equation}
where in the above we used the fact that Ind$(\mathbf{133})=36$ and
Ind$(\mathbf{56})=12$. As expected, $I_{E_7} > 0$ for
$k>1$, so indeed, we expect to always cancel the anomaly supersymmetrically
through the presence of some spacetime filling D3-branes which wrap the
non-compact two-cycle normal to the 7-brane. In this case, the extra sector is
a variant on the same SCFT obtained from order $k^{2}$ heterotic string
worldsheet theories. It would be interesting to directly calculate the chiral
contribution to the GLSM\ anomaly from this sector.

Let us make a few additional qualitative remarks. First of all, we can see
that the total number of such D3-branes will be of order $k^{2}$.
Geometrically, we can see this by lifting the gauge field flux to a four-form
flux in the dual M-theory description given by compactifying on a further
circle. Since this is proportional to $k$, the eight-form $G_{4}\wedge G_{4}$
will naturally scale as $k^{2}$. We also see that the bulk modes
contribute no Fermi multiplets, so there are no superpotential terms. Rather, we
expect there to be possibly non-trivial couplings of these bulk modes to
chiral fermions originating from our D3-brane sector.

\section{Conclusions \label{sec:CONC}}

In this paper we have used methods from string compactification to provide
UV\ completions for non-critical string theories. We gave a general class of
tools for analyzing 2D vacua arising from compactifications of type I, heterotic strings and
F-theory. In particular, we introduced a quasi-topological 8D theory to analyze vacua generated by 9-branes,
and a quasi-topological 6D theory to analyze vacua obtained from intersecting 7-branes.
One of the important points from this analysis is
that in addition to a set of sectors described by gauged linear sigma models,
there are generically spacetime filling branes which contribute
degrees of freedom to the 2D theory. These branes must be present to cancel gauge
anomalies, and are also required to eliminate the tadpole
from a non-local two-form potential. We have also presented some examples illustrating how to compute some
details of the resulting low energy effective theories, and we derived the
full 2D off-shell action obtained from the higher-dimensional gauge theory
sector of the compactification. In the remainder of this section we highlight
some avenues of future investigation.

One of the general lessons from compactifications to two dimensions is that
the GLSM sector is often accompanied by additional sectors. In some sense,
these sectors can be decoupled from
the other (gauge singlet) dynamics of the model.\ It would be interesting to study such
theories further, for example determining the operator content and correlation
functions of the system. This would clearly be important in determining the
full target space interpretation for these non-critical theories.

Indeed, one of the primary motivations for the present work was the goal of
understanding the behavior of non-critical string theories. With this in mind,
it would be quite interesting to study in detail even some simple examples of
the kind encountered here to develop a better understanding of these models.
In particular, we expect that the apparent loss of unitarity (i.e. when the
target space equations of motion become singular) is simply an indication that
we must return to our original 10D model. Establishing what sorts of
non-critical string theories admit such an embedding would be quite
interesting to develop further.

As a particular subclass of models, it would be natural to study the
special case of 6D\ SCFTs compactified on a K\"ahler surface. This gives
rise to a special class of F-theory models which should be possible to study
using the techniques presented in this paper. In particular, it should be
feasible to extract protected quantities such as the anomaly polynomial
and the elliptic genus.

We have also presented evidence of irreducible building blocks, i.e.
\textquotedblleft rigid clusters\textquotedblright\ for F-theory realizations
of 2D\ SCFTs. It would be quite instructive to obtain a full classification of
the resulting GLSM\ sectors as well as the associated F-theory geometries.
This would also likely give insight into the class of non-critical models
which admit an embedding in the physical superstring.

Though we have emphasized the role of how naturally string compactifications
combine the features of 2D\ SCFTs coupled to gravity, it would of course be
interesting to study further the possibility of fully decoupling gravity. Along
these lines, we expect a non-commutative geometric structure to
emerge in such a limit along the lines of reference
\cite{Heckman:2010pv}.

Another byproduct of our analysis is that we have also introduced a
set of quasi-topological actions for eight-dimensional gauge theory on a
Calabi-Yau fourfold (in the case of type I and heterotic strings) and
six-dimensional gauge theory on a K\"ahler threefold. We expect that the present
perspective where off-shell 2D supersymmetry is maintained at all stages
should make it possible to develop a corresponding theory of enumerative
geometric invariants. Developing this in detail would also be quite exciting.

Finally, one of the important features of super-critical string theories is the relative ease with
which novel time dependent backgrounds for effective strings readily emerge.
It would be quite exciting to combine the analysis presented here with the
general outline of ideas given in references \cite{Silverstein:2001xn, Maloney:2002rr} to
obtain examples of effective de Sitter vacua from string theory.

\newpage

\section*{Acknowledgements}

We thank A. Gadde, E. Sharpe and T. Weigand for helpful discussions.
JJH also thanks H. Verlinde for inspiring discussions several years ago.
FA, FH and JJH thank the theory groups at Columbia University and
the ITS at the CUNY\ graduate center for hospitality during the completion of this work.
JJH also thanks the CCPP at NYU for hospitality during the completion of this work.
The work of FA, FH and JJH is supported by NSF CAREER grant PHY-1452037. FA, FH and JJH
also acknowledge support from the Bahnson Fund at UNC Chapel Hill as well as the
R.~J. Reynolds Industries, Inc. Junior Faculty Development Award from the Office
of the Executive Vice Chancellor and Provost at UNC Chapel Hill.


\appendix

\section{Non-Abelian $(0,2)$ GLSMs \label{app:GLSM}}
In this Appendix we give a general discussion of two-dimensional
GLSMs with $\mathcal{N} = (0,2)$ supersymmetry. Perhaps surprisingly, we have only
been able to locate explicit Lagrangians for models with abelian gauge groups.
For this reason, we will present in some detail both the structure of the
superspace interactions, as well as the interactions in component fields. We
will make heavy use of this formalism when we turn to the off-shell actions
for 10D Super Yang-Mills theory and intersecting 7-branes.

To set our conventions, we introduce spacetime coordinates $y^{0}$ for time, and $y^{1}$ for space. In our conventions, the metric
for flat $\mathbb{R}^{1,1}$ is:
\begin{equation}
ds^2 = -(dy^0)^2 + (dy^1)^2 = - 4 dy^{+} dy^{-}.
\end{equation}
where we have introduced the lightcone coordinates:
\begin{equation}
y^{+} = \frac{1}{2} (y^{0} + y^{1}) \,\,\, \text{and} \,\,\, y^{-} = \frac{1}{2}(y^{0} - y^{1}).
\end{equation}
we choose this normalization so that our expressions for the lightcone derivatives do not contain extraneous factors of two:
\begin{equation}
\partial_{+} = \partial_{0} + \partial_{1}\,\,\, \text{and} \,\,\, \partial_{-} = \partial_{0} - \partial_{1}.
\end{equation}

Let us now turn to the $\mathcal{N} = (0,2)$ supersymmetry algebra on flat space.
We have:
\begin{align}\label{eqn:(0,2)algebra}
  \{Q_+, \overline Q_+\} &= 2 P_+ \\
  \{Q_+, Q_+\} &= \{\overline Q_+, \overline Q_+\} = 0 \\
  [Q_+, P_+] &= [\overline Q_+, P_+] = 0,
\end{align}
where $P_+ = -i \partial_{+}$.

It is helpful to give a geometric presentation of these symmetries.
Along these lines, we work in the corresponding 2D superspace
with Grassmann coordinates $\theta^+$ and $\overline{\theta}^+$,
and bosonic coordinates $y_0$ and $y_1$. In our conventions, we have:
\begin{equation}
  \int d\theta^+\,\theta^+ = 1\,\,\,\,\,\,~, \int d \theta^+ d\overline{\theta}^+  = \int d^{2} \theta ~.
\end{equation}

The supersymmetry generators and supertranslations are:
\begin{align}
\mathbf Q_+ & = \frac{\partial}{\partial \theta^+}+i\overline{\theta}^+ \partial_+, \qquad \overline{\mathbf Q}_+= -\frac{\partial}{\partial \overline{\theta}^+}-i\theta^+ \partial_+ \\
\mathbf D_+ & = \frac{\partial}{\partial \theta^+}-i\overline{\theta}^+ \partial_+, \qquad \overline{\mathbf D}_+= -\frac{\partial}{\partial \overline{\theta}^+}+i\theta^+ \partial_+,
\end{align}
where $\{\mathbf{D} , \mathbf{Q}\} = 0$.

We now introduce the various supermultiplets we shall use to build our 2D GLSM.
Along these lines, we start with the vector multiplet, and then turn to the
remaining multiplets. We assume that we have a gauge group $G$.
To avoid overloading the notation, we shall often suppress the
explicit Lie algebra indices for its algebra. We use antihermitian
generators $T^{\alpha}$ for the Lie algebra,
with commutators
\begin{equation}
  [T_a, T_b] = f_{ab}{}^c T_c
\end{equation}
with real structure coefficients $f_{ab}{}^c$. Our normalization for the generators is:
\begin{equation}
  \Tr T^\alpha T^\beta = -\delta^{\alpha \beta}\,.
\end{equation}

The gauge bosons of the 2D theory are given by $v_+$, $v_-$, and their two antichiral gauginos $\mu$, $\overline\mu$. We take all
component fields of the supermultiplets to be antihermitian. Throughout, we shall work in Wess-Zumino gauge. In this gauge, we have:
\begin{equation}
  \Xi = - i \theta^+ \overline\theta^+ v_+\,\,\,\,\,\,
  V = v_- - 2 i \theta^+ \overline\mu - 2 i \overline\theta^+ \mu + 2 \theta^+ \overline\theta^+ \mathcal{D}\,.
\end{equation}
with $\mathcal{D}$ an auxiliary field whose presence is required for the supersymmetry algebra to close off shell.

In order to write down a manifestly gauge invariant action, one has to modify the superspace derivatives $\mathbf{D}_+$ and $\overline{\mathbf{D}}_+$ to make them covariant with respect to gauge transformations. The new derivatives are:
\begin{align} \label{eq:supercovder}
  \mathcal{D}_+ &= e^{-\Xi} \mathbf{D}_+ e^{\Xi} = \mathbf{D}_+ - i \overline \theta^+ v_+ = \frac{\partial}{\partial\theta^+} - i \overline\theta^+ D_+  \\
  \overline{\mathcal{D}}_+ &= e^{\Xi} \overline{\mathbf D}_+ e^{-\Xi} = \overline{\mathbf D}_+ + i \theta^+ v_+ = - \frac{\partial}{\partial\overline\theta^+} + i \theta^+ D_+\,,
\end{align}
where $D_+$ denotes the covariant derivative:
\begin{equation}
  D_+ = \partial_+ + v_+\,.
\end{equation}
They are chosen in such a way that the anticommutation relation $\{\mathbf{D}_+, \overline{\mathbf D}_+\}$ is preserved:
\begin{equation}
  \{\mathcal{D}_+, \overline{\mathcal{D}}_+\} = 2 i D_+\,.
\end{equation}
We also gauge the partial derivative $\partial_-$ by defining
\begin{equation}
  \nabla_- = \partial_- + V = D_- - 2 i \theta^+ \overline\mu - 2 i \overline\theta^+ \mu + 2 \theta^+ \overline\theta^+ \mathcal{D}\,.
\end{equation}

Gauge transformations are parameterized by the chiral superfield $\chi$. On the vector superfield $\Xi$, they act as
\begin{equation}
  \delta_\chi \Xi = \chi + \overline\chi + [\chi - \overline\chi, \Xi]\,.
\end{equation}
Not all components of $\chi$ are independent. First, we want to keep Wess-Zumino gauge, meaning $\delta_\chi \Xi$ should not contain any terms besides $\theta^+ \overline\theta^+ \dots$. This constraint partially fixes the parameter $\chi$ to
\begin{equation}\label{eqn:resgaugetr0}
  \chi = \frac{1}{2}\rho -\frac{i}{2} \theta^+ \overline\theta^+ \partial_+ \rho
\end{equation}
with $\rho$ being the antihermitian parameter the gauge transformation
\begin{equation}
  \delta_\rho v_+ = \partial_+ \rho + [v_+, \rho]\,.
\end{equation}
This is exactly the expected transformation for a gauge connection.
Further, the ``naive'' supersymmetry transformation
\begin{equation}
  \partial_\epsilon \Xi = (\epsilon Q_+ - \overline\epsilon \overline Q_+) \Xi
\end{equation}
spoils the Wess-Zumino gauge, though we can repair this by taking the modified transformation rule:
\begin{equation}\label{eqn:compgaugetr0}
  \delta_\epsilon \Xi = (\epsilon Q_+ - \overline \epsilon \overline Q_+ + \delta \chi) \Xi = 0
    \quad \text{with} \quad
  \chi =  i \theta^+ \overline\epsilon v_+\,.
\end{equation}
Hence, we find that $\Xi$ is a singlet under supersymmetry transformations.

The remaining gauge transformations of $V$ are:
\begin{equation}
  \delta_\chi V = \partial_- (\chi - \overline\chi) - [\chi - \overline\chi, V]\,.
\end{equation}
As expected, they result in the right transformations
\begin{equation}
  \delta_\rho v_- = \partial_- v_- + [v_-, \rho]
    \quad\text{and}\quad
  \delta_\rho \mu = [\mu, \rho]
\end{equation}
for the vector potential $v_-$ and the gauginos if restricted to the residual gauge transformation \eqref{eqn:resgaugetr0}.
For the supersymmetry transformation of $V$, the compensating gauge transformations is already fixed by \eqref{eqn:compgaugetr0}:
\begin{equation}
  \delta_\epsilon V = (\epsilon Q_+ - \overline\epsilon \overline Q_+ + \delta_\chi) V
\end{equation}
resulting in
\begin{align}
  \delta_\epsilon v_- &= -2 i ( \epsilon \overline\mu + \overline\epsilon\mu ) &
  \delta_\epsilon \mu &= \epsilon( \frac{1}{2} F_{+-} - i \mathcal{D} ) \nonumber \\
  \delta_\epsilon \mathcal{D} & = \overline\epsilon D_+ \mu - \epsilon D_+ \overline\mu &
  \delta_\epsilon \overline\mu &= \overline\epsilon( \frac{1}{2} F_{+-} + i \mathcal{D}) \label{eqn:SUSYvector0}\,.
\end{align}
In the above, we introduced the field strength $F_{-+}$:
\begin{equation}
  F_{-+} = [D_-, D_+] = \partial_- v_+ - \partial_+ v_-  + [v_-, v_+]\,,
\end{equation}
which is part of the (Fermi) supermultiplet $\Upsilon$:
\begin{equation}
  \Upsilon = [\nabla_-, \overline{\mathcal D}_+] = -2 i \mu + i \theta^+ \big( F_{-+} + 2 i \mathcal{D}\big) - 2 \theta^+ \overline\theta^+ D_+ \mu\,.
\end{equation}

Let us now turn to the remaining multiplets of a $\mathcal{N} = (0,2)$ GLSM.
This will consist of a chiral multiplet (CS multiplet) and a Fermi multiplet.
A chiral multiplet $\Phi$ is defined by the condition that we have a (complex) scalar for the lowest component,
and that it obeys the condition $\overline{\mathbf{D}}_{+} \Phi = 0$. In a gauge theory, this is replaced
by the condition $\overline{\mathcal{D}}_{+} \Phi = 0$. In components, we have:
\begin{equation} \label{eq:defchiral10}
\Phi^i=\phi^i + \sqrt{2}\theta^+ \psi^i_+ - i\theta^+ \overline{\theta}^+ D_{+} \phi^i,
\end{equation}
where $i$ denotes an index indicating the representation under a gauge group.

A peculiarity of $\mathcal{N} = (0,2)$ supersymmetry in two dimensions is that we can have a multiplet
with no dynamical bosonic degrees of freedom. The aptly named Fermi multiplet is given by the expansion:
\begin{equation}
\Lambda^{m}=\lambda^{m}_- - \sqrt{2}\theta^+ \mathcal G^{m} - \sqrt{2}\overline{\theta}^+ E^{m}  - i\theta^+ \overline{\theta}^+ D_{+}\lambda_-^{m} ,
\end{equation}
where $m$ is a representation index, and the $E$'s are themselves
holomorphic functions of the chiral superfields, i.e. we have $E(\Phi)$, with a corresponding
expansion into components. The $\Lambda$'s satisfy:
\begin{equation}
\overline{\mathcal{D}}_+\Lambda^{m}= \sqrt{2} E^{m}.
\end{equation}

Having introduced the relevant multiplets, we now construct manifestly supersymmetric actions for these fields.
The action consists of D- and F-terms, i.e. we integrate over the full superspace or just half of it. In particular,
the kinetic terms descend from the D-terms, while interactions protected by supersymmetry descend from the F-terms.

Let us begin with the D-terms for the model. The D-term for the 2D gauge field involves the covariant field strength $\Upsilon$,
\begin{equation}\label{eqn:Sgauge0}
  S_\mathrm{gauge} = - \frac{1}{8 e^2} \int d^2 y d^2 \theta \,{\rm Tr} \left(\overline\Upsilon \Upsilon \right)= - \frac{1}{e^2} \int d^2 y{\rm Tr} \left( \frac{1}{4} F^2 + \frac{1}{2} \mathcal{D}^2 + i \overline\mu D_+ \mu \right)\,.
\end{equation}

Consider next a collection of chiral multiplets transforming in a representation $R$ of the gauge group $G$. We introduce a canonical pairing $(\cdot , \cdot)$ for $R^{\ast}$ and $R$, with $R^{\ast}$ the dual representation. The kinetic term is:
\begin{equation}
S_{\Phi, \text{kin}}=- \frac{i}{2}  \int d^2 y d^2 \theta \;  \left( \overline{\Phi} , \nabla_- \Phi \right) \ ,
\end{equation}
In components, we have:
\begin{equation}
S_{\Phi,\text{kin}}= \int d^2 y  \left( \left(D_{+} \overline{\phi} , D_{-} \phi \right)
+ i (\overline{\psi}_{+} , D_{-} \psi_+)
+ \sqrt{2}  ( \overline{\psi}_{+} , \overline{\mu}_{-} \phi) - \sqrt{2} (\overline{\phi} \mu_- ,  \psi_+)  - ( \overline{\phi} , \mathcal{D} \phi) \right) .
\end{equation}
Consider next the kinetic terms for the Fermi fields. Assuming we have a multiplet $\Lambda$ transforming in a representation $R$ of the gauge group the kinetic term is:
\begin{equation}
S_{\Lambda,\text{kin}} =  - \frac{1}{2} \int d^2 y  d^{2} \theta \;  \left( \overline{\Lambda} , \Lambda \right).
\end{equation}
Its expansion in components is:
\begin{equation}
S_{\Lambda,\text{kin}} =  \int d^2 y   \,  \left( i (\overline{\lambda}_{-} , D_{+}\lambda_-)   +  (\overline{\mathcal{G}}_ , \mathcal{G})  -  (\overline{E} (\overline{\phi}_i) , E(\phi^i))  - \left( \overline{\lambda}_{-} , \frac{\partial E}{\partial \phi^i} \psi_+^{i} \right) -  \left(\overline{\psi}_{+\, i} \frac{\partial \overline{E}}{\partial \overline{\phi}_i} , \lambda_{-} \right) \right).
\end{equation}

Let us now turn to the F-terms for the model. In general, these F-terms can be written as an integral over a superpotential $W$:
\begin{equation}
S_{F} = \int d^{2} y d \theta^{+} \, W + h.c..
\end{equation}
The superpotential will involve various interactions between the Fermi multiplets and the chiral multiplets:
\begin{equation} \label{eq:bulkW0}
W =  \frac{1}{\sqrt{2}} \Lambda_{m} J^{m}(\Phi^i)|_{\overline{\theta}^+=0}  ,
\end{equation}
where $J^{m}$ are holomorphic functions of the chiral fields with expansion:
\begin{equation}
J^{m}(\Phi^i)=J^{m}(\phi^i)+\sqrt{2} \theta^+ \psi^{i}_+  \frac{\partial J^{m}}{\partial \phi^i} - i \theta^+ \overline{\theta}^+ D_+ J^{m}(\phi^i).
\end{equation}
In order for supersymmetry to close off-shell, we must also require $\overline{\mathcal{D}}_+ W=0$, or:
\begin{equation}
E_{m} J^{m}=0 .
\end{equation}
In practice, this imposes further restrictions on the space of admissible couplings in a given model.

Finally, in the special case where our gauge group has abelian factors, we can also introduce a complex parameter $t$:
\begin{equation}
t =  \xi + i \frac{\eta}{2 \pi}
\end{equation}
consisting of a theta angle\footnote{Due to an unfortunate clash of notation,
we write the angle as $\eta$ rather than $\theta$.} $\eta$ and an FI parameter $\xi$.
This allows us to write an additional F-term, which for a single $U(1)$ factor is given by:
\begin{equation}
S_{FI} = - \frac{1}{4} \int d^2 y  \int d \theta^+  \, t \mathrm{Tr} \Upsilon|_{\overline{\theta}^{+} = 0} + h.c.
         = - \int d^2 y \, \xi \mathrm{Tr} \mathcal{D} - \int d^2 y \, \frac{\eta}{2 \pi} \mathrm{Tr} F .
\end{equation}

\section{10D Super Yang-Mills as a 2D GLSM \label{app:HET}}

In this Appendix, we study 10D Super Yang-Mills theory compactified on a Calabi-Yau fourfold.
This theory arises in the context of perturbative type I and heterotic string compactifications.
We treat the spacetime as fixed and non-dynamical:
\begin{equation}\label{eqn:spacestructure}
  \mathbb{R}^{1,1} \times M\,,
\end{equation}
where $M$ denotes a Calabi-Yau fourfold which we also refer to as the internal space.

As we have already discussed in section \ref{sec:PERT}, some of the advantages of presenting an off-shell formalism for this theory are that we will then be able to control the structure of some of the quantum corrections to this model. Additionally,
it will allow us to quickly read off the structure of the low energy effective action obtained by
working around a specific background for the internal gauge fields.

The extent to which we will be able to successfully arrive at such an action hinges on a few features. First of all, as we already remarked in section \ref{sec:PERT}, there is an important constraint coming from the fact that our supercharges, and thus our gauginos obey the 10D Majorana-Weyl constraint. To a certain extent, this clashes with the condition of unitarity, since it means we need to impose a non-holomorphic constraint on the fermions of our model. More concretely, this shows up in the condition that the modes of the Fermi multiplet $\Lambda_{(0,2)}$ which transforms as a $(0,2)$ differential form on $M$ must obey an on-shell unitarity constraint. A straightforward albeit ad hoc workaround for this issue is to simply assume the existence of an additional $\mathbb{Z}_2$ symmetry, and to only keep the Fermi multiplets which are even. Then, we use a non-trivial $E$-field to realize the remaining BPS equations of motion.

Alternatively, we can continue to work in terms of the full $(0,2)$ form, but then at the end impose the 10D Majorana-Weyl constraint. The advantage of proceeding in this way is that all of the internal symmetries will be manifest from the start. The disadvantage, of course, is that since we have to impose a unitarity constraint by hand, we cannot claim that the resulting supersymmetry algebra fully closes off-shell. Rather, it closes up to an overall unitarity constraint. In the interest of showing the virtues of both approaches, in this Appendix we will focus on the latter approach. That is to say, we shall assemble our Fermi multiplets into a $(0,2)$ form, and only impose the 10D Majorana-Weyl condition at the very end. The price we pay is that supersymmetry will only partially close off-shell.

As a last comment, we note that to really obtain a fully off-shell action in 2D superspace away from Wess-Zumino gauge, it is necessary to add a non-local Wess-Zumino term, which follows from reduction of the term considered in reference \cite{Marcus:1983wb} (see also \cite{ArkaniHamed:2001tb}).

\subsection{Action and Symmetries} \label{sec:10DSYM}
We start with the gauge sector of the low energy effective supergravity description of the heterotic string. It is governed by the action \cite{Green:1987mn}
\begin{equation}\label{eqn:10DSYM}
  S_\mathrm{YM} = \frac{1}{4 g^2_{YM}} \int d^{10} x \sqrt{-g} \, \big( \Tr F_{IJ} F^{IJ} + 2 i \Tr \overline\chi \Gamma^I D_I \chi \big)\,,
\end{equation}
where $\chi$ is a 10D Majorana-Weyl spinor, and $D_I$ denotes the gauge covariant derivative
\begin{equation}
  D_I \chi = \partial_I \chi + [A_I, \chi]\,
\end{equation}
and the field strength is given by:
\begin{equation}
F_{IJ} = [D_I , D_J].
\end{equation}
We suppress the indices labeling the adjoint representation of the gauge algebra and write:
\begin{equation}
  A_I = A_I^\alpha T^\alpha \quad \text{and} \quad
  \chi = \chi^\alpha T^\alpha
\end{equation}
where $T^\alpha$ are antihermitian Lie algebra generators, as per our conventions in Appendix \ref{app:GLSM}.

The action \eqref{eqn:10DSYM} is invariant under the supersymmetry variations
\begin{equation}
  \delta_\epsilon A_I = - i \overline\epsilon \Gamma_I \chi
    \quad \text{and} \quad
    \delta_\epsilon \chi = \frac{1}{2} \Gamma^{IJ} F_{IJ} \epsilon\,.
  \label{eqn:susytrafo10D}
\end{equation}
Note that the we have chosen the to be compatible with the standard conventions of the $\mathcal{N} = (0,2)$.
Both $\epsilon$ and $\chi$ are Majorana-Weyl spinors with 16 real components.
The infinitesimal gauge transformations of $A_I$ and $\chi$ read
\begin{equation}
  \label{eqn:gaugetr10D}
  \delta_\rho A_I = \partial_I \rho + [A_I, \rho] = D_I \rho \quad \text{and} \quad
  \delta_\rho \chi = [\chi, \rho]\,,
\end{equation}
where $\rho$ denotes a Lie Algebra valued parameter.

As we have already mentioned, we are interested in 10D Super Yang-Mills theory compactified on a Calabi-Yau fourfold.
Let us now take a closer look at the spinors $\epsilon$ and $\chi$. In order to discuss their representations, we first switch to flat indices by applying the vielbein $e_{\hat A}{}^I$, resulting in
\begin{equation}
  g_{\hat A\hat B} = e_{\hat A}{}^I g_{IJ} e_{\hat B}{}^J = \diag (-1, 1, \dots, 1)\,.
\end{equation}
Further, we introduces Dirac matrices which are governed by the Clifford algebra
\begin{equation}\label{eqn:cliffordalg}
  \{ \Gamma_{\hat A} , \Gamma_{\hat B} \} = 2 g_{\hat A\hat B}\,.
\end{equation}
A canonical representation for them arises from the direct products
\begin{equation}\label{eqn:gammacanonical}
\begin{array}[b]{ccrcccccccc}
  \Gamma^0 &= & -i \sigma_1 &\otimes &\mathbf{1} &\otimes &\mathbf{1} &\otimes &\mathbf{1} &\otimes &\mathbf{1}  \\
  \Gamma^1 &= &\sigma_2 &\otimes &\mathbf{1} &\otimes &\mathbf{1} &\otimes &\mathbf{1} &\otimes &\mathbf{1} \\
  \Gamma^2 &= &\sigma_3 &\otimes &\sigma_{1} &\otimes &\mathbf{1} &\otimes &\mathbf{1} &\otimes &\mathbf{1} \\
  \Gamma^3 &= &\sigma_3 &\otimes &\sigma_{2} &\otimes &\mathbf{1} &\otimes &\mathbf{1} &\otimes &\mathbf{1} \\
  \vdots &= &\vdots  & & \vdots & & \vdots & & \vdots & & \vdots \\
  \Gamma^{9} &= &\sigma_3 &\otimes &\sigma_3 &\otimes &\sigma_3 &\otimes &\sigma_3 &\otimes &\sigma_2
\end{array}
\end{equation}
of Pauli matrices $\sigma_1$, $\sigma_2$ and $\sigma_3$. In addition the Dirac matrices, we are going to use the ten dimensional chirality operator
\begin{equation}
  \Gamma_{11} = \Gamma^0 \Gamma^1 \dots \Gamma^{9}
\end{equation}
and the charge conjugation matrix
\begin{equation}\label{eqn:chargeconj10D}
  C = \sigma_2 \otimes \sigma_1 \otimes \sigma_2 \otimes \sigma_1 \otimes \sigma_2\,.
\end{equation}
It is defined by the property
\begin{equation}
  (\Gamma^I)^T = - C \Gamma^I C^{-1}\,.
\end{equation}

While the representation \eqref{eqn:gammacanonical} is complex and 32 dimensional, we are interested in a 16 dimensional real representation. Thus, we first project on spinors with positive chirality and further require the Majorana condition
\begin{equation}\label{eqn:majoranacond}
  \chi^T C^T = \chi^\dagger \Gamma^0 = \overline{\chi}
\end{equation}
to hold. Rewriting \eqref{eqn:majoranacond}, we see how complex conjugation
\begin{equation}\label{eqn:ccspinor}
  \chi^* = (\Gamma^0)^{-T} C \chi = C \Gamma^0 \chi
\end{equation}
acts on a spinor. It is possible to rotate the Dirac matrices by a unitary transformation in such a way that $C \Gamma^0$ is equivalent to the identity matrix. In this case, complex conjugation does not change a spinor at all, and so it has to be real.

For the following calculations, it is essential to make the structure \eqref{eqn:spacestructure} of the spacetime manifest. Thus, we split the ten spacetime directions into two external and eight internal ones, e.g.
\begin{equation}
  \Gamma^{\hat A} = \begin{pmatrix} \Gamma^0 & \Gamma^1 & \Gamma^A \end{pmatrix}\,.
\end{equation}
Here the index $A$ on the right hand side labels the eight different directions of the internal space. This splitting leads to the branching rule
\begin{equation}
  \mathbf{10} \rightarrow (\mathbf{2}^v,\mathbf{1}) + (\mathbf{1},\mathbf{8}^v)
\end{equation}
of SO($1,9$) into SO($1,1$)$\times$SO($8$).

Of course the spinors $\chi$ and $\epsilon$ are affected, too. They now decompose into a chiral
and an antichiral Majorana Weyl spinor of the internal space. The corresponding internal chirality operator reads
\begin{equation}
  \Gamma_{11}' =  \Gamma^2 \Gamma^3 \dots \Gamma^9
\end{equation}
and commutes with $\Gamma_{11}$. By calculating the action of the SO($1,1$) generator
\begin{equation}
  J = - \frac{i}{2} \Gamma^{01}
\end{equation}
on the 16 dimensional spinor representation, we obtain the additional branching
\begin{equation}\label{eqn:decomp16}
  \mathbf{16} \rightarrow \mathbf{8}^s_{-} + \mathbf{8}^c_{+}\,.
\end{equation}
Note that $J$ is identified with the Lorentz generator of the $(0,2)$-SUSY algebra later on. Additionally, we have
used the triality outer automorphism of $Spin(8)$ to make a convenient choice for the internal spinor assignments which is compatible with our
$\mathcal{N} = (0,2)$ conventions.

Remember that the internal space is a Calabi-Yau fourfold. Thus, it is equipped with an integrable
complex structure $J^A{}_B$, fulfilling $J^2 = - \mathbf{1}$, and a canonical holomorphic four-form $\Omega_{ABCD}$ in addition to its metric $g_{AB}$. By lowering the complex structure's first index, one obtains the K\"ahler form $\omega_{AB} = g_{AC} J^C{}_B$. We expresses it as the antisymmetric part of the tensor product
\begin{equation}
  \mathbf{8}^s \otimes \mathbf{8}^s = \mathbf{1} + \mathbf{28} + \mathbf{35}\,,
\end{equation}
involving two pure\footnote{A spinor in $d$ dimensions is called pure, if it satisfies $\overline\epsilon \Gamma_{A_1\dots A_n} \epsilon = 0$ for $1 \le n < d/2$.}
Majorana Weyl spinors $\epsilon_1$, $\epsilon_2$ and obtain
\begin{equation}
  \omega_{AB} = \overline \epsilon_2 \Gamma_{AB} \epsilon_1\,.
\end{equation}
Majorana conjugation in this equation uses the charge conjugation matrix of the internal space
\begin{equation}
  C' = \mathbf{1} \otimes \sigma_2 \otimes \sigma_1 \otimes \sigma_2 \otimes \sigma_1
\end{equation}
instead of \eqref{eqn:chargeconj10D}. Further, the canonical holomorphic form
\begin{equation}
  \Omega_{ABCD} =  \overline{\epsilon_1} \Gamma_{ABCD} \epsilon_1
\end{equation}
arises from $\epsilon_1$, too. Both the complex structure and this four-form have to be non-vanishing on the internal manifold. This requirement reduces the structure group of the internal space from $SO(8)$ to $SU(4)$. The relevant branching rules are:
\begin{align}
SO(8)  &  \supset SU(4)\times U(1)\\
\mathbf{8}^{s}  &  \rightarrow\mathbf{1}_{+2}\oplus\mathbf{1}_{-2}%
\oplus\mathbf{6}_{0}\\
\mathbf{8}^{c}  &  \rightarrow\mathbf{4}_{-1}\oplus\overline{\mathbf{4}}%
_{+1}\\
\mathbf{8}^{v}  &  \rightarrow\mathbf{4}_{+1}\oplus\overline{\mathbf{4}}_{-1}.
\end{align}
They show that the two $SU(4)$ invariant invariant spinors $\epsilon_1$ and $\epsilon_2$ are contained in $\mathbf{8}^s$.

Let us make a few comments on the geometric content of the present decomposition. Take e.g. the constituents of $\mathbf{8}^c$.
Here, complex variables are more appropriate, as evidenced by the $U(1)$ generator
\begin{equation}\label{eqn:R}
  R = \frac{1}{4} \omega_{AB} \Gamma^{AB}\,,
\end{equation}
resulting from the K\"ahler form. It is helpful to remember that the holonomy of the metric is $SU(4)$ rather than $U(4)$.
The (complexification) of the overall $U(1)$ acts as a rescaling on the holomorphic four-form.
Its action with respect to the $\mathbf{8}^c$ is diagonalized by a unitary transformation.

It is also instructive to analyze how the vector
representation $\mathbf{8}^v$ is influenced by the reduced holonomy of the CY. To do so, we express an arbitrary vector
\begin{equation}
  V^A = \overline \chi_1 \Gamma^A \chi_2\,.
\end{equation}
in terms of two spinors. Applying a Lorentz transformation $\mathcal{M}^{BC} \Gamma_{BC}$ to it results in
\begin{equation}
  \delta_\mathcal{M} V^A = (C \mathcal{M}^{BC} \Gamma_{BC} \chi_1)^T \Gamma^A \chi_2 + \overline \chi_1 \Gamma^A \mathcal{M}^{BC} \Gamma_{BC} \chi_2 = \overline \chi_2 [ \Gamma^A, \mathcal{M}^{BC} \Gamma_{BC} ] \chi_1\,.
\end{equation}
Using the Clifford algebra \eqref{eqn:cliffordalg}, this equation simplifies to
\begin{equation}
  \delta_\mathcal{M} V^A = 4 \mathcal{M}^{BC} \delta^A{}_{[B} g_{C]D} V^D\,.
\end{equation}
Combined with \eqref{eqn:R}, this result immediately tells us
\begin{equation}
  \delta_R V^A = \delta^A{}_B \omega^{BC} g_{CD} V^D = J^A{}_B V^B\,.
\end{equation}
Thus, diagonalizing $R$ is equivalent to diagonalizing the complex structure.

Let us now examine the structure of the $\mathcal{N} = (0,2)$ supersymmetry algebra obtained
by compactifying on a Calabi-Yau fourfold. To this end, we calculate the anticommutator
\begin{equation}\label{eqn:anticommepsilon12a}
  \{\delta_{\epsilon_1}, \delta_{\epsilon_2}\} A_M = -2 i \overline\epsilon_2 \Gamma^N \epsilon_1 \partial_N A_M + \dots
\end{equation}
of the SUSY variations. The dots represent a gauge transformation which is discussed later.
We will see that in superspace it arises as a compensating gauge transformation required to keep Wess-Zumino gauge.
Focussing on the contribution from the $\mathbb{R}^{1,1}$ directions, we have:
\begin{equation}\label{eqn:anticommepsilon12b}
  \{\delta_{\epsilon_1}, \delta_{\epsilon_2}\} A_M = 2 i \overline \epsilon_1  \epsilon_2 \partial_+ A_M + \dots\,.
\end{equation}
Next, we evaluate the commutator with a generic Lorentz transformation $\mathcal{M}^{IJ} \Gamma_{IJ}$
\begin{equation}
  [\delta_\mathcal{M}, \delta_\epsilon] A_M = - ( C \mathcal{M}^{IJ} \Gamma_{IJ} \epsilon)^T \Gamma_M \chi\,,
\end{equation}
where $\delta_\mathcal{M}$ acts as
\begin{equation}
  \delta_\mathcal{M} A_M = 4 \mathcal{M}^{IJ} g_{M[I} \delta_{J]}{}^N A_N
    \quad \text{and} \quad
  \delta_\mathcal{M} \chi = \mathcal{M}^{IJ} \Gamma_{IJ} \chi
\end{equation}
on vectors and spinors, respectively. Specializing to the generators $J$ and $R$, whose values
for $\epsilon$ are collected in table \ref{tab:fields}, we obtain
\begin{equation}
  [\delta_\epsilon, \delta_J] = \frac{i}{2} \delta_\epsilon
    \quad \text{and} \quad
  [\delta_\epsilon, \delta_R] = 2 i \delta_\epsilon\,.
\end{equation}
Finally, we make the substitutions
\begin{equation}
  \delta_\epsilon \mapsto Q_+\,,  \quad
  \delta_{\overline\epsilon} \mapsto \overline Q_+\,, \quad
  \partial_+ \mapsto i P_+ \,, \quad
  \delta_J \mapsto J \quad \text{and} \quad
  \delta_R \mapsto R
\end{equation}
to obtain the $\mathcal{N} = (0,2)$ supersymmetry algebra.

So far, we have seen that the gauginos can be decomposed into different irreducible representations of $SU(4)$. Following \cite{Green:1987mn}, we go a step further and identify them with differential forms on the CY fourfold. To this end, interpret the Dirac matrices $\Gamma^a$ as fermionic annihilation and $\Gamma^{\overline a}$ as fermionic creation operators with the canonical anticommutator relations
\begin{equation}
  \{\Gamma^a, \Gamma^b\} = \{\Gamma^{\overline a}, \Gamma^{\overline b}\} = 0
    \quad \text{and} \quad
  \{\Gamma^a, \Gamma^{\overline b}\} = 2 g^{a\overline b}\,.
\end{equation}
Further, take a vector $|1\rangle \sim \overline \epsilon$, where $\overline \epsilon$ is the $\mathbf{1}_{-2}$ part of $\mathbf{8}^s$, with $\langle 1 | 1 \rangle = 1$. It is annihilated by all creation operators. Thus, it is the vacuum state on which we build all other representations. The guess for $\chi$ in terms of creation operators would be
\begin{equation}
  \chi = \frac{1}{\sqrt{2}} \big( -\frac{1}{2!} \lambda_{\overline a\overline b} \Gamma^{\overline a\overline b} + i \psi_{\overline a} \Gamma^0 \Gamma^{\overline a} + \sqrt{2} \overline\mu \big) |1 \rangle\,.
\end{equation}
In order to match with the conventions in the literature, we have to insert the given prefactors in the expansion. The two components in this equation are directly connected to the representations
\begin{equation}\label{eqn:ansatzchi0}
  \overline \mu \sim \mathbf{1}_{-2}\,, \quad
  \lambda_{\overline a\overline b} \sim \mathbf{6}_0
    \quad  \text{and} \quad
  \psi_{\overline a} \sim \mathbf{4}_{-1}\,,
\end{equation}
in the $\mathbf{8}^s$ and the $\mathbf{8}^c$, respectively. But what about the remaining spinors?  They have positive $U(1)$ charge and cannot be associated the antiholomorphic differential forms directly. Solving this puzzle, we have to take into account that $\chi$ is a Majorana-Weyl spinor and thus has to fulfill a reality condition. The naive ansatz \eqref{eqn:ansatzchi0} fails to do so. We have to expand it with complex conjugate $\mu$, $\overline{\psi}_a$ of $\overline\mu$, $\psi_{\overline a}$. These are exactly the missing
\begin{equation}
  \mu \sim \mathbf{1}_{2}
    \quad \text{and} \quad
  \psi_a \sim \overline{\mathbf 4}_1
\end{equation}
spinors. Hence, the complete expansion of $\chi$ reads
\begin{equation}
  |\chi\rangle = \frac{1}{\sqrt{2}} \big(- \sqrt{2} \frac{1}{4!} \mu \overline\Omega_{\overline a\overline b\overline c\overline d} \Gamma^{\overline a\overline b\overline c\overline d} + i \frac{2}{3!} \overline{\psi}_a  \overline\Omega^{a}{}_{\overline b\overline c\overline d} \Gamma^0 \Gamma^{\overline b\overline c\overline d} - \frac{1}{2!} \lambda_{\overline a\overline b} \Gamma^{\overline a\overline b} + i \psi_{\overline a} \Gamma^0 \Gamma^{\overline a} + \sqrt{2} \overline\mu \big) |1 \rangle\,. \label{eqn:ansatzchi}
\end{equation}
It fulfills the Majorana condition
\begin{equation}
  \big( C \Gamma^0 |\chi\rangle \big)^* = |\chi\rangle \label{eqn:majorana1}\,.
\end{equation}
In explicitly checking this statement, we encounter two types of terms
\begin{align}
  \big( C \Gamma^0 \Gamma^0 \Gamma^{\overline a_1 \dots \overline a_N} |1\rangle \big)^* & = - \frac{1}{4!} \overline\Omega_{\overline b_1\overline b_2 \overline b_3 \overline b_4} \Gamma^0 \Gamma^{a_1 \dots a_N} \Gamma^{\overline b_1\overline b_2\overline b_3\overline b_4} |1\rangle \quad \text{and} \nonumber\\
  \big( C \Gamma^0 \Gamma^{\overline a_1 \dots \overline a_N} |1\rangle \big)^* & = - \frac{1}{4!} \overline\Omega_{\overline b_1\overline b_2 \overline b_3 \overline b_4} \Gamma^{a_1 \dots a_N} \Gamma^{\overline b_1\overline b_2\overline b_3\overline b_4} |1\rangle\,,\label{eqn:majorana2}
\end{align}
where we used
\begin{equation}
  \big( C \Gamma^0 |1\rangle \big)^* = \frac{1}{4!} \overline\Omega_{\overline a\overline b\overline c\overline d} \Gamma^{\overline a\overline b\overline c\overline d}\,.
\end{equation}
With the property
\begin{equation}
  \Gamma^{a} \Gamma^{\overline b_1\overline b_2 \dots \overline b_N} =
  2 N g^{a [\overline b_1} \Gamma^{\overline b_2 \dots \overline b_N]} +
    (-1)^N \Gamma^{\overline b_1\overline b_2 \dots \overline b_N} \Gamma^a
\end{equation}
of the Dirac matrices, we further derive the identity
\begin{equation}
  \frac{1}{2^N} \overline\Omega_{\overline b_1 \dots \overline b_M} \Gamma^{a_1 \dots a_N} \Gamma^{\overline b_1 \dots \overline b_M} |1\rangle =
  \frac{M!}{(M-N)!} \overline\Omega^{a_1 \dots a_N}{}_{\overline b_{N+1} \dots \overline b_M} \Gamma^{\overline b_{N+1} \dots \overline b_{M}} |1\rangle\,. \label{eqn:commutegamma}
\end{equation}
which in combination with \eqref{eqn:majorana2} and \eqref{eqn:ansatzchi} gives exactly \eqref{eqn:majorana1}. For $\lambda_{\overline a\overline b}$, we obtain the additional constraint
\begin{equation}
  2 \overline{\lambda}_{ab} \overline\Omega^{ab}{}_{\overline c\overline d} = - \lambda_{\overline c\overline d}\,.
\end{equation}
Again, $\overline{\lambda}_{ab}$ denotes the Hermitian conjugate of $\lambda_{\overline a\overline b}$.

Here we encounter an important issue. Since $\mathbf{6}_0$ is the only real representation of SU(4) we encountered, such a constraint is natural when trying to express it in terms of a complex quantity. Let us finally comment on the prefactor $1/\sqrt{2}$ in the expansion \eqref{eqn:ansatzchi}. It is required because we combine two 2D Majorana-Weyl spinors into one Weyl spinor.

It simplifies the notation considerable to introduce the differential forms
\begin{equation}
  \psi_{(0,1)} = \psi_{\overline a} d\overline z^{\overline a}\,, \quad
  \overline{\psi}_{(1,0)} = \overline{\psi}_a d z^a\,, \quad
  \lambda_{(0,2)} = \frac{1}{2} \lambda_{\overline a\overline b} d\overline z^{\overline a}\wedge d\overline z^{\overline b}
    \quad \text{and} \quad
  \lambda_{(2,0)} = \frac{1}{2} \overline{\lambda}_{ab} d z^a \wedge d z^b\,.
\end{equation}
According to the Majorana-Weyl condition:
\begin{equation}\label{eqn:numajorana}
  * ( \lambda_{(0,2)} \wedge \Omega ) = - \overline{\lambda}_{(0,2)}\,.
\end{equation}
On the fourfold, the Hodge star is defined as
\begin{align}
  * \varphi = &\frac{ \det g} {p!q!(4-p)!(4-q)!} \epsilon^{m_1 \dots m_p}{}_{\overline  j_1\dots\overline j_{4-p}} \epsilon^{\overline n_1\dots\overline n_q}{}_{i_1\dots i_{4-q}} \nonumber\\
  &\cdot \varphi_{m_1\dotsm_p \overline n_1\dots\overline n_q} d z^i_1\wedge\dots \wedge d z^{i_{4-q}} \wedge d \overline z^{j_1} \wedge \dots \wedge d \overline z^{j_{4-p}}\,.
\end{align}
It maps a ($p,q$)-form
\begin{equation}
  \varphi = \frac{1}{p!q!} \varphi_{i_1\dots i_p\overline j_1 \dots \overline j_q} d z^{i_1}\wedge\dots\wedge z^{i_p}\wedge d \overline z^{\overline j_1}\wedge \dots \wedge d \overline z^{\overline j_q}
\end{equation}
to a $(4-q,4-p)$-form, while complex conjugation
\begin{equation}
  \overline{\varphi} = \frac{1}{p!q!} \overline\varphi_{j_1\dots j_q \overline i_1 \dots \overline i_p} d z^{j_1}\wedge\dots\wedge z^{j_q}\wedge d \overline z^{\overline i_1}\wedge \dots \wedge d \overline z^{\overline i_p}
\end{equation}
relates $(p,q)$-forms with $(q,p)$-forms. The internal components of the vector potential $A_a$ and $A_{\overline a}$ can be combined the differential forms
\begin{equation}
  A_{(0,1)} = A_{\overline a} d \overline z^{\overline a}
    \quad \text{and} \quad
  A_{(1,0)} = A_a d z^a\,.
\end{equation}

Table \ref{tab:fields} summarizes all the fields we discussed so far and states its charges under $J$, $R$ and the chirality operator
\begin{equation}
  \Gamma_3' = \Gamma^0 \Gamma^1\,.
\end{equation}
To reproduce the correct number (16) of degrees of freedom (DOF), remember that the complex fields $A_{(0,1)}$, $\overline \mu$, $\overline \epsilon$ and their complex conjugates $A_{(1,0)}$, $\mu$ and $\epsilon$ are not independent from each other. The canonical choice is to only consider $A_{(0,1)}$, $\overline \mu$, $\overline \epsilon$ as dynamic.
\begin{table}
  \centering
  \begin{tabular}{|c||cccc|ccccc|cc|}\hline
    Field & $A_0$ & $A_1$ & $A_{(1,0)}$ & $A_{(0,1)}$ &
      $\mu$ & $\overline\mu$ & $\lambda_{(0,2)}$ &
      $\psi_{(1,0)}$ & $\psi_{(0,1)}$ &
      $\epsilon$ & $\overline\epsilon$ \\
    \hline
    DOF & $\phantom{-}1$ & $1$ & $8$ & $\phantom{-}8$ &
      $\phantom{-}2$ & $\phantom{-}2$ & $\phantom{-}6$ &
      $\phantom{-}8$ & $\phantom{-}8$ &
      $\phantom{-}2$ & $\phantom{-}2$ \\
    $J$ & $-1$ & $1$ & $0$ & $\phantom{-}0$ &
      $\phantom{-}\frac{1}{2}$ & $\phantom{-}\frac{1}{2}$ & $\phantom{-}\frac{1}{2}$ &
      $-\frac{1}{2}$ & $-\frac{1}{2}$ &
      $\phantom{-}\frac{1}{2}$ & $\phantom{-}\frac{1}{2}$ \\
    $R$ & $\phantom{-}0$ & $0$ & $1$ & $-1$ &
      $-2$ & $\phantom{-}2$ & $\phantom{-}0$ &
      $-1$ & $\phantom{-}1$ &
      $\phantom{-}2$ & $-2$ \\
    $\Gamma_3'$ & \phantom{-}-- & -- & -- & \phantom{-}-- &
      $-1$ & $-1$ & $-1$ & $\phantom{-}1$ & $\phantom{-}1$ &
      $-1$ & $-1$\\\hline
  \end{tabular}
  \caption{Field content of the 10D vector multiplet packaged in terms of differential forms of the internal space. Here, $J$ refers to the
  spin, $R$ to the $U(1)$-charge, and $\Gamma_{3}^{\prime}$ to the chirality of the fermionic states.}
\label{tab:fields}
\end{table}
Further, we have to restrict $\lambda_{(0,2)}$ according to the condition \eqref{eqn:numajorana}. Out of the 12 real degrees of freedom a (0,2)-form has, only 6 real degrees of freedom survive.

We repeat the steps outlined for $\chi$ in the last subsection for the parameter of supersymmetry transformations $\epsilon$, giving rise to
\begin{equation}
  |\epsilon\rangle = \big(-\epsilon \frac{1}{4!} \overline\Omega_{\overline a\overline b\overline c\overline d} \Gamma^{\overline a\overline b\overline c\overline d} + \overline \epsilon\big) |1\rangle\,.
\end{equation}
Let us now calculate the supersymmetry transformations for the different fields in table~\ref{tab:fields}. We start with
\begin{equation}
  \delta_\epsilon A_{\overline a} = - i \langle \epsilon | C^T \Gamma_{\overline a} |\chi\rangle  = \frac{1}{\sqrt{2}}\langle \epsilon | C^T (\Gamma^0)^{-1} \Gamma_{\overline a} \Gamma^{\overline b} \psi_{\overline b} |1\rangle = \sqrt{2} \langle \overline \epsilon | 1 \rangle \psi_{\overline a} = \sqrt{2} \epsilon \psi_{\overline a}\,,
\end{equation}
where we used that $C^T (\Gamma^0)^{-1}$ is the same as complex conjugation according to \eqref{eqn:ccspinor}. Hence, it transforms $|\epsilon\rangle$ in its complex conjugated
\begin{equation}
  |\overline\epsilon\rangle = \big(-\overline\epsilon \frac{1}{4!} \overline\Omega_{\overline a\overline b\overline c\overline d} \Gamma^{\overline a\overline b\overline c\overline d} + \epsilon\big) |1\rangle\,.
\end{equation}

Using the gamma matrices
\begin{equation}
  \Gamma_+ = \Gamma_0 + \Gamma_1
    \quad \text{and} \quad
  \Gamma_- = \Gamma_0 - \Gamma_1,
\end{equation}
we calculate the supersymmetry variation
\begin{equation}
  \delta_\epsilon A_{\pm} = - i \langle \epsilon | C^T \Gamma_\pm |\chi\rangle = - i \langle \overline\epsilon | \Gamma^0 (- \Gamma^0 \pm \Gamma^1) |\chi\rangle = - i \langle \overline\epsilon |  1 \pm \Gamma_3' |\chi\rangle\
\end{equation}
of $A_{\pm}$ directly from \eqref{eqn:susytrafo10D}. This equation tells us that only fermions with positive/negative chirality $\Gamma_3'$ in two dimensions contribute. Thus $\delta_\epsilon A_+$ vanishes, while
\begin{equation}
  \delta_\epsilon A_- = -2 i ( \epsilon \overline\mu + \overline\epsilon\mu)\,.
\end{equation}
To derive the supersymmetry transformation
\begin{equation}
  \delta_\epsilon \lambda_{\overline a\overline b} = \sqrt{2} (2  \epsilon \overline F_{ab} \overline\Omega^{ab}{}_{\overline a\overline b} -  \overline{\epsilon} F_{\overline a\overline b})
\end{equation}
of $\lambda_{\overline a\overline b}$, we remember \eqref{eqn:commutegamma}. From $\Gamma^{+-} = -\frac{1}{2} \Gamma'_3$, we further conclude
\begin{equation}
  \delta_\epsilon \overline\mu =  \overline\epsilon \big( \frac{1}{2}F_{+-} - F_{a\overline b} g^{a \overline b} \big)\,.
\end{equation}
Finally, there is
\begin{equation}
  \delta_\epsilon \psi_{\overline a} = -i \sqrt{2} \overline\epsilon F_{+\overline a}\,,
\end{equation}
which follows by applying the identity
\begin{equation}
  - \Gamma^0 \Gamma^{\overline a \pm} |1\rangle = \frac{1}{2} \Gamma^0 (\Gamma^0 \pm \Gamma^1) \Gamma^{\overline a} |1\rangle = \frac{1}{2} ( 1 \pm \Gamma_3' ) \Gamma^{\overline a} |1\rangle \,.
\end{equation}

\subsection{2D Action}\label{sec:offshellaction}

We now construct the two-dimensional off-shell action for our 10D Super Yang-Mills theory. The main idea will be to assemble the fields of the
higher-dimensional theory into supermultiplets of a 2D GLSM. Along these lines, we formally label the collection of such fields by points
of the internal manifold. The discussion is quite similar to that given in Appendix \ref{app:GLSM}, so we shall be somewhat brief. Essentially,
we need to take the components of the 10D gauge field $A_{+}$ and $A_{-}$ with legs along $\mathbb{R}^{1,1}$, and identify them with the components
$v_{+}$ and $v_{-}$ of the respective supermultiplets $\Xi$ and $V$. Additionally, we have a gauge field strength supermultiplet $\Upsilon$. We shall also encounter a supermultiplet transforming as the $(0,1)$ component of the internal gauge field, as well as a Fermi multiplet which transforms as an adjoint valued $(0,2)$ differential form. One important point about organizing the mode content in this way is that although most of the 2D vector multiplets labelled by points of the internal manifold will implicitly pick up a mass and so should be counted as massive (rather than massless) vector multiplets, the super Higgs mechanism naturally pairs these with the $(0,1)$-form chiral multiplets which we also explicitly track. Therefore, it is appropriate to work in terms of the massless basis of fields adopted here. As a last general comment, we note that the 2D gauge coupling for the zero modes will be controlled by:
\begin{equation}
  \frac{1}{e^2} = \frac{\mathrm{Vol}(M)}{g^2_{YM}}
\end{equation}
In what follows, however, we will be integrating over the internal space, so we have more than just the zero modes.

Inspired by $\nabla_-$, we also define the chiral and anti-chiral covariant derivatives:
\begin{equation}
  \mathbb{D}_{\overline a} = \partial_{\overline a} +  \mathbb{A}_{\overline a}
    \quad\text{and}\quad
  \overline{\mathbb D}_a = \partial_a +  \overline{\mathbb A}_a = \overline{\mathbb{D}_{\overline a}}\,,
\end{equation}
which satisfy:
\begin{equation}
  [\mathbb{D}_{\overline a}, \overline{\mathcal D}_+] = [ \overline{\mathbb D}_a, \mathcal{D}_+] = 0\,.
\end{equation}
Their connections contain the internal components of $A_I$ and the fermions in the $\mathbf{8}^c$.
\begin{align}
  \mathbb{A}_{\overline a} &= A_{\overline a} + \sqrt{2}  \theta^+ \psi_{\overline a} - i \theta^+ \overline\theta^+ F_{+\overline a} \\
  \overline{\mathbb A}_a &= A_a + \sqrt{2}  \overline\theta^+ \psi_a + i \theta^+ \overline\theta^+ \overline{F}_{+a}.
\end{align}
Here, $F_{+ \overline a} = [D_{+} , D_{\overline{a}}]$,
and so quite naturally, the field strength with one leg
along the internal space, and one along $\mathbb{R}^{1,1}$ will, upon squaring
give us the kinetic term for the scalar components of this superfield.

Under a 10D gauge transformation parameterized by a collection of adjoint valued
chiral superfields $C$ labelled by internal points of $M$, we have the standard rule for the internal
gauge fields and their transformation:
\begin{equation}
\mathbb{D}_{\overline{a}} \mapsto e^{-C} \mathbb{D}_{\overline{a}} e^{+C}.
\end{equation}
Upon expanding (recall we have anti-hermitian Lie algebra generators) $C = (\chi - \overline{\chi}) + ...$, one recovers:
\begin{equation}
  \delta_\chi \mathbb{A}_{\overline a} = \partial_{\overline a} (\chi - \overline\chi) + [\mathbb{A}_{\overline a}, \chi - \overline\chi]\,.
\end{equation}
Hence calculating the supersymmetry transformations
\begin{equation}
  \delta_\epsilon \mathbb{A}_{\bar a} = (\epsilon Q_+ - \bar\epsilon \bar Q_+ + \delta_\chi) \mathbb{A}_{\bar a}\,,
\end{equation}
we obtain
\begin{equation}
  \delta_\epsilon A_{\bar a} = \sqrt{2} \epsilon \psi_{\bar a}
    \quad \text{and} \quad
  \delta_\epsilon \psi_{\bar a} = - i \sqrt{2} \bar\epsilon F_{+\bar a}\,,
\end{equation}
which perfectly match with the 10D results.

In analogy with the field strength $\Upsilon$ in the 2D directions,
we also have a field strength in the internal directions given by:
\begin{equation}
  \mathbb{F}_{AB} =  [\mathbb{D}_A, \mathbb{D}_B]\,.
\end{equation}
We hasten to add that only the combination with just anti-holomorphic indices defines a chiral superfield.
Expanding in components, we have:
\begin{align}
  \mathbb{F}_{\overline a\overline b} &= F_{\overline a\overline b} + \sqrt{2}  \theta^+ \big( D_{\overline a} \psi_{\overline b} - D_{\overline b} \psi_{\overline a} \big) - i \theta^+ \overline\theta^+ D_+ F_{\overline a\overline b} \\
  \mathbb{F}_{a\overline b} &= F_{a\overline b} - \sqrt{2} \theta^+ D_a \psi_{\overline b} - \sqrt{2}  \overline\theta^+ D_{\overline b} \psi_a - i \theta^+ \overline\theta^+ \big( D_a F_{+\overline b} + D_{\overline b} \overline{F}_{+a} + 2 i \{\psi_a, \psi_{\overline b}\} \big)\,.
\end{align}
The kinetic term for the chiral superfield $\mathbb{D}_{\overline{a}}$ is:
\begin{equation}
  L_{\mathbb{D}} = \frac{1}{2} \int_{M} g^{a \overline b} \Tr( \overline{\mathbb{D}}_a [\nabla_- , \mathbb{D}_{\overline b}] )\,.
\end{equation}
which gives rise to the component action:
\begin{equation}
  S_{\mathbb{D}} = - \int d^2 y \int_{M} g^{a \overline{b}}\Tr \left( \overline{F}_{+ a} F_{- \overline b}
  +  i \overline{\psi}_{a} D_- \psi_{\overline b}  - i \sqrt{2} \mu \overline{D}_{a} \psi_{\overline{b}}
    - i \sqrt{2} \overline{\mu} D_{\overline{b}} \overline{\psi}_{a}
    - \mathcal{D} F_{a\overline b} \right).
\end{equation}
By combining this contribution with that coming from the term proportional to $\overline{\Upsilon} \Upsilon$,
we can solve the equations of motion for the auxiliary field:
\begin{equation}\label{eqn:sold}
  \mathcal{D} =  2 g^{a\overline b} F_{a\overline b}\,.
\end{equation}
Plugging \eqref{eqn:sold} into the supersymmetry variations, we obtain
\begin{equation}
  \delta_\epsilon \mu = \epsilon ( \frac{1}{2} F_{+-} - i g^{a\overline b} F_{a\overline b})
    \quad \text{and} \quad
  \delta_\epsilon \overline\mu =  \overline\epsilon ( \frac{1}{2} F_{+-} + i g^{a\overline b} F_{a\overline b})\,,
\end{equation}
which reproduces the variations from the 10D theory.

Finally, there is the Fermi multiplet which has the fermions $\lambda_{\overline a\overline b}$ as top component. As we have already remarked
in section \ref{sec:10DSYM} and also at the beginning of the Appendix, there is a tradeoff here between maintaining a proper count
of the off-shell degrees of freedom (i.e. by imposing a $\mathbb{Z}_2$ symmetry on the geometry), or by working in terms of a
$(0,2)$ differential form on the Calabi-Yau fourfold. In the former case, the component expansion for the superfield is:
\begin{equation}
\Lambda^{(even)}_{\overline a\overline b} = \lambda^{(even)}_{\overline a\overline b} - \sqrt{2} \theta^+ \mathcal{G}^{(even)}_{\overline a\overline b} - \sqrt{2} \overline\theta^+ \mathbb{F}^{(even)}_{\overline a\overline b}
- i \theta^+ \overline\theta^+ D_+ \lambda^{(even)}_{\overline a\overline b} \,,
\end{equation}
where the $E$-field has been chosen for the Fermi multiplet so that it is given by $\mathbb{F}^{(even)}_{\overline a\overline b}$.
In the latter case, we simply set to zero the contribution from the E-field, and in this case we have:
\begin{equation}
\Lambda_{\overline a\overline b} = \lambda_{\overline a\overline b} - \sqrt{2} \theta^+ \mathcal{G}_{\overline a\overline b}
- i \theta^+ \overline\theta^+ D_+ \lambda_{\overline a\overline b} \,.
\end{equation}
The primary disadvantage of the latter case is that our superspace action will not respect the correct counting of degrees of freedom
when compared with the 10D Majorana-Weyl constraint. The advantage, of course, is that the symmetries of the internal geometry are more manifest from the start. We shall indeed adhere to the latter version in this Appendix, but it is important to keep in mind that at all stages of our analysis, we can make the substitution of superfields $\Lambda \rightarrow \Lambda^{(even)}$, and keep only the $\mathbb{Z}_2$ invariant interaction terms to obtain a fully off-shell presentation in two dimensions of 10D Super Yang-Mills.

Working in terms of the full $\mathcal{N} = (0,2)$ superfield, its kinetic terms are given by:
\begin{equation}
  S_\Lambda = \frac{1}{2} \int d^2 y d^{2} \theta \int_M \, g^{a\overline c} g^{b\overline d}\Tr \overline{\Lambda}_{ab} \Lambda_{\overline c\overline d} = - \int d^2 y \Tr \left( \overline{\mathcal{G}}^{\overline a\overline b} \mathcal{G}_{\overline a\overline b} + i \overline{\lambda}^{\overline a\overline b} D_+ \lambda_{\overline a\overline b} \right)
\end{equation}

So far, we only considered D-terms. As we have already explained in section \ref{sec:10DSYM}, the appropriate superpotential is:
\begin{equation}
  W_{top} = - \int_M \Omega^{\overline a\overline b\overline c\overline d} \Tr(\Lambda_{\overline a\overline b} \mathbb{F}_{\overline c\overline d} ).
\end{equation}
So, we obtain the F-term interactions:
\begin{equation}
  S_W = \int d^2 y d \theta^{+}\, W + \mathrm{h.c.} = \int d^2 y \int_M \Omega^{\overline a\overline b\overline c\overline d} \Tr ( \mathcal{G}_{\overline a\overline b} F_{\overline c\overline d} + \lambda_{\overline a\overline b} D_{\overline c} \psi_{\overline d} ) + \mathrm{h.c.}\,.
\end{equation}
Solving the equation of motion for the auxiliary field $\mathcal{G}$ yields:
\begin{equation}
  \mathcal{G}_{\overline a\overline b} =  \overline\Omega_{\overline a\overline b}{}^{cd} \overline{F}_{cd}\,.\label{eqn:solG}
\end{equation}

\subsection{Summary}
Let us summarize the results of the last section. In terms of superfields, the complete two-dimensional
$\mathcal{N} = (0,2)$ supersymmetric action can be written in terms of differential forms as:
\begin{align}
  S_\mathrm{tot} & = S_{D} + S_{F} \\
  S_{D} & = - \frac{1}{g^2_{YM}} \int d^2 y d^{2} \theta \int_M \, \Tr \left( \frac{1}{8}* \overline{\Upsilon} \wedge \Upsilon -
  \frac{i}{2}* \overline{\mathbb{D}_{(0,1)}} \wedge \nabla_- \mathbb{D}_{(0,1)} - \frac{1}{2}  * \overline{\Lambda} \wedge \Lambda \right) \\
  S_{F} & = - \frac{1}{\sqrt{2}}\frac{1}{g^{2}_{YM}} \int d^2 y d \theta^{+} \int_M \, \Tr \left( \Omega\wedge\Lambda_{(0,2)}\wedge\mathbb{F}_{(0,2)} \right) + h.c.
\end{align}
where here, we present the Fermi multiplets as $(0,2)$ differential forms on the Calabi-Yau fourfold and the 10D Majorana-Weyl constraint
is then imposed ``by hand''. Alternatively, when there is a $\mathbb{Z}_2$ symmetry available we can make the action fully off-shell by making
the substitution $\Lambda_{(0,2)} \mapsto \Lambda_{(0,2)}^{(even)}$, and keeping only the $\mathbb{Z}_{2}$ invariant F-terms. Observe that nothing
is projected out of the D-terms since they are always $\mathbb{Z}_2$ invariant.
Finally, the expansion into components is:
\begin{align}
  S = - \frac{1}{g^2_{YM}} \int d^2 y \int_M \, \Tr\Big( \frac{1}{4} &* F_{(0,0)}\wedge F_{(0,0)} + \frac{1}{2}*\mathcal{D}\wedge \mathcal{D}
  +i *\overline{\mu}\wedge D_+ \mu + * \overline{F_{(0,1)}} \wedge F_{(0,1)}  \nonumber \\[-0.5em]
  & + i * \overline{\psi} \wedge D_- \psi - i \sqrt{2} (* \omega\wedge \overline{\mu} \overline{\partial}_A \overline{\psi} + \mathrm{h.c})- *\omega\wedge F_{(1,1)} \mathcal{D} + \nonumber \\[0.2em]
    & -  *\overline{\mathcal{G}}\wedge\mathcal{G} + i * \overline{\lambda} \wedge D_+ \lambda  \nonumber\\[0.2em]
    & (- \Omega \wedge \mathcal{G} \wedge F_{(0,2)} -  \Omega\wedge\lambda\wedge\overline{\partial}_{A} \psi + h.c. ) \Big)\,,
\end{align}
where we introduced the differential forms
\begin{equation}
  F_{(0,0)} = F_{-+}\,, \quad
  F_{(0,1)} = F_{-\overline a} d z^{\overline a}\,, \quad
  F_{(0,2)} = \frac{1}{2} F_{\overline a\overline b} d z^{\overline a} \wedge d z^{\overline b}
    \quad \text{and} \quad
  \mathcal{G} = \frac{1}{2} \mathcal{G}_{\overline a\overline b} d z^{\overline a} \wedge d z^{\overline b}\,.
\end{equation}
After integrating out the auxiliary fields $\mathcal{D}$ and $\mathcal{G}$, we obtain the 10D BPS equations of motion:
\begin{equation}
  F_{(0,0)} = 0\,, \quad
  F_{(0,1)} = 0\,, \quad
  *\omega \wedge F_{(1,1)} = 0
    \quad \text{and} \quad
  F_{(0,2)} = 0\,.
\end{equation}

\section{Intersecting 7-Branes as a 2D GLSM \label{app:FTH}}

In the previous Appendix we presented the action for 10D Super Yang-Mills theory, but written in terms
of a two-dimensional $\mathcal{N} = (0,2)$ GLSM. Our plan in this Appendix will be to follow a similar procedure in the case of intersecting
7-branes. An important feature of this construction is that the ``ad hoc'' $\mathbb{Z}_2$ symmetry introduced by hand
in the case of the 10D theory is automatically implemented for intersecting 7-branes.

\subsection{Explicit Decomposition of the Field Content}
\label{sec:decomp}
To begin, we recall the decomposition of the bulk modes of 10D Super Yang-Mills theory into modes of 8D Super Yang-Mills theory on the
spacetime:
\begin{equation}
\mathbb{R}^{1,1} \times X
\end{equation}
with $X$ a K\"ahler threefold. Following
conventions as in section \ref{sec:NPERT}, we have:
\begin{eqnarray} \label{eq:fingaugedec}
SO(1,9) &\rightarrow & SO(1,1) \times SU(4) \times U(1)_R\nonumber\\
A_{I} &\rightarrow & \begin{cases} A_{\pm}=A_0\pm A_1 \quad \leftrightarrow \quad  \mathbf{1_{\pm \pm 0}}, \\ A_{\overline{a}}= \frac{A_{2a}-i A_{2a+1}}{\sqrt{2}} \;\; a=1,\ldots, 3, \quad \leftrightarrow \quad  \mathbf{\overline{3}_{-1}} \; \leftrightarrow \; A_{(0,1)}= A_{\overline{a}} \, d\overline{z}^{\overline{a}}   \ ,\\ \phi\equiv \frac{A_{8}+i A_{9}}{\sqrt{2}} \quad \leftrightarrow \quad  \mathbf{1_3} \; \leftrightarrow\; \phi_{(3,0)}=\phi_{abc}\, dz^a \wedge dz^b \wedge dz^c \ ,
\end{cases}
\end{eqnarray}
where $(z_1,z_2,z_3)$ is a local basis of coordinates for $X$. All the fields are also adjoint valued, since $A_M^{\alpha} T^{\alpha}$ where $T^{\alpha}$ are the generator of the gauge algebra in the adjoint representation.

The ten-dimensional gaugino $\chi$ decomposes and organizes supermultiplets as stated in section \ref{sec:NPERT}. In order to decompose the ten-dimensional supersymmetry variations into variations of the 2D $\mathcal{N} = (0,2)$ theory, we give here an explicit basis of ten-dimensional gamma matrices and the relative decomposition of the 10D gaugino $\chi$ in components. A ten dimensional Clifford algebra represented by gamma matrices, $\{\Gamma^I,\Gamma^J\}=2g^{IJ}$, decomposes in the following way, according to $\mathbb R^{9,1} \rightarrow \mathbb{R}^{1,1} \times X \times \mathbb C$,
\begin{align}
& \Gamma^0 = i \sigma_2 \otimes \mathbb I_8 \otimes \mathbb I_2,\\
& \Gamma^1 =  \sigma_1 \otimes \mathbb I_8 \otimes \mathbb I_2,\\
&  \Gamma^{1+m} = \sigma_3 \otimes \gamma^i \otimes \mathbb I_2,\\
&  \Gamma^{8} = \sigma_3 \otimes \gamma \otimes \mathbb \sigma_2,\\
&  \Gamma^{9} = \sigma_3 \otimes \gamma \otimes \mathbb \sigma_1,
\end{align}
where $m=1,\cdots, 6$, and $\mathbb{I}_d$ is the $d$-dimensional identity matrix. The six-dimensional gamma matrices are given by
\begin{align}
&\gamma^1=\sigma_2 \otimes \mathbb I_2 \otimes\mathbb I_2, \\
&\gamma^2=\sigma_1 \otimes \mathbb I_2 \otimes\mathbb I_2, \\
&\gamma^3=\sigma_3 \otimes \sigma_2 \otimes\mathbb I_2, \\
&\gamma^4=\sigma_3 \otimes \sigma_1 \otimes\mathbb I_2, \\
&\gamma^5=\sigma_3 \otimes \sigma_3 \otimes \sigma_2, \\
&\gamma^6=\sigma_3 \otimes \sigma_3 \otimes \sigma_1, \\
&\gamma=\sigma_3 \otimes \sigma_3 \otimes \sigma_3.
\end{align}
In light-cone coordinates $y^{\pm}=\frac{y^0\pm y^1}{2}$, and complex coordinates on $X$ and $\mathbb C$, $z^{i}=\frac{x^{2a} + i x^{2a+1}}{\sqrt{2}}$ and $z^{\bot}=\frac{x^{8} + i x^{9}}{\sqrt{2}}$, the gamma matrices transform as
\begin{align}
& \Gamma^{\pm} =\frac{\Gamma^0\pm \Gamma^1}{2},\\
& \Gamma^a=  \frac{\Gamma^{2a}+ i\Gamma^{2a+1}}{\sqrt{2}},\\
&  \Gamma^{\overline{a}} =  \frac{\Gamma^{2a}- i\Gamma^{2i+1}}{\sqrt{2}},\\
&  \Gamma^{\bot} ,=\frac{\Gamma^{8}+ i\Gamma^{9}}{\sqrt{2}}, \\
&  \Gamma^{\overline{\bot}} = \frac{\Gamma^{8}- i\Gamma^{9}}{\sqrt{2}}.
\end{align}
where $a=1,\cdots, 3$. The ten-dimensional spinors which are singlets under the global symmetry are given in the following notation
\begin{equation}
\epsilon=   |\downarrow\downarrow\downarrow\downarrow\uparrow \,\rangle, \qquad \overline{\epsilon}=   | \downarrow\uparrow\uparrow\uparrow\downarrow\, \rangle , \qquad |\uparrow\, \rangle= \left( \begin{array}{c}1 \\ 0  \end{array}\right), \qquad |\downarrow\, \rangle= \left( \begin{array}{c} 0 \\ 1  \end{array}\right) ,
\end{equation}
where $\overline{\epsilon}=B \epsilon^*$, and $B= \mathbb{I}_2 \otimes i \sigma^2 \otimes \sigma^2 \otimes \sigma^2 \otimes\sigma^1$.
Finally let us describe the fermionic fields in terms of the components of a ten-dimensional spinor:
\begin{subequations} \label{eq:fieldsspincomp}
\begin{eqnarray}
&&\mathbf{1_{-,0}}\leftrightarrow \mu_{-} \sim -|\downarrow \downarrow \downarrow\downarrow \uparrow \, \rangle, \\
&&\mathbf{3_{+,-1}} \leftrightarrow \psi_{+}^{0,1} \sim | \uparrow \uparrow \uparrow \downarrow \downarrow \, \rangle, \label{eq:3(+-1)} \\
&&\mathbf{3_{-,-2}} \leftrightarrow \lambda^{0,2}_{-} \sim |\downarrow \uparrow \downarrow \downarrow \downarrow \, \rangle, \label{eq:3(+-2)} \\
&& \mathbf{1_{+,3}}  \leftrightarrow \chi^{3,0}_+ \sim | \uparrow \uparrow \uparrow \uparrow \uparrow \,\rangle,\\
&&\mathbf{\overline{1}_{-,0}}\leftrightarrow \overline{\mu}_{-} \sim |\downarrow \uparrow \uparrow \uparrow \downarrow \,\rangle, \\
&&\mathbf{\overline{3}_{+,1}} \leftrightarrow \overline{\psi}_{+}^{1,0} \sim -| \uparrow \uparrow \downarrow \downarrow \uparrow \,\rangle, \label{eq:3(+1)} \\
&&\mathbf{\overline{3}_{-,2}} \leftrightarrow \overline{\lambda}^{2,0}_{-}\sim | \downarrow \downarrow \uparrow \uparrow \downarrow \, \rangle, \label{eq:3(+2)} \\
&& \mathbf{\overline{1}_{+,-3}}  \leftrightarrow \overline{\chi}^{0,3}_+\sim -|\uparrow \downarrow \downarrow \downarrow \downarrow \, \rangle,
\end{eqnarray}
\end{subequations}
where the other elements of the triplet in \eqref{eq:3(+-1)}, \eqref{eq:3(+-2)}, \eqref{eq:3(+1)} and \eqref{eq:3(+2)} are all the permutations of the arrow from the second to the fourth places. All these fermionic fields are adjoint valued since the ten-dimensional gaugino transforms as $\chi^{\alpha} T^{\alpha}$ with adjoint generators of the Lie algebra, $T^{\alpha}$. The chiral $\Gamma$ in ten-dimension is $\Gamma=(\sigma^3)^{\otimes 5}$, which exactly matches with \eqref{eq:fieldsspincomp} having the same ten-dimensional chirality. Moreover one can check that $\Gamma^{a \overline{b}}$, with $a\neq b$ satisfy the $SU(3)$ algebra on the $\mathbf{3}$ and $\mathbf{\overline{3}}$, representations. The explicit generators of the $U(1)$'s that we have in section \ref{sec:NPERT}, $J_X= \Gamma^{1\overline{1}} + \Gamma^{2 \overline{2}} + \Gamma^{3\overline{3}}$, and $R=\mathbb{I}_{16} \otimes \sigma^3$.

Let us briefly digress and discuss the action of CPT conjugation on the modes of our model.
First of all, let us see how it acts on a basis of ten-dimensional
gamma matrices, $\Gamma^{I}$. The time-reversal symmetry behaves as
follows $y_0 \mapsto -y_0$. In terms of Gamma matrices, it translates to
\begin{equation}
T \Gamma^0 T^{-1}= -\Gamma^0, \qquad T \Gamma^I T^{-1}= \Gamma^I, \quad M=1,\ldots,9\ ,
\end{equation}
which allows us to choose
\begin{equation}
T=-\Gamma^0\Gamma \ ,
\end{equation}
where $\Gamma$ is the chiral operator. Parity instead, $x_M \mapsto -x_M$ with $M=1,\ldots,9$. In terms of Gamma matrices we have
\begin{equation}
P \Gamma^0 P^{-1}= \Gamma^0, \qquad P \Gamma^I P^{-1}= -\Gamma^I, \quad M=1,\ldots,9\ ,
\end{equation}
and we choose
\begin{equation}
P=\Gamma^0, \Rightarrow PT=\Gamma\ .
\end{equation}
The charge conjugation matrix $C = B^{\ast} \Gamma^{0}$ is defined by introducing a matrix $B$, such that
\begin{equation}
B\Gamma^I B^{-1}= (\Gamma^{I})^*\ .
\end{equation}
Finally, we need to see how CPT acts on our decomposed fields:
\begin{subequations}
\begin{eqnarray}
\mathbf{1_{+,3}} &\overset{{\rm CPT}}{\longleftrightarrow}& \mathbf{1_{+,-3}}\\
\mathbf{3_{+,-1}} &\overset{{\rm CPT}}{\longleftrightarrow}& \mathbf{\overline{3}_{+,1}}\\
\mathbf{1_{-,0}} &\overset{{\rm CPT}}{\longleftrightarrow}& \mathbf{1_{-,0}}\\
\mathbf{3_{-,-2}} &\overset{{\rm CPT}}{\longleftrightarrow}& \mathbf{\overline{3}_{+,2}}
\end{eqnarray}
\end{subequations}
Following the same logic we can see that CPT acts on the gauge field, $A_M$ in the following way
\begin{subequations}
\begin{eqnarray}
\mathbf{1_{0,3}} &\overset{{\rm CPT}}{\longleftrightarrow}& \mathbf{1_{0,-3}}\\
 \mathbf{\overline{3}_{0,-1}}&\overset{{\rm CPT}}{\longleftrightarrow}& \mathbf{3_{0,1}} \ ,
\end{eqnarray}
\end{subequations}
and it trivially maps $A_{\pm}$ into themselves.

\subsection{Supersymmetry Variations}
The ten-dimensional supersymmetry variations for 10D SYM are given by
\begin{subequations}
\begin{eqnarray}
&&\delta_{\epsilon} \Psi = \frac{1}{2}F_{IJ}\Gamma^{IJ} \epsilon, \label{eq:10Dvarferm}\\
&&\delta_{\epsilon} A_I= -i \overline{\epsilon} \Gamma_I \chi\ . \label{eq:10Dvargauge}
\end{eqnarray}
\end{subequations}
The ten-dimensional field strength decomposes as follows
\begin{align}
F_{IJ}\Gamma^{IJ}=& \bigl( F_{+-}\Gamma^{+-} +(F_{+a}\Gamma^{+a}+F_{+\overline{a}}\Gamma^{+\overline{a}}) + (F_{ab}\Gamma^{ab} + F_{\overline{a}\overline{b}}\Gamma^{\overline{a}\overline{b}}) + 2 F_{a\overline{b}}\Gamma^{a\overline{b}}  \\
& (F_{+\perp}\Gamma^{+\perp}+F_{+\overline{\perp}}\Gamma^{+\overline{\perp}}) +  (F_{a\perp}\Gamma^{a\perp} + F_{\overline{a}\overline{\perp}}\Gamma^{\overline{a}\overline{\perp}}+F_{a\overline{\perp}}\Gamma^{a\overline{\perp}} + F_{\overline{a}\perp}\Gamma^{\overline{a}\perp}) + 2 F_{\perp\overline{\perp}}\Gamma^{\perp\overline{\perp}} \bigr). \nonumber
\end{align}
Now we can get the variations of the $(0,2)$ theory, by plugging in the decompositions \eqref{eq:fingaugedec} and \eqref{eq:fieldsspincomp}, where the spinors $\epsilon$ and $\overline{\epsilon}$ correspond to the (static) singlets, $\mathbf{1_{-,0}}$. We also use the decomposition of gamma matrices and fermions given in section \ref{sec:decomp}. The variations on the bosonic fields coming from the decomposition in \eqref{eq:10Dvargauge} read
\begin{subequations} \label{eq:susyvarbos}
\begin{align}
& \delta_{\varepsilon} A_- = -2i  \varepsilon \overline{\mu}_- \ , &\qquad &\overline{\delta}_{\overline{\varepsilon}} A_- = 2i \overline{\varepsilon}\mu_- \ ,\\
&\delta_{\varepsilon} A_+ = 0\ ,  &\qquad &\overline{\delta}_{\overline{\varepsilon}} A_+=0\ ,\\
& \delta_{\varepsilon} A_{ a} =0\ ,  &\qquad &\overline{\delta}_{\overline{\varepsilon}}  A_{a}  =\sqrt{2}    \overline{\varepsilon}\overline{\psi}_{+\, a}\ ,\\
& \delta_{\varepsilon} A_{\overline{a}}=\sqrt{2}  \varepsilon \psi_{+\, \overline{a}}\ ,  &\qquad &\overline{\delta}_{\overline{\varepsilon}} A_{\overline{a}}=0\ ,\\
&\delta_{\varepsilon} \phi_{ abc} =\sqrt{2}  \varepsilon \chi_{+\, abc}\ ,  &\qquad &\overline{\delta}_{\overline{\varepsilon}}  \phi_{abc} =0\ ,\\
&\delta_{\varepsilon}\overline{\phi}_{ \overline{a}\overline{b}\overline{c}}=0\ ,  &\qquad &\overline{\delta}_{\overline{\varepsilon}}\overline{\phi}_{\overline{a}\overline{b}\overline{c}} =\sqrt{2}  \overline{\varepsilon}  \overline{\chi}_{+\, \overline{a}\overline{b}\overline{c}}\ ,
\end{align}
\end{subequations}
where $D_{+}=\partial_0 + \partial_1 + A_+$, and we have redefined $\Gamma^i \rightarrow i \Gamma^i$, $\Gamma^{\perp} \rightarrow i \Gamma^{\perp}$.

In order to fix a bit of notation, let us introduce the covariant derivative $\overline{\partial}_A=\overline{\partial}+A_{(0,1)}$ on $X$, the field strengths on $X$ and $\mathbb R^{1,1}$,
\begin{align}
&F_{(0,2)}=[\overline{\partial}_A, \overline{\partial}_A]= (\overline{\partial}_{\overline{a}} A_{\overline{b}}+ [A_{\overline{a}}, A_{\overline{b}}]) d\overline{z}^{\overline{a}}\wedge d\overline{z}^{\overline{b}}=F^{\alpha}_{\overline{a}\overline{b}} T^{\alpha} d\overline{z}^{\overline{a}}\wedge d\overline{z}^{\overline{b}} ,\\
&\overline{\partial}_A \phi_{(3,0)}=[\overline{\partial}_A, \phi_{(3,0)}]= (\overline{\partial}_{[\overline{a}} \phi_{\overline{b}\overline{c}\overline{d}]}  + A_{[\overline{a}},  \phi_{bcd]}) d\overline{z}^{\overline{a}}\wedge dz^b \wedge dz^c\wedge dz^d,\\
&F_{+-}= [D_+,D_-]=(\partial_{[+} A_{-]}+ [A_{+}, A_{-}]),\\
&F_{+i}= D_+ A_i = \partial_+ A_a + [A_+,A_a] ,\quad  F_{+a}= D_+ A_{\overline{a}}=\partial_+ A_{\overline{a}} + [A_+,A_{\overline{a}}],
\end{align}
where all the fields carry also an adjoint index $\alpha$ contracted with the generators of the gauge Lie algebra, $T^{\alpha}$. All of this extend also for the fermions, in fact, the commutators will be extended later to the superfields, moreover, they will mostly used when we commute two derivative operators, and dropped when we have a covariant derivative acting on an adjoint valued gauge field.

Decomposing \eqref{eq:10Dvarferm} we get the following supersymmetry variations:
\begin{subequations} \label{eq:susyvarferm}
\begin{align}
& \delta_{\varepsilon} \mu_- = \varepsilon \left( \frac{1}{2}F_{-+} + i \mathcal{D} \right) \ ,  &\qquad &\overline{\delta}_{\overline{\varepsilon}} \mu_- = 0\ ,\\
&\delta_{\varepsilon} \overline{\mu}_- = 0\ ,  &\qquad &\overline{\delta}_{\overline{\varepsilon}} \overline{\mu}_- = - \overline{\varepsilon} \left( \frac{1}{2} F_{-+} -  i \mathcal{D} \right)\ ,\\
& \delta_{\varepsilon} \overline{\psi}_{+\, a} = -i \sqrt{2} \varepsilon 	 F_{+a}\ ,  &\qquad &\overline{\delta}_{\overline{\varepsilon}} \overline{\psi}_{+\, a} =0\ ,\\
& \delta_{\varepsilon}\psi_{+\, \overline{a}} =0\ ,  &\qquad &\overline{\delta}_{\overline{\varepsilon}} \psi_{+\, \overline{a}}  =-i \sqrt{2} \overline{\varepsilon} 	 F_{+\overline{a}}\ ,\\
& \delta_{\varepsilon} \lambda_{-\, \overline{a}\overline{b}} = - \sqrt{2} \varepsilon \left(\overline{\partial}^{\dagger}_{ A} \overline{\phi} \right)_{\overline{a}\overline{b}}\ ,  &\qquad &\overline{\delta}_{\overline{\varepsilon}}\lambda_{-\, \overline{a}\overline{b}} = +\sqrt{2}  \overline{\varepsilon} F_{\overline{a}\overline{b}}\ ,\\
& \delta_{\varepsilon}\overline{\lambda}_{-\,ab} = - \sqrt{2} \varepsilon  F_{ab}\ ,  &\qquad &\overline{\delta}_{\overline{\varepsilon}} \overline{\lambda}_{-\, ab}  =  \sqrt{2}  \overline{\varepsilon}\left(\partial^{\dagger}_{A}  \phi \right)_{ab}\ ,\\
& \delta_{\varepsilon}\chi_{+\, abc} =0\ ,  &\qquad &\overline{\delta}_{\overline{\varepsilon}} \chi_{+\, abc}  =-i  \sqrt{2}\overline{\varepsilon}  D_{+}\overline{\phi}_{ \overline{a}\overline{b}\overline{c}}\ ,\\
& \delta_{\varepsilon} \overline{\chi}_{+\, \overline{a}\overline{b}\overline{c}}=-i\sqrt{2} \varepsilon  D_{+}\overline{\phi}_{ \overline{a}\overline{b}\overline{c}}\ ,  &\qquad &\overline{\delta}_{\overline{\varepsilon}}\overline{\chi}_{+\, \overline{a}\overline{b}\overline{c}} =0\ ,
\end{align}
\end{subequations}
where, for example, $\partial_A^{\dagger} \phi_{(3,0)}= \omega \llcorner (\overline{\partial}_A^{\dagger} \phi_{(3,0)})=g^{a\overline{b}} (\overline{\partial}_A)_{\overline{b}} \phi_{ acd} dz^{c}\wedge dz^{d}$.  Moreover on-shell, we have:
\begin{equation}
\mathcal D= -\ast_X \left( \omega \wedge \omega \wedge F_{(1,1)} + [\phi, \overline{\phi}]\right),
\end{equation}
where in components $[\phi, \overline{\phi}]=\phi_{[abc} \overline{\phi}_{\overline{abc}]} dz^a\wedge dz^b \wedge dz^c \wedge d\overline{z}^{\overline{a}}\wedge d\overline{z}^{\overline{b}}\wedge d\overline{z}^{\overline{c}}$ and $F_{(1,1)}=[\partial_A,\overline{\partial}_A]$. $\ast_X$ is the Hodge-dual operator on $X$. These are just the on-shell variations, we need to extend them by adding the auxiliary fields $\mathcal{G}$ and $\mathcal{D}$ in \eqref{eq:susyvarferm},
\begin{subequations} \label{eq:susyvarfermaux}
\begin{align}
& \delta_{\varepsilon} \mu_- = \varepsilon \left(\frac{1}{2} F_{-+} +  i\, \mathcal{D}\right),  &\qquad &\overline{\delta}_{\overline{\varepsilon}} \mu_- = 0\ ,\\
&\delta_{\varepsilon} \overline{\mu}_- = 0\ ,  &\qquad &\overline{\delta}_{\overline{\varepsilon}} \overline{\mu}_- = - \overline{\varepsilon}\left( \frac{1}{2} F_{-+} - i\,  \mathcal{D}\right)\ ,\\
& \delta_{\varepsilon} \overline{\psi}_{+\, a} = - i\sqrt{2}	 \varepsilon F_{+a}\ ,  &\qquad &\overline{\delta}_{\overline{\varepsilon}} \overline{\psi}_{+\, a} =0\ ,\\
& \delta_{\varepsilon}\psi_{+\, \overline{a}} =0\ ,  &\qquad &\overline{\delta}_{\overline{\varepsilon}} \psi_{+\, \overline{a}}  = - i\sqrt{2}	\overline{\varepsilon} F_{+\overline{a}}\ ,\\
& \delta_{\varepsilon} \lambda_{-\, \overline{a}\overline{b}} = -\sqrt{2} \varepsilon \mathcal{G}_{\overline{b}\overline{c}}\ ,  &\qquad &\overline{\delta}_{\overline{\varepsilon}}\lambda_{-\, \overline{a}\overline{b}} =\sqrt{2} \overline{\varepsilon} E_{\overline{a}\overline{b}}\ ,\\
& \delta_{\varepsilon}\overline{\lambda}_{-\, ab} = -  \sqrt{2} \varepsilon \overline{E}_{ a b}\ ,  &\qquad &\overline{\delta}_{\overline{\varepsilon}} \overline{\lambda}_{-\, ab}  = \sqrt{2} \overline{\varepsilon} \overline{\mathcal{G}}_{ab}\ ,\\
& \delta_{\varepsilon}\chi_{+\, abc} =0\ ,  &\qquad &\overline{\delta}_{\overline{\varepsilon}} \chi_{+\, abc}  = - i \sqrt{2}\varepsilon  D_{+}\phi_{ abc}\ ,\\
& \delta_{\varepsilon} \overline{\chi}_{+\, \overline{a}\overline{b}\overline{c}}= - i \sqrt{2}  \overline{\varepsilon}D_+\overline{\phi}_{ \overline{a}\overline{b}\overline{c}}\ ,  &\qquad &\overline{\delta}_{\overline{\varepsilon}}\overline{\chi}_{+\, \overline{a}\overline{b}\overline{c}} =0\ ,
\end{align}
\end{subequations}
where\footnote{Here the $E$ is just the top bosonic component of a field that will be promoted to a superfield in the superspace formalism later on.} $E=  F_{(0,2)}$.
In order to close the ($0,2$) supersymmetry algebra,
\begin{eqnarray}
[ \delta_{\varepsilon}, \delta_{\varepsilon} ]=[\overline{\delta}_{\overline{\varepsilon}},\overline{\delta}_{\overline{\varepsilon}} ]=0, \qquad [\delta_{\varepsilon}, \overline{\delta}_{\overline{\varepsilon}} ]=2 i  \varepsilon \overline{\varepsilon} D_+,
\end{eqnarray}
The auxiliary field variations are
\begin{subequations} \label{eq:susyvaraux}
\begin{align}
& \delta_{\varepsilon} \mathcal{D} =\varepsilon D_+ \overline{\mu}_-\ ,  &\qquad &\overline{\delta}_{\overline{\varepsilon}}  \mathcal{D} = \overline{\varepsilon} D_+ \mu_- \ ,\\
&\delta_{\varepsilon} \overline{\mathcal{G}}_{ab} =  \sqrt{2} \varepsilon (\partial_{A} \overline{\psi}_+^{(1,0)})_{ab} - \sqrt{2} i \varepsilon  D_+ \overline{\lambda}_{-\, ab}\ ,  &\qquad &\overline{\delta}_{\overline{\varepsilon}}\overline{\mathcal{G}}_{ ab }= 0 \ ,\\
& \delta_{\varepsilon} \mathcal{G}_{\overline{a}	\overline{b}} = 0 \ , &\qquad &\overline{\delta}_{\overline{\varepsilon}}  \mathcal{G}_{\overline{a}	\overline{b}} = + \sqrt{2} \overline{\varepsilon} (\overline{\partial}_{A} \psi_{+}^{(0,1)})_{\overline{ab}} + \sqrt{2} i \overline{\varepsilon} D_+ \lambda_{-\, \overline{a}\overline{b}} \ ,
\end{align}
\end{subequations}
where we know the explicit expression for the variation $E$,
\begin{equation}
\delta_{\varepsilon} E= + \sqrt{2} \varepsilon \overline{\partial}_{A} \psi_{+}^{(0,1)}\ , \qquad \delta_{\overline{\varepsilon}} \overline{E}= - \sqrt{2} \overline{\varepsilon} \partial_{A} \overline\psi_{+}^{(1,0)}\  .
\end{equation}

\subsection{Superfields}
In the previous section we have derived the supersymmetry variations of a $(0,2)$ QFT starting from the variations for 10D SYM considering 7-branes wrapping a K\"{a}hler threefold $X$. We are now ready to organize the fields in supermultiplets, and to do so we use the superspace formalism.
In appendix \ref{app:GLSM} we have defined the supersymmetry generators as well as the covariant derivative in superspace \eqref{eq:supercovder}. Let us now write down the corresponding multiplets for our bulk 7-brane theory.

First of all, just as in the analysis of Appendix \ref{app:HET}, we have a collection of superfields which are labelled by points of the internal
manifold $X$. This includes the vector multiplets, a Fermi multiplet $\Lambda_{(0,2)}$ with non-trivial $E$-field, and chiral multiplets $\Phi_{(3,0)}$ and $\mathbb{D}_{(0,1)}$, the superfield associated with the anti-holomorphic component of the internal covariant derivative. Indeed, as we explained in section \ref{sec:NPERT}, the field content descends directly from that of the associated 9-brane model. Borrowing our discussion from Appendix \ref{app:HET}, let us therefore focus on the few features of the field content which are distinct from the 9-brane theory.

We have a chiral multiplet $\Phi_{(3,0)}$ with expansion in components:
\begin{equation} \label{eq:defchiral1}
\Phi_{(3,0)}=\phi_{(3,0)}+ \sqrt{2}\theta^+ \chi_{+,(3,0)} - i\theta^+ \overline{\theta}^+ D_{+} \phi_{(3,0)}.
\end{equation}
In addition to the rather similar expansion for $\mathbb{D}_{(0,1)}$, the Fermi multiplet is a $(0,2)$ form with a non-trivial
$E$-field:
\begin{equation} \label{eq:deffermi}
\Lambda_{(0,2)}=\lambda^{(0,2)}_- - \sqrt{2}\theta^+ \mathcal G_{(0,2)}- \sqrt{2}\overline{\theta}^+ E_{(0,2)} - i\theta^+ \overline{\theta}^+ D_{+}\lambda_{-,(0,2)},
\end{equation}
where the superfield $E_{(0,2)}$ is given by
\begin{align} \label{eq:E1}
E_{(0,2)} = \mathbb F_{(0,2)}= [\mathbb{D}_{(0,1)},\mathbb{D}_{(0,1)}] = ( F_{(0,2)} + \sqrt{2} \theta^+ \overline{\partial}_{\overline A} \psi_{+,(0,1)} - i \theta^+ \overline{\theta}^+ [[\overline{\partial}_{\overline A}, D_{+}],\overline{\partial}_{\overline A}]) \ .
\end{align}

\subsection{Non-Abelian Bulk Twisted Action}
\label{sec:act}
We are now ready to write down the effective action of 7-branes wrapping a K\"ahler threefold $X$. As a notational device for writing
the kinetic terms for superfields, we introduce a pairing $(\cdot , \cdot)$ for bundle valued differential forms which are Serre dual to one another. In addition to the Hermitian metric of the K\"ahler threefold, this also requires us to introduce a Hermitian pairing on the associated
bundle. Whenever we write such a pairing, it will implicitly be a top differential form which can be integrated over the manifold. When we turn to
modes localized on a K\"ahler surface $S$, we shall employ a similar notation. Finally, we shall also introduce a holomorphic pairing
$\langle \cdot , \cdot \rangle$  which only makes use of the complex structure of the associated bundles. For the bulk modes, this is implicitly
captured by a simple trace over the adjoint representation.

We begin with the kinetic terms for the various fields.
The kinetic term for the chiral field $\Phi_{(3,0)}$ is given by
\begin{equation}
S_{\Phi}=- \frac{i}{2}  \int d^2 y d^{2} \theta \int_{X} \;  \left( \,\overline{\Phi_{(3,0)}}\,,\, [  \nabla_-, \Phi_{(3,0)}]\,\right).
\end{equation}
Expanding in components yields:
\begin{align}
S_{\Phi}= - \int_{\mathbb{R}^{1,1}} d^2 y \int_{X}  & {\rm Tr} \biggr(  \left(D_{+} \overline{\phi}_{(0,3)} \wedge D_{-} \phi_{(3,0)} \right) + i  \overline{\chi}^{0,3}_+ \wedge D_{-}
 \chi^{3,0}_+ \nonumber  \\
&+ \sqrt{2}  \left( \overline{\mu}_-[  \phi_{(3,0)} ,\overline{\chi}^{0,3}_+] + \mu_- [\chi^{3,0}_+,  \overline{\phi}_{(0,3)} ]  \right)
+ \mathcal{D}   [ \phi_{(3,0)} ,\overline{\phi}_{(0,3)} ] \biggl),
\end{align}
where we used the properties of cyclicity of the trace in the pairing as well as integration by parts.

The kinetic term for $\mathbb{D}_{(0,1)}$ is:
\begin{equation}
S_{\mathbb{D}}= - \frac{1}{2} \int d^2 y d^{2} \theta \int_{X} \;   \left(\overline{\mathbb D_{(0,1)}} , [\nabla_-,\mathbb{D}_{(0,1)}]\right).
\end{equation}
The expansion in component fields is:
\begin{align}
S_{\mathbb{D}}= - \int d^2 y \int_{X} \omega \wedge \omega \wedge \;  &   {\rm Tr}\biggl( F_+^{(0,1)} \wedge \overline{F}_-^{(1,0)}    + i\,  \overline{\psi}^{(1,0)}_+ \wedge D_{-} \psi^{(0,1)}_+    \nonumber\\
&- \sqrt{2} \, \left( (\overline{\partial}_{A} \overline{\mu}_-) \wedge \overline{\psi}^{(1,0)}_+ +    \psi^{(0,1)}_+ \wedge (\partial_{A}\mu_-) \right) + F_{(1,1)} \mathcal{D} \biggr).
\end{align}
The kinetic term for the Fermi field $\Lambda_{(0,2)}$ is given by
\begin{equation}
S_{\Lambda}= -\frac{1}{2}\int d^2 y d^{2} \theta \int_{X} \;  \left(\overline{\Lambda_{(0,2)}}, \Lambda_{(0,2)}\right),
\end{equation}
and when we plug in \eqref{eq:deffermi}, its expansion in components is
\begin{align}
S_{\Lambda}=  - \int d^2 y d^{2} \theta \int_{X} \omega \wedge \;  & {\rm Tr} \biggl( - i\, (D_{+}\lambda_-^{(0,2)}) \wedge  \overline{\lambda}^{(2,0)}_-  + \, \mathcal{G}_{(0,2)} \wedge \overline{\mathcal{G}}_{(2,0)}  -  F_{(0,2)}\wedge F_{(2,0)} \nonumber \\
&  \sqrt{2} (\overline{\partial}_{A} \psi_+^{(0,1)} \wedge \overline{\lambda}_-^{(2,0)} + \lambda_-^{(0,2)} \wedge \partial_A \overline{\psi}_+^{(1,0)}) \biggr).
\end{align}

Finally, the action includes a contribution from $\Upsilon$ given by integrating $\frac{1}{8} \overline{\Upsilon} \Upsilon$ over $X$.

\subsubsection{Superpotential Terms}
The bulk superpotential is given by
\begin{align} \label{eq:bulkW}
W_{X} = - \frac{1}{\sqrt{2}} \int_{X}  \; {\rm Tr} (\Lambda_{(0,2)}\wedge \mathbb{D}_{(0,1)} \Phi_{(3,0)}) \ .
\end{align}
In the nomenclature of $\mathcal{N} = (0,2)$ supersymmetric models, this
amounts to setting $J(\Phi)=  [\mathbb{D}_{(0,1)}, \Phi^{3,0}] $.
In components, the resulting contribution to the action from the F-terms is:
\begin{equation}
S_{F} = \int d^2 y \int_{X} \, \, {\rm Tr} (\mathcal{G}_{(0,2)} \wedge \overline{\partial}_A \phi_{(3,0)}  + \lambda^{(0,2)}_- \wedge \overline{\partial}_{A} \chi_+^{(3,0)} + \lambda^{(0,2)}_- \wedge \,[\phi_{(3,0)},\psi^{(0,1)}_+]) + h.c,
\end{equation}
where we get an additional constraint by requiring that $W$ is a chiral quantity,
\begin{equation}
{\rm Tr} (E_i \cdot J^i)= \langle E_i, J_i \rangle = {\rm Tr} ( \mathbb{F}_{(0,2)} \wedge  [\mathbb{D}_{(0,1)}, \Phi_{(3,0)}]  ).
\end{equation}
This vanishes on-shell, but would give a topological condition off-shell in order for supersymmetry to be manifestly preserved.
So inevitably we must couple to some of the background geometric moduli.

\subsection{Localized Surface Defects}
As we have already discussed in section \ref{sec:NPERT}, one of the important features of intersecting 7-branes is that some of the matter fields
localize on intersections, i.e. from the intersection of $X_1$ and $X_2$. On general grounds, we expect there to be a hypermultiplets worth of degrees of freedom localized on the surface. That is to say, we expect there to be two chiral multiplets and two Fermi multiplets localized on the surface. These organize according to ``generalized bifundamental'' representations of $G_{1} \times G_{2}$, where $G_{i}$ denotes the corresponding bulk gauge group. Denote the representation of $G_{1} \times G_{2}$ by $r_1 \times r_2$, the corresponding bundle will be $\mathcal{R}_1 \otimes \mathcal{R}_2$.

Our goal in this subsection will be to understand the action for these surface defects using the general Katz-Vafa collision rules discussed for example in \cite{Katz:1996xe} and \cite{Beasley:2008dc}. With this in mind, we start with the action for an isolated bulk 7-brane with gauge group $G$, and we consider the effects of activating a background value for the $(3,0)$-form $\Phi_{(3,0)}$. Then, there will be
localized modes trapped on the intersection of K\"ahler threefolds which intersect along a K\"ahler surface. We obtain the action for the localized modes by starting from the bulk action, and expanding to second order in the fluctuations. The third order fluctuations are associated with interactions between three localized terms.

The superpotential describing the defect theory on the surface intersection, $S$, is given in terms of the localized matter fields:
\begin{subequations} \label{eq:bundasssurf}
\begin{align}
&(\delta \mathbb{D}_{(0,1)\;{\rm surface}}=  Q)\; \in K_S^{1/2} \otimes \mathcal{R}_1 \otimes \mathcal{R}_2,\\
&(\delta \Phi_{{\rm surface}}= Q^c) \; \in K_S^{1/2} \otimes \mathcal{R}_1^{\vee} \otimes \mathcal{R}_2^{\vee},\\
&(\delta \Lambda_{(0,2) \, {\rm surface}}=  \Psi) \; \in \Omega^{0,1}_S ( K_S^{1/2} \otimes \mathcal{R}_1 \otimes \mathcal{R}_2).
\end{align}
\end{subequations}
In the above, we have included the contributions from the propagating bulk modes. Another way to arrive at the same mode count is to work in terms of the bulk topological term $W_{top,X}$, and include variations with respect to a all bulk modes.
The expansion in component fields is:
\begin{align} \label{eq:defsuperexp}
&Q=\sigma + \sqrt{2}\theta^+ \eta + \ldots \, ,\\
&Q^c=\sigma^c + \sqrt{2}\theta^+ \eta^c + \ldots\, ,\\
&\Psi=\xi - \sqrt{2}\theta^+ \mathcal{K} -  \sqrt{2}\overline{\theta}^+ E \ldots\, ,\\
&\Psi^c=\xi^c - \sqrt{2}\theta^+ \mathcal{K}^c -  \sqrt{2}\overline{\theta}^+ E^{c} \ldots\, ,
\end{align}
where $\sigma$ is a boson, $\xi, \eta$ are fermions, $\mathcal K$ and $\mathcal{K}^c$ are auxiliary fields. Here, we hasten to
add that $\Psi^c$ is not an independent degree of freedom. In the case of the localized modes, we must also include the
$E$-fields for the Fermi multiplet, which is in turn captured by the contribution to $W_{top}$.
See section \ref{sec:NPERT} for further discussion on this point.

The kinetic term for the defect theory is then given by
\begin{align}
S_{{\rm def.}\; {\rm kinetic}}= \int d^2 y  d^{2} \theta \int_S  \; \biggl(- \frac{i}{2}  \left( \overline{Q} , \nabla_- Q \right)  - \frac{i}{2}\left(\overline{Q^c} , \nabla_- Q^c \right) - \frac{1}{2} \left(  \overline{\Psi}, \Psi \right) \biggr)
\end{align}
where $(\cdot , \cdot)$ is the canonical pairing introduced earlier. The
expansion into component fields is entirely straightforward, and follows
the rules laid out in Appendix \ref{app:GLSM}.

To explicitly count the modes localized on the K\"ahler surface, it is helpful to return to the
bulk action and study the fermionic modes which are localized as a result of having a non-trivial
profile for $\phi_{(3,0)}$. We begin by looking at the part of the action that includes all the fermions:
\begin{align} \label{eq:fermact}
S_{{\rm ferm.}}= - \sqrt{2} \int_{\mathbb R^{1,1} \times X} & \,\mathrm{Tr}\biggl(  \overline{\mu}_-[  \phi_{(3,0)} ,\overline{\chi}^{(0,3)}_+] + \mu_- [\chi^{(3,0)}_+,  \overline{\phi}_{(0,3)} ]  - \,  (\overline{\partial}_{A} \overline{\mu}_-) \wedge \overline{\psi}^{(1,0)}_+  -    \psi^{(0,1)}_+ \wedge (\partial_{A}\mu_-)  \nonumber \\
&  + \omega \wedge \overline{\partial}_{A} \psi_+^{(0,1)} \wedge \overline{\lambda}_-^{(2,0)} + \omega \wedge \lambda_-^{(0,2)} \wedge \partial_A \overline{\psi}_+^{(1,0)} \biggr)  \nonumber \\
& \left(- \lambda^{(0,2)}_- \wedge \overline{\partial}_{A} \chi_+^{(3,0)}-\lambda^{(0,2)}_- \wedge \,[\phi_{(3,0)},\psi^{(0,1)}_+] + h.c. \right).
\end{align}
Now, to see how fermions localize on a surface $S$, switch on a background value for $\phi$. In a small neighborhood of $S$, we can use the local holomorphic coordinates $(z_1,z_2,z_3)$ on $X$. Let us assume that a section of the canonical bundle of $X$, $K_X$, exists, then
\begin{equation}\label{eq:backvalphi}
\phi=\phi_0 t, \qquad t \in {\rm ad}(G_X), \qquad \phi_0 \in H^0(K_{X},X),
\end{equation}
and since $\phi_0$ is a section of the canonical bundle, which locally is parameterized by $z_3$, we have that
\begin{equation} \label{eq:phiback}
\phi_{(3,0)}= t z_3 \; dz_1\wedge dz_2 \wedge dz_3,
\end{equation}
where $S$ corresponds to the locus where $z_3=0$.
For ease of exposition, we assume that the expectation value \eqref{eq:phiback} breaks $G_X$ to $\Gamma_X \times U(1) \subset G_X$. We would like now to solve the equations of motion for the fermions in a neighborhood of the surface $S$. To do so we look at the fermionic action written in \eqref{eq:fermact}. This basically follows the same analysis spelled out in great detail in reference \cite{Beasley:2008dc}, so we shall simply summarize the main points.

By varying this action we indeed find trapped zero modes localized along the vanishing locus of the holomorphic three-form $\phi_{3,0}$. The modes take the following schematic form for the fermions descending both from the chiral multiplet and the Fermi multiplet:
\begin{equation}
\overline{\partial}_{\overline{z}_3}( \overline{{\rm ferm}})\; + z_3 \; \widetilde{{\rm ferm}}=0, \qquad
\overline{\partial}_{\overline{z}_3}( \overline{\widetilde{{\rm ferm}}})\; + z_3 \; {\rm ferm}=0.
\end{equation}
This leads to a Gaussian profile for the zero modes with falloff of the form $\sim \exp(- c |z_3|^3)$ for $z_{3} \neq 0$. The quantity $c$ depends
on details of the geometry such as the K\"ahler metric for $X$ as well as the Hermitian pairing for the various bundles of the 7-brane theory.
We find trapped fermionic modes which are part of the chiral multiplets $Q$ and $Q^c$,
and another trapped mode $\Psi$ which fills out a Fermi multiplet and transforms
as a $(0,1)$ differential form on $S$. Observe that
in the flat space limit, we get a 4D $\mathcal{N} = 2$ hypermultiplet's
worth of degrees of freedom.

We can also determine the bundle assignments for our localized modes using this analysis. Since
$S$ is defined by a section of the canonical bundle of $X$, $K_X$, $\phi_0 \in H^0(X,K_X)$.
Given this, we can now write the following twisted Koszul sequence:
\begin{equation}
0 \longrightarrow N^{\vee}_{S/X} \otimes K_X \longrightarrow K_X \longrightarrow K_X|_S \longrightarrow 0,
\end{equation}
resulting in $N_{S/X}=K_X|_S$. By the adjunction formula we also know that
\begin{equation}
K_S=K_X|_S \otimes N_{S/X},
\end{equation}
and by construction, we have $N_{S/X}=K_X|_S$, and hence
\begin{equation} \label{eq:cannorm2}
K_S^{1/2}=N_{S/X}.
\end{equation}
Recalling the bundle cohomologies in \eqref{eq:bundasssurf}, we conclude
that the massless fermions localized along $S$  can be identified
with the zero-modes of the fermions in the defect theory $\eta, \xi$.

\section{Brief Review of Chern Classes \label{app:CHERN}}

At various stages in our analysis, we have used some basic elements about the
structure of Chern classes, especially as it pertains to bundles on Calabi-Yau
fourfolds and general K\"ahler threefolds and K\"ahler surfaces. In this Appendix
we collect some of these formulae.

We recall (for instance from\cite{Bott:1982df}) that the Pontryagin classes for
a real vector bundle $E$ are related to the Chern classes of the complexified bundle $E\otimes\mathbb{C}$ as:
\begin{equation}
p_{i}(E)=(-1)^{i}c_{2i}(E\otimes \mathbb{C} )~.
\end{equation}
If $E=\mathcal{E}\oplus{{\mathcal{E}}}^{\vee}$ with ${{\mathcal{E}}}^{\vee}$
the dual bundle where $\mathcal{E}$ is a complex vector bundle, then%
\begin{equation}
c(E\otimes \mathbb{C})=c(\mathcal{E}\oplus{{\mathcal{E}}}^{\vee})=c(\mathcal{E}%
)c({{\mathcal{E}}}^{\vee})~
\end{equation}
Since $c_{i}({{\mathcal{E}}}^{\vee})=(-1)^{i}c_{i}(\mathcal{E})$, it is easy
enough to make the expansion%
\begin{align}
p_{1}(E)  &  =-2c_{2}(\mathcal{E})+c_{1}(\mathcal{E})^{2}~,\\
~p_{2}(E)  &  =2c_{4}(\mathcal{E})+c_{2}(\mathcal{E})^{2}-2c_{1}%
(\mathcal{E})c_{3}(\mathcal{E})~
\end{align}
We are most interested in the case $c_{1}(\mathcal{E})=0$, which leads to%
\begin{align}
p_{1}(E)  &  =-2c_{2}(\mathcal{E})~,\\
p_{2}(E)  &  =2c_{4}(\mathcal{E})+c_{2}(\mathcal{E})^{2}~,
\end{align}
and the remarkable identity%
\begin{equation}
4p_{2}(E)-p_{1}(E)^{2}=8c_{4}(\mathcal{E})~.
\end{equation}
In particular, when we apply this to the tangent bundle of a Calabi Yau
fourfold $T_{CY_{4}}$ we obtain%
\begin{equation}
4p_{2}(T_{CY_{4}})-p_{1}(T_{CY_{4}})^{2}=8\chi(CY_{4}),
\end{equation}
where we have used the fact that the Euler characteristic of a Calabi-Yau
fourfold is given by:%
\begin{equation}
\chi(CY_{4})=\underset{CY_{4}}{\int}c_{4}(T_{CY_{4}}).
\end{equation}
We note that this identity holds more generally for any eight manifold that
admits a nowhere vanishing spinor~\cite{Becker:1996gj}.

We shall often have occasion to calculate the bundle valued cohomology groups
$H_{\overline{\partial}}^{i}(M,\mathcal{E})$ for $M$ a K\"ahler manifold with
$\mathcal{E}$ a holomorphic vector bundle. Here, we typically need to know the
dimensions $h^{i}(M,\mathcal{E})$ for $i=0,...,$dim$_{\mathbb{C}}M$. Though
the dimensions can depend on the geometric and vector bundle moduli, some
specific combinations are protected by a topological index formula. For
example, there is a holomorphic Euler characteristic:%
\begin{equation}
\chi(M,\mathcal{E})=\underset{i=0}{\overset{\text{dim}_{\mathbb{C}}M}{\sum}%
}(-1)^{i}h^{i}(M,\mathcal{E})=\underset{M}{\int}\operatorname{ch}%
(\mathcal{E})\operatorname{Td}(M)
\end{equation}
where the final equality follows from the Hirzebruch-Riemann-Roch index
formula, and we have introduced the Chern character and Todd class of a
general bundle:%
\begin{align}
\operatorname{ch}(\mathcal{E})  &  =\text{rk}(\mathcal{E)+}c_{1}%
(\mathcal{E)+}\frac{1}{2}\left(  c_{1}^{2}(\mathcal{E})-2c_{2}(\mathcal{E}%
)\right)  +\frac{1}{3!}\left(  c_{1}^{3}(\mathcal{E})-3c_{2}(\mathcal{E}%
)c_{1}(\mathcal{E})+3c_{3}(\mathcal{E})\right) \\
&  +\frac{1}{4!}\left(  c_{1}^{4}(\mathcal{E})-4c_{2}(\mathcal{E})c_{1}%
^{2}(\mathcal{E})+4c_{3}(\mathcal{E})c_{1}(\mathcal{E})+2c_{2}^{2}%
(\mathcal{E})-4c_{4}(\mathcal{E})\right)  +...\\
\text{Td}(\mathcal{E})  &  =1+\frac{1}{2}c_{1}(\mathcal{E)+}\frac{1}{12}%
(c_{1}^{2}(\mathcal{E})+c_{2}(\mathcal{E}))+\frac{1}{24}(c_{1}(\mathcal{E}%
)c_{2}(\mathcal{E}))\\
&  +\frac{1}{720}\left(  -c_{1}^{4}(\mathcal{E})+4c_{1}^{2}(\mathcal{E}%
)c_{2}(\mathcal{E})+c_{1}(\mathcal{E})c_{3}(\mathcal{E})+3c_{2}^{2}%
(\mathcal{E})-c_{4}(\mathcal{E})\right)  +...
\end{align}

\subsection{Special Case: Calabi-Yau Fourfolds}

In our specific applications to compactifications of type I and heterotic
strings, we will specialize further to the case of stable irreducible
holomorphic vector bundles on an irreducible Calabi-Yau fourfold. In these cases, we can
set the first Chern class to zero, and we get the simplified formulae:%
\begin{align}
\operatorname{ch}(\mathcal{E})  &  =\text{rk}(\mathcal{E)}-c_{2}%
(\mathcal{E})+\frac{1}{2}c_{3}(\mathcal{E})+\frac{1}{12}(c_{2}^{2}%
(\mathcal{E})-2c_{4}(\mathcal{E}))\\
\text{Td}(\mathcal{E})  &  =1+\frac{1}{12}c_{2}(\mathcal{E})+\frac{1}%
{720}\left(  3c_{2}^{2}(\mathcal{E})-c_{4}(\mathcal{E})\right)  .
\end{align}
We shall also encounter the holomorphic Euler characteristics:%
\begin{equation}
\chi_{i}(CY_{4})\equiv\chi(CY_{4},\Omega_{CY_{4}}^{(0,i)}%
)=\underset{j=0}{\overset{4}{\sum}}(-1)^{j}h^{j,i}(CY_{4})=\underset{CY_{4}%
}{\int}\operatorname{ch}(\Omega_{CY_{4}}^{(0,i)})\operatorname{Td}(CY_{4}).
\end{equation}
The resulting expression in terms of Chern classes is (see e.g.
\cite{Klemm:1996ts}):%
\begin{align}
\chi_{0}(CY_{4})  &  =\frac{1}{720}\underset{CY_{4}}{\int}\left(  3c_{2}%
^{2}-c_{4}\right) \\
\chi_{1}(CY_{4})  &  =\frac{1}{180}\underset{CY_{4}}{\int}\left(  3c_{2}%
^{2}-31c_{4}\right) \\
\chi_{2}(CY_{4})  &  =\frac{1}{120}\underset{CY_{4}}{\int}\left(  3c_{2}%
^{2}+79c_{4}\right)  .
\end{align}

We can also simplify the various relations between $c_{2}$ and $c_{4}$. For
example, evaluating the holomorphic Euler characteristic for the bundle
$\mathcal{E}=\mathcal{O}_{CY_{4}}$ the structure sheaf and using $\chi
(CY_{4},\mathcal{O}_{CY_{4}})=2$, we immediately find the relation:%
\[
3c_{2}(M_{8})^{2}-c_{4}(M_{8})=1440~.
\]
Setting $\mathcal{E}=\mathcal{T}_{CY_{4}}$ leads to the further relation:%
\[
\chi(CY_{4},\mathcal{T}_{CY_{4}})=8-\frac{1}{6}\chi(CY_{4}),
\]
i.e. we find a simple relation between the holomorphic Euler characteristic
and the Euler characteristic of the manifold. Using the above, we can also
show much as in reference \cite{Sethi:1996es}, that $\chi(CY_{4})$ is
divisible by $6$.

\subsection{Special Case: K\"{a}hler Threefolds and Surfaces}

In our discussion of intersecting 7-branes, it is also helpful to recall
some general features of index formulae for a general K\"{a}hler threefold $X$
and a K\"ahler surface $S$. Specializing the index formula to a vector bundle
over each such space, we have:%
\begin{align}
\chi(X,\mathcal{E})=\underset{X}{\int}  &  \left(
\begin{array}
[c]{l}%
\frac{\text{rk}(\mathcal{E)}}{24}(c_{1}(X)c_{2}(X))+\frac{1}{12}%
c_{1}(\mathcal{E)}(c_{1}^{2}(X)+c_{2}(X))\\
+\frac{1}{4}\left(  c_{1}^{2}(\mathcal{E})-2c_{2}(\mathcal{E})\right)
c_{1}(X\mathcal{)+}\frac{1}{3!}\left(  c_{1}^{3}(\mathcal{E})-3c_{2}%
(\mathcal{E})c_{1}(\mathcal{E})+3c_{3}(\mathcal{E})\right)
\end{array}
\right) \\
\chi(S,\mathcal{E})  &  =\underset{S}{\int}\left(  \frac{\text{rk}%
(\mathcal{E)}}{12}(c_{1}^{2}(S)+c_{2}(S))+\frac{1}{2}c_{1}(\mathcal{E)}%
c_{1}(S\mathcal{)+}\frac{1}{2}\left(  c_{1}^{2}(\mathcal{E})-2c_{2}%
(\mathcal{E})\right)  \right)  .
\end{align}

\section{Normalizing the Green-Schwarz Contribution \label{app:NORM}}

In our discussion of compactifications of the perturbative type I\ and
heterotic string theory, we saw that the number of spacetime filling
1-branes is controlled by the contribution from the Green-Schwarz term:%
\begin{equation}
S_{eff}\supset{{\boldsymbol{a}}}\int B_{2}\wedge X_{8}(F,R).
\end{equation}
In this Appendix we present a general argument for fixing the overall
normalization of this term.

We recall the basic story of the gauge anomaly. Consider a Weyl fermion in
$d=2n$-dimensions coupled in a representation ${{\boldsymbol{r}}}$ to a
background Yang-Mills gauge field with field strength $F$.\footnote{We continue to work with anti-Hermitian generators, so that, for instance,
the Chern-Simons three-form below has no factor of $i=\sqrt{-1}$.} The gauge
anomaly is encoded in the transformation of the one-loop effective action
$W[A]$: under $\delta_{v}A=-Dv=-(dv+{[A,v]})$ the change in the effective
action is given by~\cite{AlvarezGaume:1984dr}:%
\begin{equation}
\delta_{v}W[A]=-\int\operatorname{Tr}vD_{\mu}\frac{\delta W[A]}{\delta A_{\mu
}}=\frac{i^{n}}{(2\pi)^{n}(n+1)!}\int Q_{2n}^{1}(v,A)~.
\end{equation}
The local quantity $Q_{2n}^{1}$ is fixed by descent in the familiar way. Given
the anomaly polynomial $I_{2n+2}^{W}$, we have the local relations%
\begin{align}
I_{2n+2}^{W}  &  =dQ_{2n+1}~,\\
\delta_{v}Q_{2n+1}  &  =dQ_{2n}^{1}~
\end{align}
So, as far as the gauge anomaly goes, to fix the normalization we just need to
specify $I_{2n+2}^{W}$. Fortunately, we know this:%
\begin{equation}
I_{2n+2}^{W}=\operatorname{tr}_{{{\boldsymbol{r}}}}F^{n+1}~.
\end{equation}

We now specialize to the case of a Majorana-Weyl fermion in ten spacetime
dimensions. We have just one thing to do in this case: multiply the Weyl
answer by $1/2$. Thus,%
\begin{equation}
I_{12}^{MW}=\frac{1}{2}\operatorname{tr}_{{{\boldsymbol{r}}}}F^{6}%
\end{equation}
In our notation above, a Majorana-Weyl fermion in the adjoint representation
has%
\begin{equation}
I_{12}^{MW}=\frac{1}{2}\operatorname{Tr}F^{6}%
\end{equation}
This will therefore lead to a variation of the effective action by%
\begin{equation}
\delta_{v}W[A]=\frac{1}{2}\frac{i}{(2\pi)^{5}6!}\int Q_{5}^{1}(v,A)~.
\end{equation}

We now compare with $I_{12}(R,F)$, the anomaly polynomial of the 10D heterotic supergravity theory.
Setting $R=0$ yields:
\begin{align}
I_{12}(R  &  =0,F)=-\frac{1}{90}\operatorname{Tr}F^{2}\operatorname{Tr}%
F^{4}+\frac{1}{27000}\left(  \operatorname{Tr}F^{2}\right)  ^{3}~\\
&  =-\frac{48}{90}\left[  \frac{1}{48}\operatorname{Tr}F^{2}\operatorname{Tr}%
F^{4}-\frac{1}{14400}\left(  \operatorname{Tr}F^{2}\right)  ^{3}\right] \\
&  =-\frac{8}{15}\operatorname{Tr}F^{6}~.
\end{align}
In the last line we used the factorization condition%
\begin{equation}
\operatorname{Tr}F^{6}=\frac{1}{48}\operatorname{Tr}F^{2}\operatorname{Tr}%
F^{4}-\frac{1}{14400}\left(  \operatorname{Tr}F^{2}\right)  ^{3}~.
\end{equation}

Let ${{\widehat{I}}}_{12}$ be the correctly normalized polynomial. By this we
mean that the ${{\widehat{Q}}}_{10}^{1}$ constructed from ${{\widehat{I}}}$ by
descent shows up in the variation of the effective
action with constant:
\begin{equation}
\delta W=\frac{i}{(2\pi)^{5}6!}\int{{\widehat{Q}}}_{10}^{1}~.
\end{equation}
From our comparison we see that ${{\widehat{I}}}_{12}=-\frac{15}{16}I_{12}$,
so that we also have ${{\widehat{Q}}}_{10}^{1}=-\frac{15}{16}Q_{10}^{1}$, and
therefore%
\begin{equation}
\delta W=\frac{i}{(2\pi)^{5}6!}\left(  -\frac{15}{16}\right)  \int Q_{10}%
^{1}~.
\end{equation}
We also know that factorization allows us to write $Q_{10}^{1}$ in a simple
way:%
\begin{equation}
Q_{10}^{1}=Q_{2}^{1}X_{8}~,
\end{equation}
where~\footnote{Here ${{\epsilon}}$ is the gauge parameter, $v$ is the Lorentz
parameter, and $\omega$ is the spin connection.}%
\begin{equation}
Q_{2}^{1}=-\frac{1}{30}\operatorname{Tr}{{\epsilon}}dA+\operatorname{tr}%
vd\omega~.
\end{equation}

The final piece of information we need is that if the Euclidean worldsheet
has the coupling%
\begin{equation}
S_{\text{string}}\supset\frac{i}{2\pi\alpha^{\prime}}\int\phi^{\ast}(B)~,
\end{equation}
then cancellation of worldsheet anomalies requires that we set%
\begin{equation}
\delta B=\frac{\alpha^{\prime}}{4}\left[  \operatorname{tr}vd\omega-\frac
{1}{30}\operatorname{Tr}{{\epsilon}}dA\right]  =\frac{\alpha^{\prime}}{4}%
Q_{2}^{1}~.
\end{equation}
So, we now see that the full one-loop effective action that includes Green-Schwarz term
has gauge variation%
\begin{equation}
\delta W=\left[  \frac{i}{(2\pi)^{5}6!}\left(  -\frac{15}{16}\right)
+{{\boldsymbol{a}}}\frac{\alpha^{\prime}}{4}\right]  \int Q_{2}^{1}X_{8}~.
\end{equation}
Gauge invariance thus fixes the constant ${{\boldsymbol{a}}}$ to be%
\begin{equation}
{{\boldsymbol{a}}}=\frac{i}{2\pi\alpha^{\prime}}\frac{1}{192(2\pi)^{4}}~.
\end{equation}
We observe from our studies above that in every case case except for the
irreducible $\operatorname{E}_{8}\times\operatorname{E}_{8}$ case%
\begin{equation}
\frac{1}{192(2\pi)^{4}}\underset{CY_{4}}{\int}X_{8}\in{\mathbb{Z}}~.
\end{equation}
Placing $N$ fundamental space-filling strings will lead to the additional
two-dimensional coupling (in Euclidean signature)%
\begin{equation}
\frac{iN}{2\pi\alpha^{\prime}}\underset{2D}{\int}B~.
\end{equation}
So, to cancel the tadpole we will need:%
\begin{equation}
N=-\frac{1}{192(2\pi)^{4}}\underset{CY_{4}}{\int}X_{8}~.
\end{equation}
To preserve supersymmetry we need $N\geq0$, and therefore%
\begin{equation}
\frac{1}{192(2\pi)^{4}}\underset{CY_{4}}{\int}X_{8}\leq0~.
\end{equation}
This is satisfied in many but not all cases.

\newpage

\bibliographystyle{utphys}
\bibliography{2Dcpct}

\end{document}